\documentclass[usegraphicx,usenatbib]{mn2e}
\usepackage{amsmath}
\usepackage{amssymb}
\usepackage{mathptmx}
\usepackage{txfonts}
\usepackage[T1]{fontenc}
\usepackage{ae,aecompl}
\usepackage{times}

\usepackage{graphicx}	
\usepackage{amsmath}	
\usepackage{amssymb}	
\usepackage{times}
\title[The complex, dusty narrow-line region of NGC 4388]{The complex, dusty narrow-line region of NGC 4388: Gas-jet interactions, outflows, and extinction revealed by near-IR spectroscopy}

\author[Rodr\'iguez-Ardila et al.]{A. Rodr\'iguez-Ardila$^{1}$\thanks{E-mail:
aardila@lna.br}; R. E. Mason$^{2}$; L. Martins$^{3}$; C. Ramos Almeida$^{4,16}$; R. A. Riffel$^{5}$;
\newauthor
R. Riffel$^{6}$; P. Lira$^{7}$; O. Gonz\'alez Mart\'{\i}n$^{8}$; N. Z. Dametto$^{6}$; H. Flohic$^{9}$; L. C. Ho$^{10,11}$;
\newauthor
D. Ruschel-Dutra$^{6}$; K. Thanjavur$^{12}$; L. Colina$^{13}$; R. M. McDermid$^{14,15}$; E. Perlman$^{15}$; 
\newauthor
C. Winge$^{2}$; \\
$^{1}$Laborat\'orio Nacional de Astrof\'isica/MCTI, Rua dos Estados Unidos, 154, Bairro das Na\c c\~oes, Itajub\'a, MG, Brazil\\
$^{2}$Gemini Observatory, Hawaii, Northern Operations Center, 670 North A?ohoku Place, Hilo, HI 96720, USA.\\
$^{3}$NAT-Universidade Cruzeiro do Sul, Rua Galv\~ao Bueno, 868, S\~ao Paulo, SP, Brazil\\
$^{4}$Instituto de Astrof\'isica de Canarias, Calle V\'{\i}a L\'actea, s/n, E-38205, La Laguna, Tenerife, Spain\\
$^{5}$Universidade Federal de Santa Maria, Departamento de F\'isica/CCNE, 97105-900, Santa Maria, RS, Brazil\\
$^{6}$Universidade Federal do Rio Grande do Sul, Instituto de F\'{\i}sica, CP 15051, Porto Alegre 91501-970, RS, Brazil\\
$^{7}$Departamento de Astronom\'ia, Universidad de Chile, Casilla 36-D, Santiago, Chile\\
$^{8}$Instituto de Radioastronom\'ia y Astrof\'isica (IRAF-UNAM), 3-72 (Xangari), 8701,
Morelia, Mexico\\
$^{9}$University of the Pacific, Department of Physics, 3601 Pacific Avenue, Stockton, CA 95211, USA\\
$^{10}$Kavli Institute for Astronomy and Astrophysics, Peking University, Beijing 100871, China\\
$^{11}$Department of Astronomy, Peking University, Beijing, China\\
$^{12}$Department of Physics \& Astronomy, University of Victoria, Victoria, BC, V8W 2Y2, Canada\\
$^{13}$Centro de Astrobiolog\'ia (CAB, CSIC-INTA), Carretera de Ajalvir, 28850 Torrej\'on de Ardoz, Madrid, Spain \\
$^{14}$Department of Physics and Astronomy, Macquarie University, Sydney NSW 2109, Australia;\\ 
$^{15}$Department of Physics \& Space Sciences, Florida Institute of Technology, 150 West University Boulevard, Melbourne, FL 32901, USA\\
$^{16}$Departamento de Astrof\'isica, Universidad de La Laguna, E-38205, La Laguna, Tenerife, Spain
}

\date{Accepted 2016 October 12. Received 2016 September 28; in original form 2015 September 26}

\pubyear{2016}

\begin{document}
\label{firstpage}
\pagerange{\pageref{firstpage}--\pageref{lastpage}}
\maketitle

\begin{abstract}
We present Gemini/GNIRS spectroscopy of the Seyfert~2 galaxy NGC\,4388, 
with simultaneous coverage from 0.85 - 2.5 $\mu$m. Several spatially-extended 
emission lines are detected for the first time, both in the obscured and unobscured portion of 
the optical narrow line region (NLR), allowing us to assess 
the combined effects of the central continuum source, outflowing gas and shocks generated 
by the radio jet on the central 280~pc gas.  The H\,{\sc i} and 
[Fe\,{\sc ii}] lines allow us to map the extinction affecting the 
NLR. We found that the nuclear region is heavily obscured, with E(B-V) $\sim$1.9~mag. 
To the NE of the nucleus and up to 
$\sim$150~pc, the extinction remains large, $\sim$1 mag or larger,
consistent with the system of dust lanes seen in optical imaging. 
We derived position-velocity diagrams for the most prominent
lines as well as for the stellar component. Only the molecular gas and the stellar
component display a well-organized pattern consistent with disk rotation. Other emission
lines are kinematically perturbed or show little evidence of rotation. Extended 
high-ionization emission of sulfur, silicon and calcium is observed to distances of at 
least 200~pc both NE and SW of the nucleus. We compared flux ratios between these lines with
photoionization models and conclude 
that radiation from the central source alone cannot explain the observed high-ionization 
spectrum. Shocks between the radio-jet and the ambient gas are very likely 
an additional source of excitation.  We conclude that NGC~4388 is a prime laboratory 
to study the interplay between all these mechanisms.
\end{abstract}




\section{Introduction}

Active galactic nuclei (AGNs) commonly have outflows and jets, and these structures/phenomena 
may strongly influence the surroundings of the AGN. 
While direct jet-gas interactions affect the velocity field of the local medium, 
they also produce fast, auto-ionizing shocks which can significantly influence (or even dominate) 
the observed emission-line strengths and kinematics. In nearby AGN, the warm (T$\sim 10^4$\, K), 
ionized gas in the narrow-line region (NLR) can be resolved on scales of a few tens of parsecs in 
the optical and near-infrared (NIR). They are therefore useful laboratories for determining the 
extent and kinematics of the various species in the gas, and the role of shocks in producing the 
integrated emission-line spectrum \citep{Emonts05,ROA06,Morganti13}.

Because extinction by dust is lower by a factor of $\sim$10 relative to the optical, NIR spectroscopy 
allows us to probe depths unreachable at shorter 
wavelengths. Seyfert\, 2 galaxies are 
the preferred targets as the line of sight to the nucleus is blocked by the intervening 
dust and molecular material, allowing the study of the environment of the AGN
without the effect of dilution caused by the bright central source.
 
In this context, NGC\,4388, a highly inclined \citep[$i~\sim 78\degr$;]{Veilleux99} 
Seyfert\,2 galaxy in the Virgo cluster \citep{Helou+}, is a prime 
target for studying possible feedback effects of the supermassive black hole (SMBH) on the distribution 
and kinematics of ionized gas and stars. To start with, 
it was one of the first galaxies in which a conically-shaped NLR
was detected \citep{Pogge88,Yoshida02}. WFPC2/HST observations reported by \citet{Schmitt+}
in the [O\,{\sc iii}] filter confirmed the V-shaped NLR with opening angle of 90$\degr$ 
toward the south, extended over 
560~pc in this direction. In the perpendicular direction, [O\,{\sc iii}]  was detected over 720~pc. 
Most of the [O\,{\sc iii}] emission comes from regions south of the nucleus, except for some emission corresponding 
to the counter-cone, which is obscured by the host galaxy, at the NE side of the nucleus.  
Galactic-scale outflows as well as a 
rich complex of highly ionized gas that extends $\sim$4\,kpc above the disk were detected 
by \citet{Veilleux99}. \citet{Matt+}  detected soft X-ray emission extending 
over 4.5\,kpc in observations with ROSAT. Later, \citet{Iwasawa+} and \citet{Bianchi+} 
using Chandra, found that the soft X-ray emission is coincident in extent and overall 
morphology with [O\,{\sc iii}]~$\lambda$5007~\AA. In the radio, \citet{Stone+} and 
\citet{Hummel+} found that NGC\,4388 is double-peaked with a primary peak on the nucleus 
and a secondary peak 230~pc southwest of it. They also report a plume of radio 
plasma to the north of the optical nucleus. The good overall match between optical emission-line and radio 
morphology reported by \citet{Falcke+}  led them
to suggest that NGC\,4388 is an example of an interaction between a radio jet and ambient gas. 
  
An additional property that makes NGC\,4388 interesting is the presence of water maser
emission from a circumnuclear disk, allowing accurate measurement of the mass of its SMBH. 
\citet{Kuo+}, using VLBI, studied the kinematics of the 
water maser emission and derived a BH mass of 8.5$\pm~0.2\times 10^6$~M$\odot$. Moreover, \citet{Greene13}
found a stellar nuclear disk with PA of 75$\degr$, radius of $\sim$200~pc, and a
100~pc-scale jet oriented at a PA of 24$\degr$.

In the NIR, NGC\,4388 has not been studied very extensively, with most 
spectroscopic studies reported in the $K-$ and $J-$bands only 
\citep{Winge00,Knop+,Lutz+,IAH04,vlaan+,Greene14}. They all indicate 
that NGC\,4388 is a complicated system with much spatial structure at all wavelengths. Very recently, 
\citet{Greene14} found, by means of $K-$band IFU data, a drop in 
the stellar velocity dispersion in the inner $\sim$100~pc, interpreted as the signature of a 
dynamically cold central component. They also report [Si\,{\sc vi}]~1.963$\mu$m and Br$\gamma$ 
oriented at PA$\sim30\degr$, aligned with the jet on similar scales, and also with the 
[O\,{\sc iii}] emission that traces the narrow line region.

Because of the mounting evidence of jet-gas interactions and
rich circumnuclear structures/environment in NGC\,4388, we are interested in studying 
the inner 500~pc of this source to more closely examine the physical conditions of the atomic, molecular and 
ionized gas.  Our aim is to carry out, for the first time, a simultaneous analysis of the spectral 
region 0.84$-2.5~\mu$m. It includes the wavelength
intervals where the brightest NIR lines in AGNs are located (i.e., [S\,{\sc iii}]~0.953~$\mu$m, 
He\,{\sc i}~1.083~$\mu$m, [Si\,{\sc vii}]~2.483~$\mu$m), and these lines have not yet been observed in NGC~4388.
We perform detailed 1D mapping along PA = 64$\degr$ of the most relevant 
NIR spectroscopic properties, searching for features of the morphology and gas kinematics
that will help us to understand the origin and nature of the nuclear/circumnuclear gas, 
and its role within the AGN and outflows already detected in this source.

This paper is structured as follows. In Sect~\ref{sec:obs} we describe the observations
and data reduction. In Sect.~\ref{sec:nir_spec} we describe the most important NIR features
detected in the spectra, analyses the main spectral (line and continuum) properties
including the extinction affecting the gas and the main excitation mechanisms that 
produce the observed lines. 
Sect~\ref{sec:kinematics} discusses the kinematics of the neutral, low, medium and high ionization 
gas as well as the kinematics of the stellar component. Sect~\ref{sec:coronal} deals with the main 
ionization mechanisms of the high-excitation gas. Sect~\ref{sec:final}
contains the main conclusions found from our work. Throughout this paper we
adopt the Tully-Fisher distance to NGC\,4388 of 19~Mpc \citep{Kuo+}, which translates
into a spatial scale of 92 pc/\arcsec. Emission lines with wavelengths shortwards of 
1~$\mu$m will be quoted in Angstroms (\AA) while those longwards of that value will be quoted
in microns ($\mu$m).

 \section[]{Observations and Data Reduction} 
 \label{sec:obs}

\begin{figure*}
\includegraphics[scale=0.46]{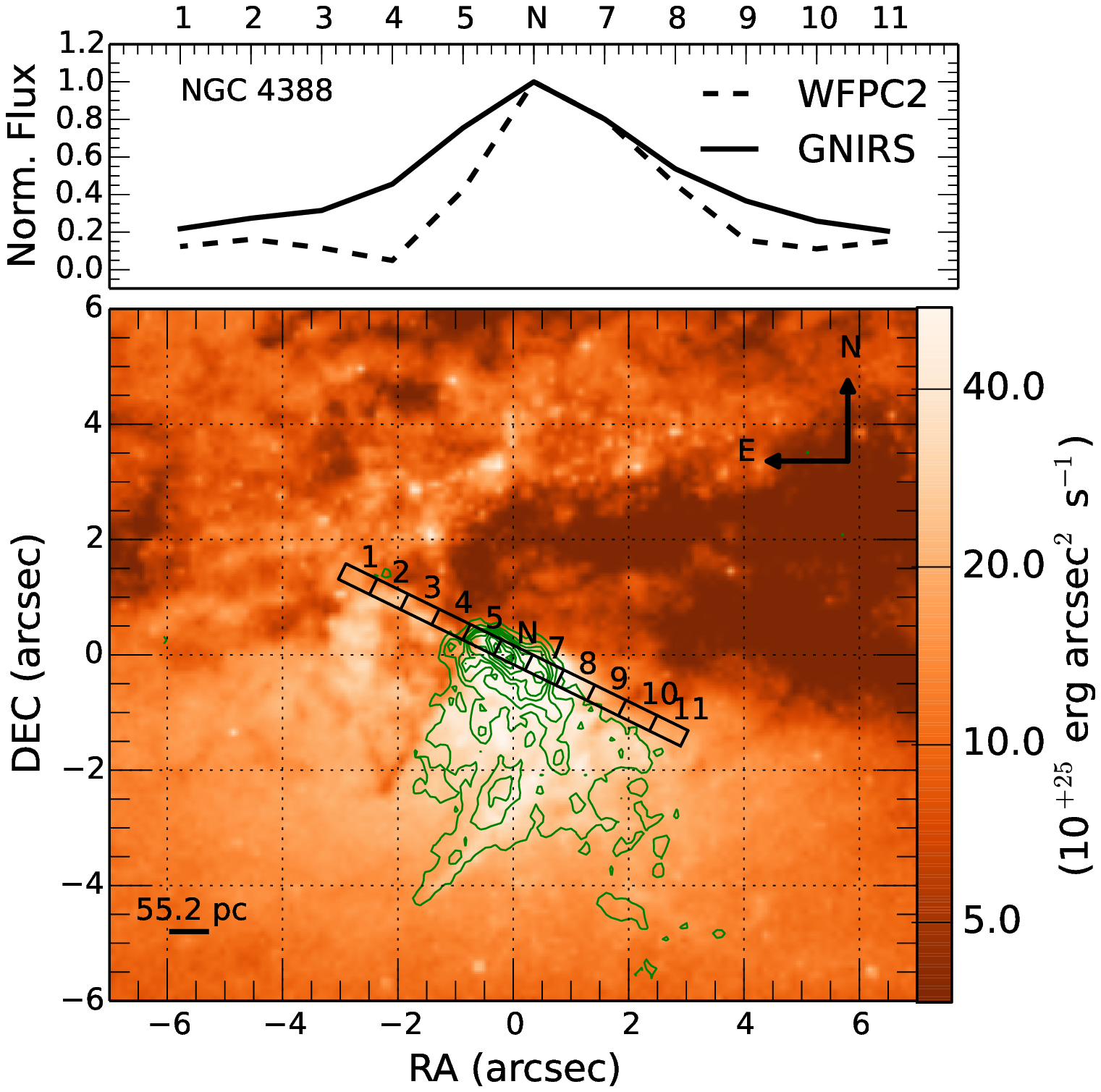}
\includegraphics[scale=0.37]{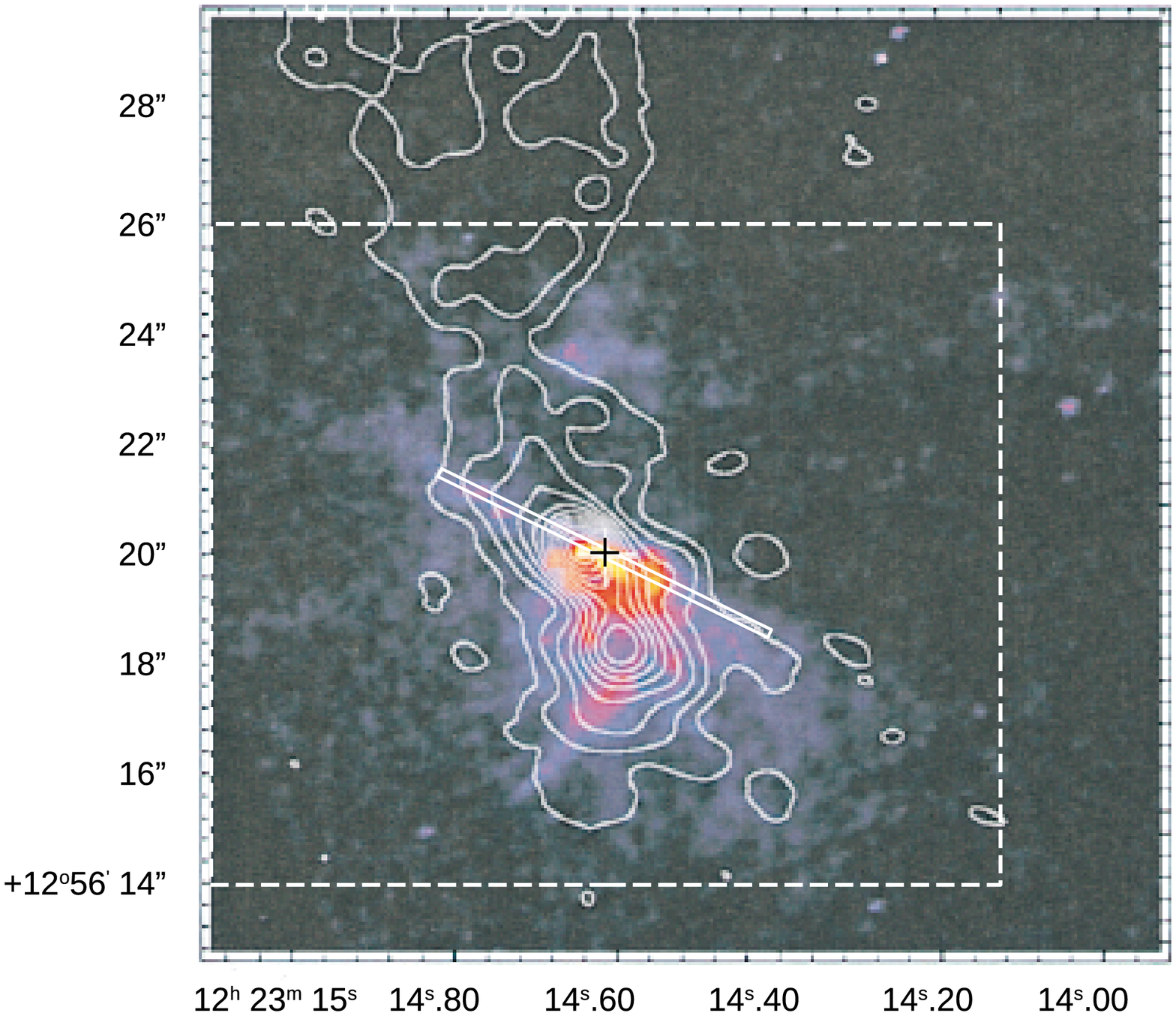}
\caption{Left: Continuum WFPC2/HST in the F606W filter overlaid to [O\,{\sc iii}]\,$\lambda$5007
contours (described in Falcke et al. 1998). The GNIRS slit is positioned at an angle
of PA=64$\degr$ and illustrates approximately the region covered by
the NIR spectra. ``N'' marks the position of the nuclear aperture while the numbers 1 to 5 
and 7 to 11 denote the off-nuclear extractions. The two curves on top show the light profile
along the spatial direction of the slit for the GNIRS spectrum (black solid
line) and in the [O\,{\sc iii}] image at the same position angle as the slit
(dashed line). North is up and East is to the left. The spatial scale is
indicated by the bar at the bottom left corner and represents the size
of the aperture window used in extracting the spectra. Right: H$\alpha$ image
from WFPC2/HST overlaid to radio contours as originally published by
\citet{Falcke+}. The thick white box represents the GNRIS slit while the dashed
square marks approximately the field covered by the image to the left. 
The black cross is the position of the optical nucleus.}
\label{fig:slitpos}
\end{figure*}

NGC\,4388 was observed as part of a set of NIR spectroscopic observations of galaxies from 
the Palomar nearby galaxy survey \citep{Ho97,Mason15}. 
The spectra were obtained using the 
cross-dispersed mode of the Gemini Near-Infrared Spectrograph
(GNIRS) on the Gemini North 8.1 m telescope\footnote{Program ID: GN-2013A-Q-16}. This configuration provides
a continuous spectral coverage from $\sim$8400~\AA\ to 2.48 $\mu$m at
a spectral resolution of $\sim$1200 with a spatial scale of 0.15\arcsec/pixel.
The 0.3$\arcsec \times 7\arcsec$ slit was set at a position angle PA = 64$\degr$ 
east of north and centered on the peak of the 1.6 $\mu$m emission.
The seeing during the galaxy observation was 0.6\arcsec\ as measured
from the telluric A1V standard HIP\,58616, observed right before the
galaxy at a similar airmass. Left panel of Figure~\ref{fig:slitpos} 
shows the slit position overlaid on the WFPC/HST F606W (5935\,\AA) image of
NGC\,4388. The contours correspond to WFPC/HST observations
of [O\,{\sc iii}]~$\lambda$5007, described in \citet{Falcke+}.

The observations used an object-sky-sky-object
pattern, with the sky position 50\arcsec\ away from the
galaxy nucleus, free of extended emission or background stars. Four
individual on-source integrations of 240~s each were carried out. 

The spectral reduction, extraction, and wavelength and flux calibration 
procedures were performed using version 1.9 of the ``XDGNIRS'' code detailed in
\citet{Mason15}. Briefly, the processing consists of removing cosmic ray-like features, 
dividing by a flat field, subtracting sky emission, and rectifying the tilted, curved spectra.
Wavelength calibration is achieved using argon arc spectra, and then a spectrum of each order 
is extracted, divided by a standard star to cancel telluric absorption lines, and roughly 
flux-calibrated using the telluric standard star spectrum. The pipeline merges the different 
spectral orders for each extraction
window into a single 1D spectrum from 0.84~$\mu$m to 2.48~$\mu$m. 
In all cases the agreement in flux between the overlapping regions of two consecutive
orders was very good, and scaling factors of $<$ 3\%  were necessary.

The brightest emission lines clearly extend all along the slit,
so 10 off-nuclear extractions were made 
in the spatial direction, as shown in Figure~\ref{fig:slitpos}.
``N'' represents the region where the nuclear spectrum was
extracted, centred at the peak of the continuum emission. The labels 1 to 11 
mark the different extractions, with 1 to 5 to the NE and 7 to 11 to the SW. 
The aperture size of the extraction window used in all cases was 
0.6\arcsec, similar to the seeing measured from the telluric standard. 

\begin{figure*}
\includegraphics[width=150mm]{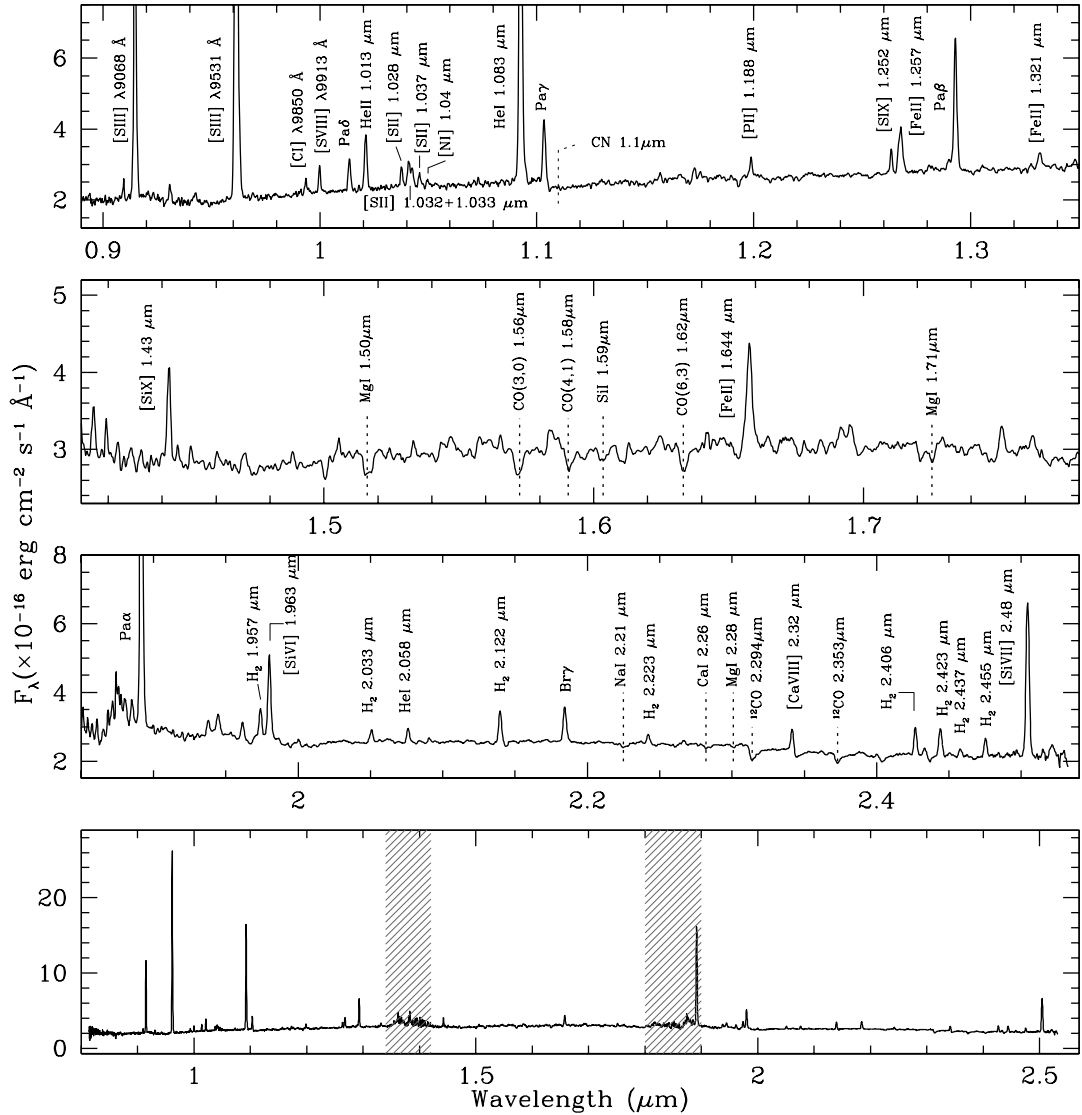}
\caption{Nuclear spectrum of NGC\,4388 in the galaxy frame for the $Y+J$ bands (upper panel),
$H-$band (second panel) and $K-$band (third panel), with the most conspicuous
emission lines identified. The wavelengths of the lines are in the laboratory frame.
Absorption lines/bands are marked by dotted lines. The last pannel 
shows the overall continuum shape and brightest emission features detected in this
source in the wavelength interval covered by GNIRS. The shaded regions
correspond to regions of low atmospheric transmission.}
\label{fig:XD_nuclear}
\end{figure*}

\begin{figure*}
\includegraphics[width=140mm]{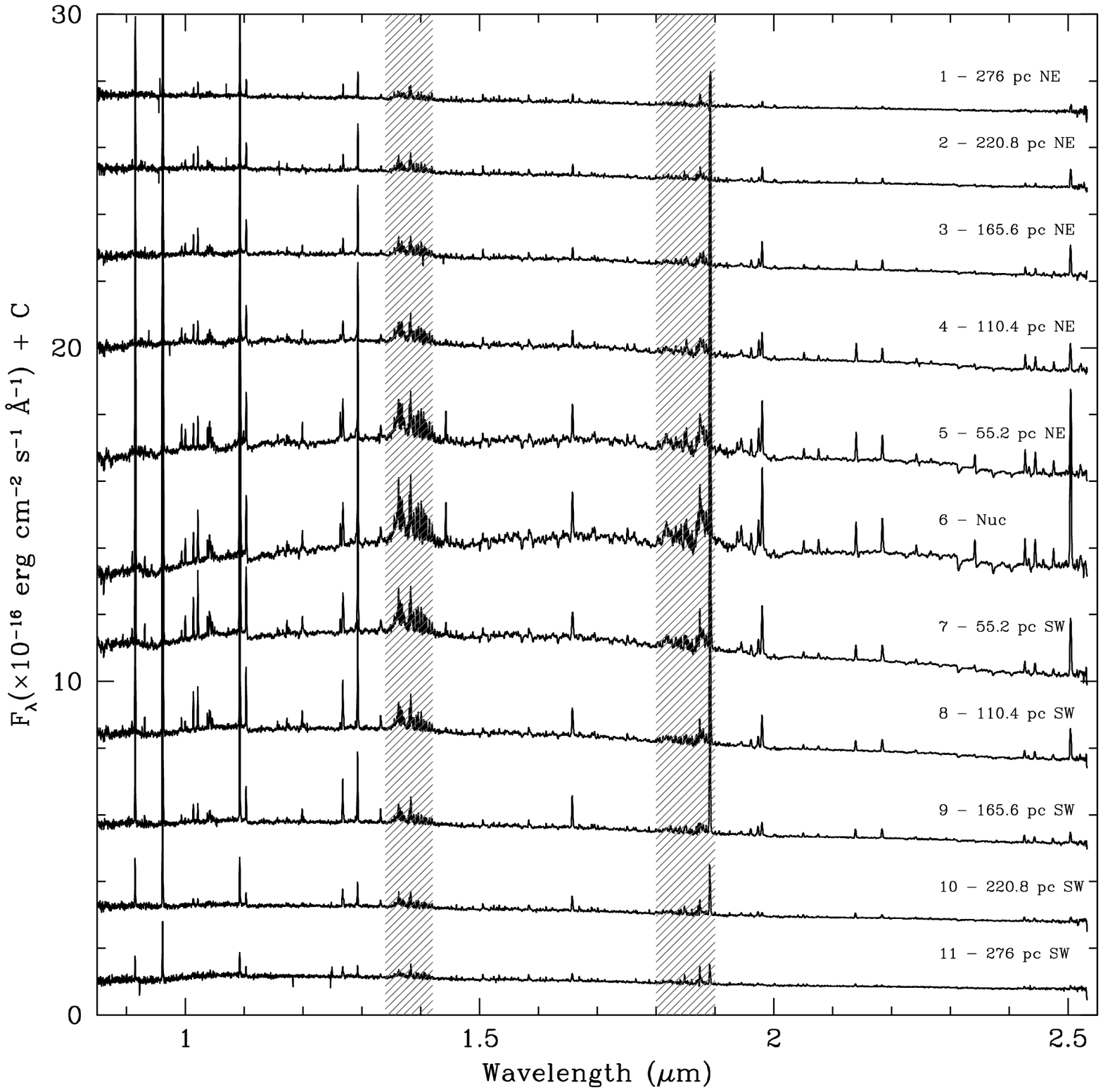}
\caption{Nuclear and off-nuclear spectra extracted for NGC\,4388. The shaded regions
correspond to regions of low atmospheric transmission. The numbers at the
right border correspond to the different appertures identified in Figure~\ref{fig:slitpos}, 
followed by the distance (in parsecs) from the nuclear aperture (denoted by ''NUC'')
to the centre of the corresponding extraction region.}
\label{fig:XD_allspectra}
\end{figure*}

Figure~\ref{fig:XD_nuclear} shows the nuclear spectrum in the galaxy frame 
with the most conspicuous emission lines identified in the laboratory frame. 
Figure~\ref{fig:XD_allspectra} displays all extractions made along the spatial 
direction, also in the galaxy frame.
A Galactic extinction of $A_{\rm v}$ of 0.09 from \citet{Schlegel+} was
found for this source. Because it is negligible, we have not corrected for this
effect.

\section[]{The Near-infrared spectrum of NGC 4388}
\label{sec:nir_spec}

Figures~\ref{fig:XD_nuclear} and~\ref{fig:XD_allspectra} reveal that NGC\,4388 displays an outstanding
emission line spectrum with prominent lines of [S\,{\sc iii}]\,$\lambda\lambda$9068,9531~\AA,
He\,{\sc i}\,1.083~$\mu$m, [Fe\,{\sc ii}]\,1.257$\mu$m, Pa$\beta$, 
H$_2$\,2.122~$\mu$m, [Si\,{\sc vi}]\,1.963~$\mu$m and [Si\,{\sc vii}]\,2.48~$\mu$m. 
These lines all extend from the nucleus to the NE and SW ends of the slit.
Extended emission of [Fe\,{\sc ii}], Pa$\beta$,
H$_2$\,2.122~$\mu$m  and [Si\,{\sc vi}] have previously been reported 
for this AGN \citep{Knop+,vlaan+,Greene14}. Our data, however, reveal that 
not only are those lines spatially resolved but also [S\,{\sc iii}], [C\,{\sc i}],
He\,{\sc i}, [S\,{\sc ix}] and [Si\,{\sc vii}],
which are detected at distances of up to 280~pc NE and SW
of the nucleus. 

The presence of high ionization lines in the nuclear and off-nuclear spectra of 
NGC\,4388 is also evident from Figures~\ref{fig:XD_nuclear} and~\ref{fig:XD_allspectra}. 
[Si\,{\sc vii}]\,2.48~$\mu$m, a coronal line
with IP = 205\,eV is the second brightest forbidden line after [S\,{\sc iii}]\,0.953~$\mu$m. 
In addition to [Si\,{\sc vii}], we also report here first detections of 
[S\,{\sc viii}]\,0.991~$\mu$m, [Si\,{\sc x}]\,1.43~$\mu$m and [Al\,{\sc ix}]\,2.048$\mu$m.
The latter two are rather compact, with no evidence of extended emission. 
[S\,{\sc ix}]\,1.252~$\mu$m and [Ca\,{\sc viii}]\,2.32\,$\mu$m are also detected
in our data in the nuclear and off-nuclear apertures. These lines were previously 
reported by \citet{Knop+} and \citet{Greene14}, respectively.

In addition to the emission lines, the continuum emission of NGC\,4388
displays stellar absorption features of CO and CaT in the extreme
red and blue portions of the NIR spectra, respectively.
Moreover, absorption lines of Mg\,{\sc i}, CO, and Si\,{\sc i} are also evident in the $H$-band. 
The CN band at 1.1\,$\mu$m is prominent in the nuclear and circumnuclear apertures, indicating the 
presence of red giant and/or asymptotic giant branch stars \citep{Riffel15}.

\begin{figure*}

    \includegraphics[scale=0.4]{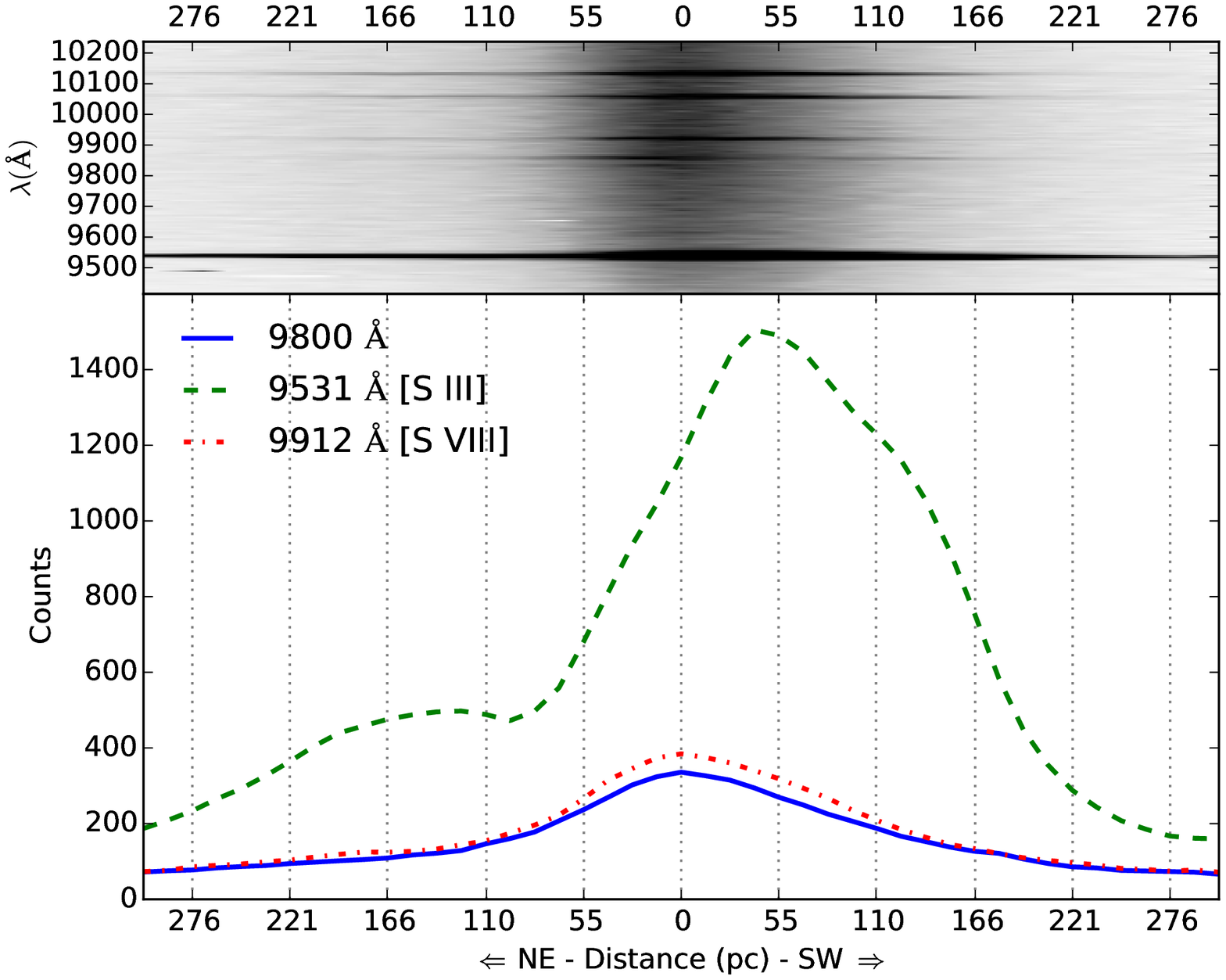}
    \includegraphics[scale=0.4]{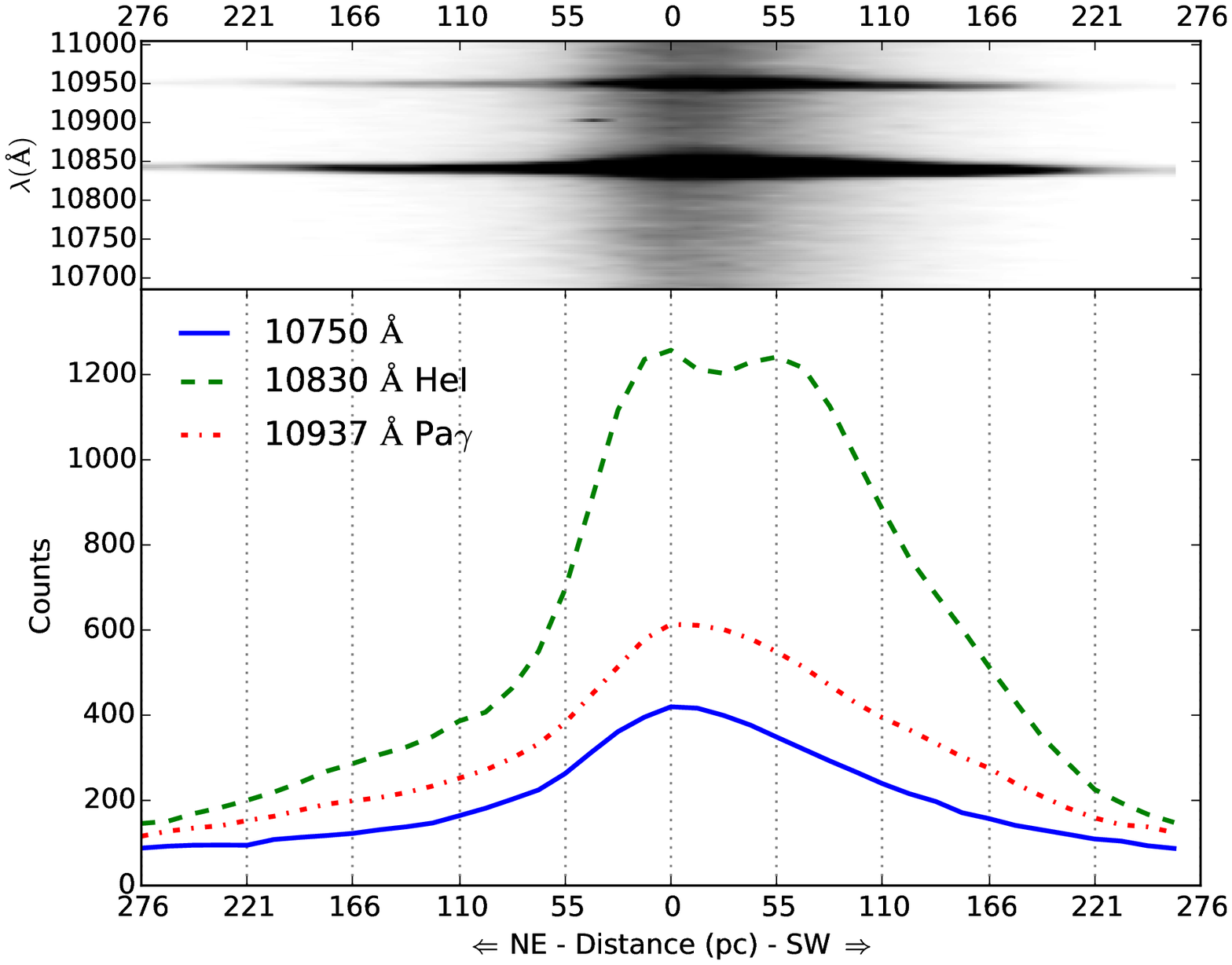}
     \\ 
    \includegraphics[scale=0.4]{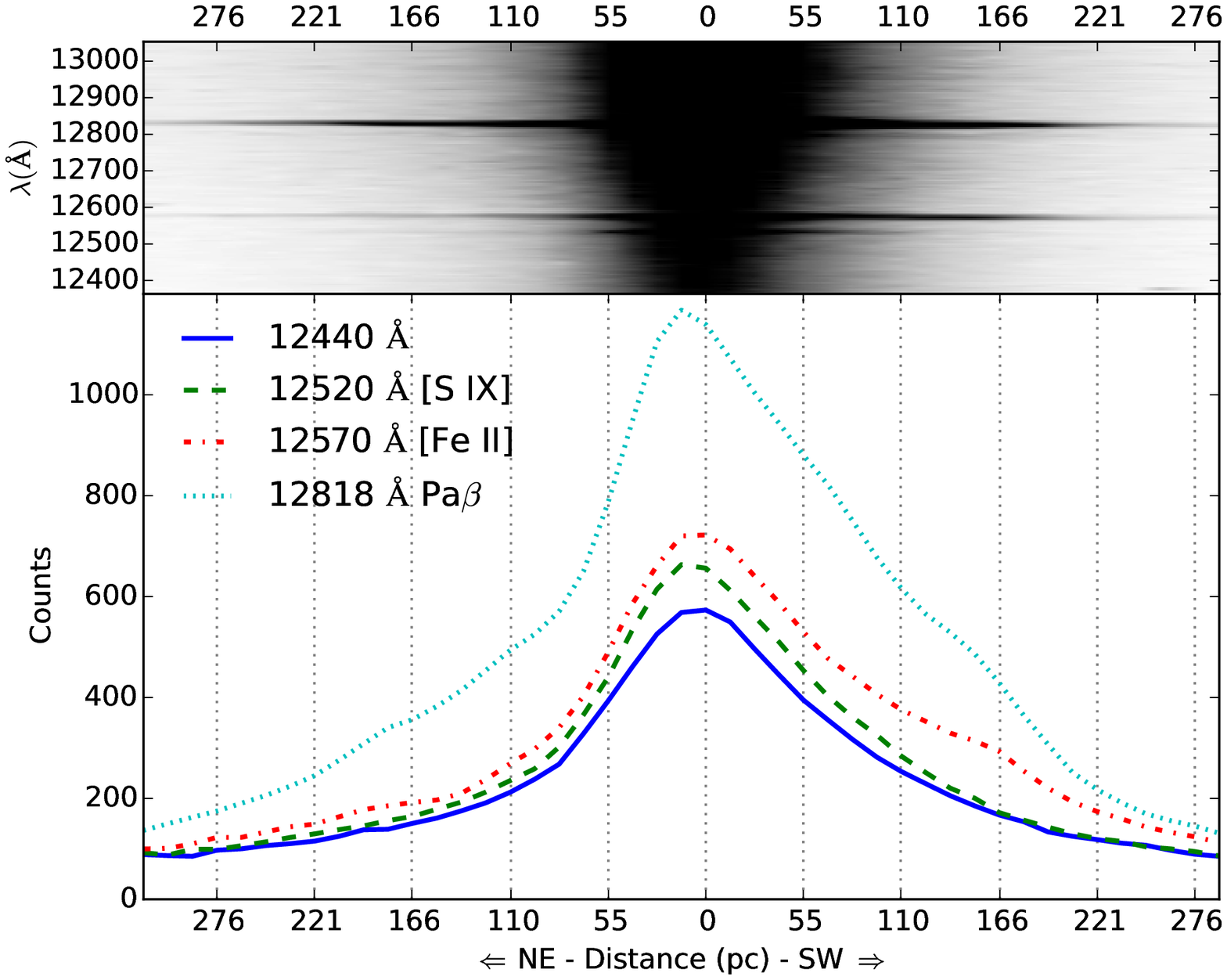}
    \includegraphics[scale=0.4]{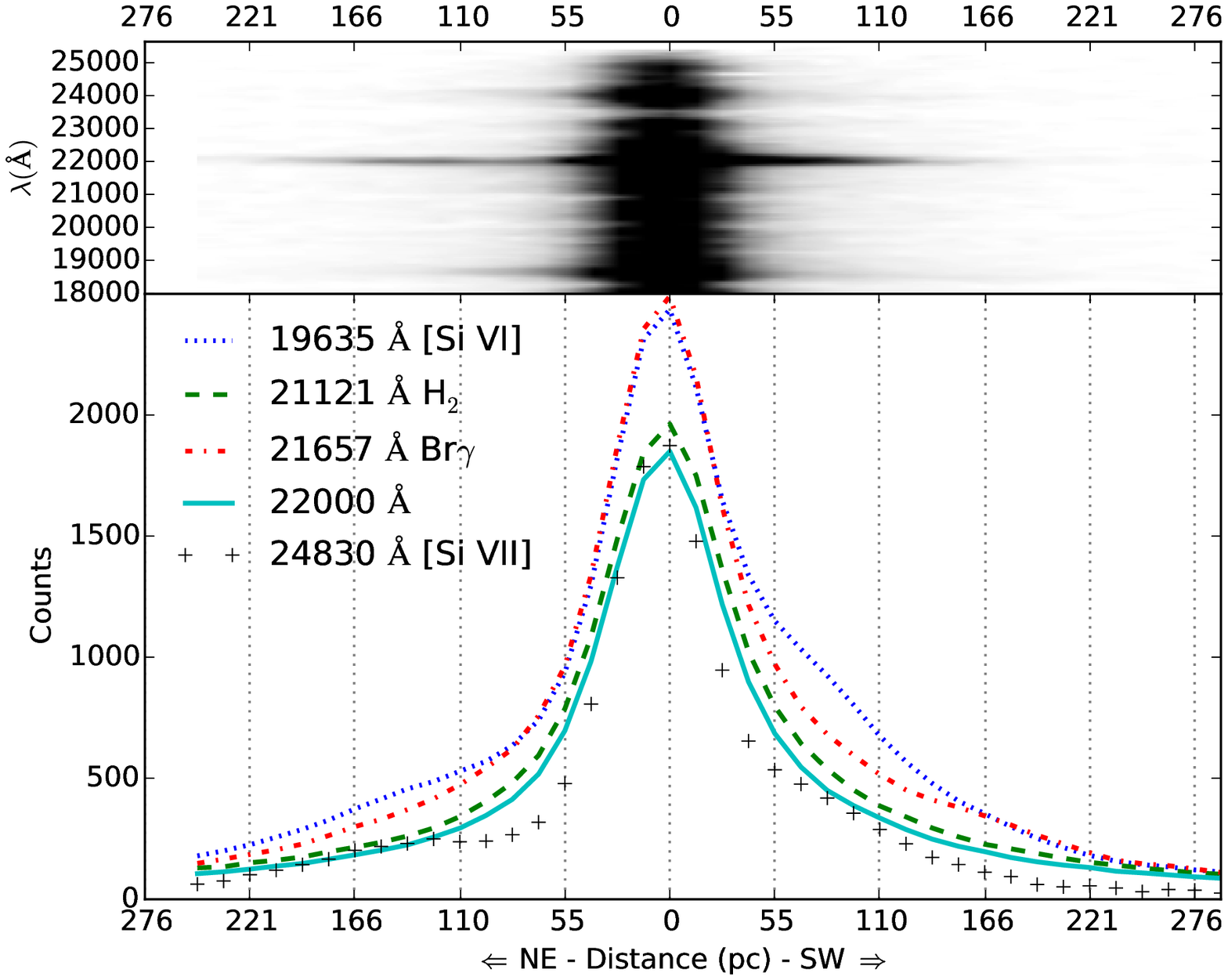}
    \caption{Light distribution across the slit in the 2D data for different emission lines and
continua observed in NGC\,4388. Upper left panel: [S\,{\sc iii}], [S\,{\sc viii}] and
continuum at 9800~\AA. Upper right panel: He\,{\sc i}, Pa$\gamma$ and continuum at 1.07$\mu$m. 
Bottom left panel: [Fe\,{\sc ii}], Pa$\beta$, [S\,{\sc ix}] and continuum at 1.24$\mu$m. 
Bottom right panel: [Si\,{\sc vi}], H$_2$, [Si\,{\sc vii}], Br$\gamma$ and continuum at 2.2$\mu$m.
In all panels, NE is to the left and SW is to the right. The dotted vertical lines in each panel
mark the position of the centre each aperture extracted along the slit.  Notice that the 2D combined 
frames used to produce these plots are not flux-calibrated. Therefore, 
relative intensities between different emission lines are not reliable here.}
\label{fig:perfil_luz}    
\end{figure*}

Figure~\ref{fig:perfil_luz}
displays the light distribution along the spatial direction of the brightest emission lines 
detected in this object. For comparison, the light profile of the continuum at
different wavelengths is also shown. It can be seen that the NLR of NGC\,4388 has a complex, irregular
structure. For most lines, the brightest emission coincides with the peak of the continuum light 
(and we identify this location as the active nucleus). The exception is the [S\,{\sc iii}]\,$\lambda$9531 line, 
which reaches its maximum $\sim$50~pc SW of the continuum peak. In fact, many of the lines show secondary 
peaks $\sim$50~pc SW of the nucleus. Such peaks are particularly prominent in He\,{\sc i}, [Si\,{\sc vi}] and 
[Si\,{\sc vii}]. A third peak is observed at $\sim$150~pc SW from the nucleus, most noticeable 
in Pa$\beta$, [Fe\,{\sc ii}] and [Si\,{\sc vii}] but also detected in other lines. A fourth peak 
of emission at $\sim$150~pc NE of the nucleus is detected, quite prominent in [S\,{\sc iii}] and 
also observed in He\,{\sc i}, H\,{\sc i}, [Fe\,{\sc ii}], [Si\,{\sc vi}] and [Si\,{\sc vii}].

The asymmetry of the line distributions, with excess emission towards the SW compared with the NE, 
is most pronounced at shorter wavelengths. This reflects 
the dusty nature of NGC\,4388, whose inner few hundred parsecs from
the centre to the NE are highly obscured at optical wavelengths. Strong dust lanes crossing 
the nuclear and circumnuclear region are evident in the WFPC/HST images of 
\citet{Schmitt+}. The fact that the NIR region is significantly less affected
by dust allows us to map this hidden region in a variety of emission lines, 
something that has not been possible using optical spectroscopy.

The very rich nuclear and extended emission line spectrum found in the NIR for NGC\,4388 
is not surprising. \citet{Pogge88}, using optical imaging and spectrophotometry, 
described the complexity of the extended, ionized gas clouds surrounding the nucleus, 
reaching distances on the kiloparsec scale, well above the plane of the galaxy. 
When compared to other well-known Seyfert nuclei, NGC\,4388 has one of the most 
richly structured circumnuclear regions yet observed. Our data offer a unique
opportunity to extend and compare previous results on the kinematics and ionization structure
of NGC\,4388 based on molecular and low-ionization gas to the high-ionization gas 
covering a wide range of ionization potentials. 

\subsection{Emission-line fluxes}
\label{lines}

In order to accurately measure weak emission lines, 
it is necessary to remove the stellar continuum. As we only wish to obtain a good representation of 
the stellar spectrum, rather than extract information about the stellar population itself, we fit the 
spectrum with the IRTF library of empirical stellar spectra \citep{rayner+09}. This library contains
0.8 - 5.0~$\micron$ spectra of 210 stars of spectral type F, G, K, M and S/C. We used a subset of 
60 stars, removing similar spectra of stars with the same spectral 
types. To this we added theoretical spectra of hotter stars 
(T$_eff$ = 9000K, 10000K and 20000K, log$g$=3.0 and 4.5 dex, solar metallicity -
calculated as in Coelho 2014, (private communication) which might be important if a younger stellar population is present in the galaxy.
The stellar population modeling was done using the spectral synthesis code 
STARLIGHT \citep{cid+04, cid+05a,cid+05b,asari+07}. In addition to the stars we included a power-law in the form of 
$F_\nu \propto \nu^{-1.5}$ to represent the AGN featureless continuum.
Extinction is modelled by STARLIGHT as due to foreground dust, and parametrized 
by the $V-$band extinction A$_{\rm v}$. We use the \citet{Cardeli+} extinction law. 
To accurately model the stellar continuum, the emission lines are masked out of the fit. 
We also masked the regions of strong telluric absorption, 
where their correction was not possible, and the bluest part of the spectrum 
($\lambda \leq 9300 \AA$), where the flux calibration can be most uncertain.

\begin{figure}
\includegraphics[width=90mm]{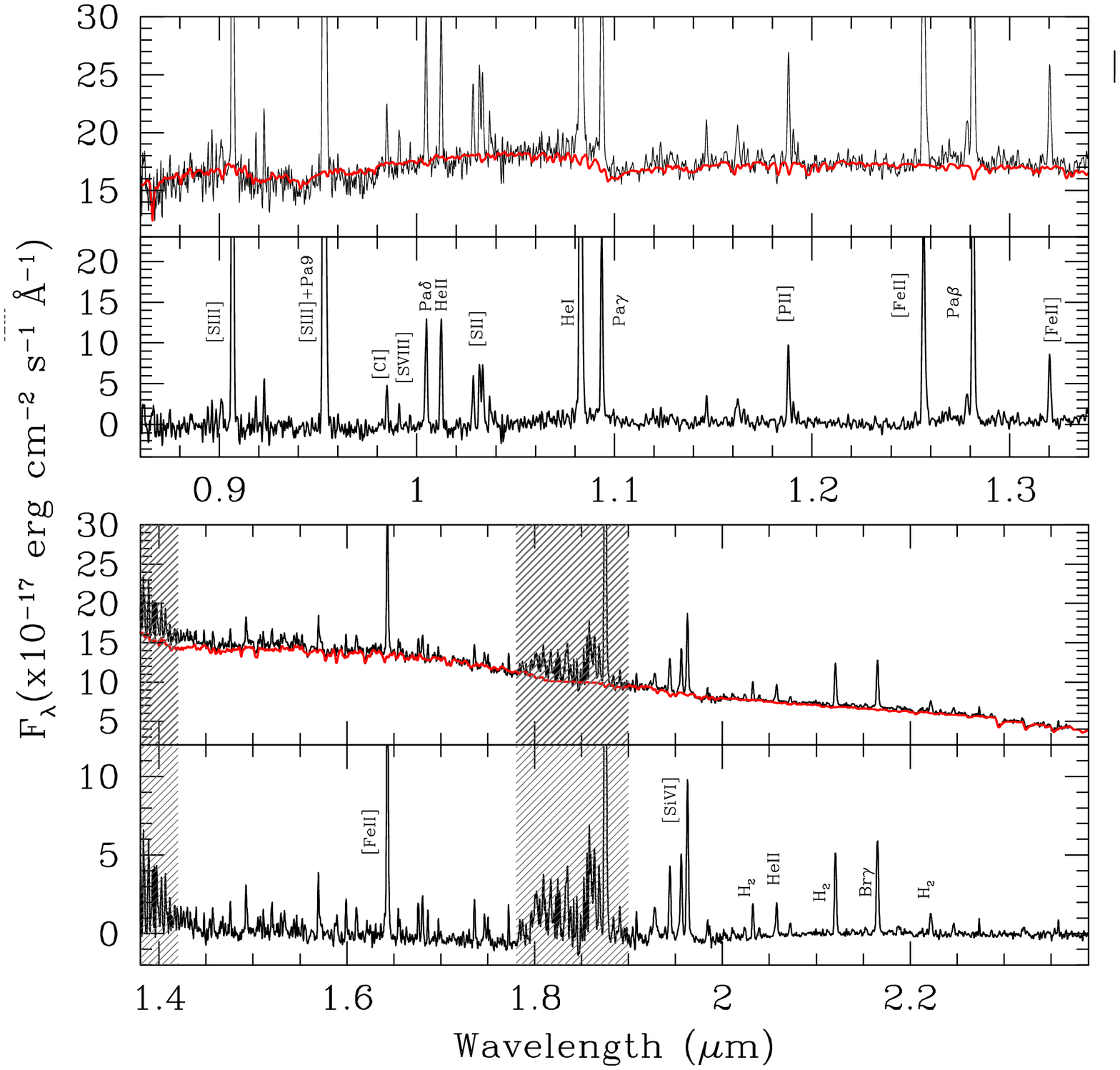}
\caption{Example of the subtraction of the stellar population for the aperture centred at 166~pc SW of the nucleus.  
The observed spectrum (black line) overlaid to the stellar population template (red line) are shown 
followed by the spectrum free of stellar absorption features. The shadow areas correspond to regions
of poor atmospheric transmission. $z+J-$bands are in the upper pannel while $H+K-$bands is in
lower panels.}
\label{fig:substellar}
\end{figure}

Examples of the fit obtained can be seen in Figure~\ref{fig:substellar}, where the 
stellar template derived for the aperture 9, centred at 166~pc SW, can be observed. 
The observed spectrum (black line) is overlaid on the stellar 
population template (red line) followed by the residual spectrum after removing the stellar
population. The most conspicuous nebular emission features are marked.

\begin{figure}
\includegraphics[width=90mm]{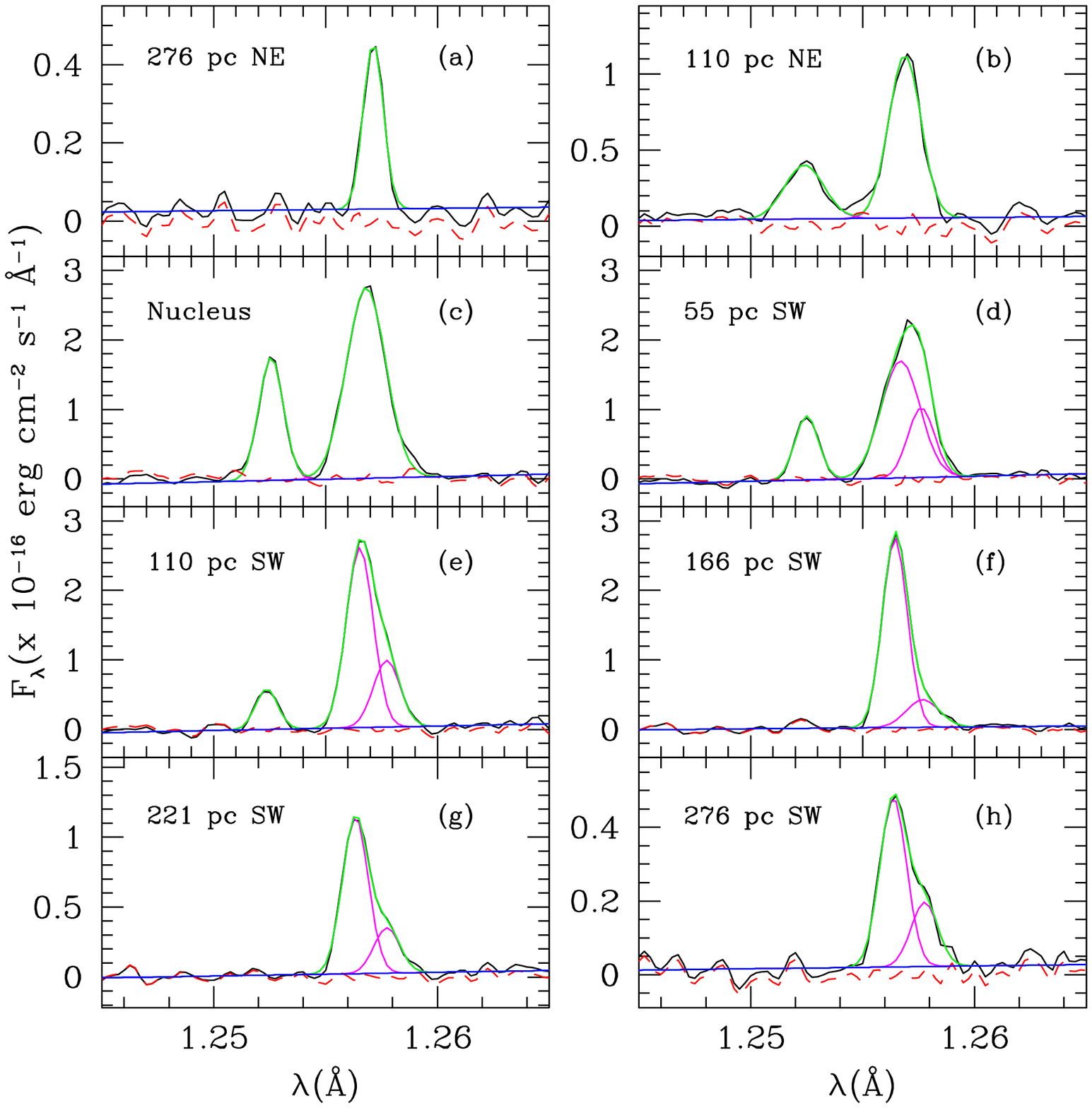}
\caption{Examples of the Gaussian fitting for the [Fe\,{\sc ii}]\,1.257\,$\mu$m line. Panels 
(a) and~(b) display the fitting done at 276~pc and 110~NE of the nucleus. In both apertures a single Gaussian 
was necessary. Panel~(c) shows the fit done in nucleus. Panels~(d) to~(h) show the results for five apertures 
(all to the SW) where two Gaussians were required. In all cases, individual Gaussian components are in magenta. 
The green line shows the total fit. The dashed-red line is the residual after subtrating the fit. The black line 
is the observed data and the local fit to the residual continuum appears in blue. The line to left of 
[Fe\,{\sc ii}]\,1.257\,$\mu$m is [S\,{\sc ix}]~1.252\,$\mu$m.
}
\label{fig:deblending}
\end{figure}

We measured line fluxes on the starlight-subtracted spectra using the LINER routine \citep{Pogge93},
a $\chi$-squared minimization algorithm that can fit simultaneously up to eight profile 
functions to a given line or set of blended lines. In addition, LINER also provides values for the
peak position and the full width at half maximum (FWHM) of each profile function fit.
In NGC\,4388 one Gaussian was necessary to represent the observed 
profiles for most lines. The exception was [Fe\,{\sc ii}]~1.257\,$\mu$m, which required two Gaussians in the extractions
from 55~pc SW from the centre and outwards. Examples of the Gaussian fit for [Fe\,{\sc ii}]~1.257\,$\mu$m
are shown in Figure~\ref{fig:deblending}. The two uppermost panels, labeled (a) and (b), show the fit
done in the apertures centred at 276~pc and 110~pc NE from the nucleus, respectively. Panel~(c) shows the result for
the nucleus. Note that only one component is evident in these spectra. In contrast, panels~(d) to~(h)
display the best fit for the apertures where two Gaussian profiles were necessary to represent the iron line. 
Note the presence of the red component, which carries up to one third of the
total [Fe\,{\sc ii}]~1.257\,$\mu$m flux at some apertures. [S\,{\sc ix}]~1.252\,$\mu$m 
is well-represented by a single Gaussian even at the positions where two components are 
employed for [Fe\,{\sc ii}]~1.257\,$\mu$m (55~pc and 110~pc SW). [Fe\,{\sc ii}]~1.644\,$\mu$m, the second 
most brightest [Fe\,{\sc ii}] line, was fit in all apertures by a single-Gaussian. The lack of
the red component in that line is probably due to the fact that it is intrinsically weaker than
[Fe\,{\sc ii}]~1.257\,$\mu$m (by a factor of $\sim$30\%).

The fluxes of most lines changed little after subtracting the stellar component.  The exceptions 
are [Ca\,{\sc viii}]\,2.322~$\mu$m,
Pa$\gamma$ and He\,{\sc i}\,1.083~$\mu$m. The former is severely affected by the CO bandhead at 
2.324~$\mu$m while the latter two lines sit over the broad CN-band at 1.1$\mu$m. 

The starlight-subtracted, integrated fluxes of the most conspicuous
lines are reported in table~\ref{tab:fluxes}. The errors quoted are 1-$\sigma$ although
a line was considered detected if it was above the 3-$\sigma$ errors of the adjacent continuum. 
Upper limits are 3-$\sigma$ representative.

\begin{table*}  
 \centering
  \caption{Integrated emission line fluxes in units of 10$^{-15}$~erg\,cm$^{-2}$\,s$^{-1}$ measured in 
  the nuclear and off-nuclear apertures in NGC\,4388. \label{tab:fluxes}}
  \begin{tabular}{@{}lcccccccccc@{}}
  \hline
N$^{*}$ & R  & [S\,{\sc iii}] & [C\,{\sc i}] & [S\,{\sc viii}] &  Pa$\delta$ & He\,{\sc ii} & [S\,{\sc ii}] & He\,{\sc i} &  Pa$\gamma$ & [S\,{\sc ix}] \\
 & (pc)  & $\lambda$9531 & $\lambda$9851 & $\lambda$9913 &  1.005~$\mu$m & 1.0124~$\mu$m & 1.032~$\mu$m &  1.083~$\mu$m & 1.0937~$\mu$m & 1.252~$\mu$m \\
 \hline
1 & -276	&	7.21$\pm$0.03	&	$<$0.1		&	0.14$\pm$0.03	&	0.27$\pm$0.03	&	0.40$\pm$0.03	&	0.30$\pm$0.06	&	1.84$\pm$0.03	&	0.59$\pm$0.03 &   $<$0.1	\\
2 & -220.8	&	12.51$\pm$0.04	&	0.14$\pm$0.03	&	0.30$\pm$0.04	&	0.55$\pm$0.03	&	0.75$\pm$0.03	&	0.82$\pm$0.12	&	3.49$\pm$0.03	&	1.01$\pm$0.03 &   0.13$\pm$0.02	\\
3 & -165.6	&	17.53$\pm$0.05	&	0.39$\pm$0.04	&	0.53$\pm$0.03	&	0.95$\pm$0.03	&	1.18$\pm$0.03	&	1.17$\pm$0.15	&	6.26$\pm$0.05	&	1.73$\pm$0.05 &   0.21$\pm$0.02	\\
4 & -110.4	&	21.41$\pm$0.06	&	0.59$\pm$0.05	&	0.65$\pm$0.07	&	1.29$\pm$0.06	&	1.32$\pm$0.06	&	1.75$\pm$0.20	&	11.09$\pm$0.10	&	2.59$\pm$0.10 &   0.55$\pm$0.07	\\
5 & -55.2	&	36.27$\pm$0.11	&	1.33$\pm$0.07	&	1.17$\pm$0.08	&	1.86$\pm$0.09	&	1.88$\pm$0.08	&	4.41$\pm$0.34	&	27.40$\pm$0.13	&	3.64$\pm$0.13 &   1.85$\pm$0.10	\\
6 & 0   	&	66.67$\pm$0.10	&	1.13$\pm$0.08	&	1.91$\pm$0.08	&	2.71$\pm$0.10	&	3.89$\pm$0.10	&	6.55$\pm$0.40	&	42.61$\pm$0.15	&	5.82$\pm$0.16 &   2.41$\pm$0.15	\\
7 & 55.2	&	71.20$\pm$0.06	&	0.56$\pm$0.08	&	1.88$\pm$0.08	&	3.35$\pm$0.10	&	4.93$\pm$0.09	&	5.52$\pm$0.23	&	41.65$\pm$0.11	&	6.03$\pm$0.11 &   1.16$\pm$0.09	\\
8 & 110.4	&	60.87$\pm$0.08	&	0.88$\pm$0.07	&	1.02$\pm$0.08	&	2.93$\pm$0.07	&	3.61$\pm$0.07	&	4.43$\pm$0.21	&	29.16$\pm$0.09	&	5.25$\pm$0.09 &   0.73$\pm$0.07	\\
9 & 165.6	&	29.51$\pm$0.08	&	0.46$\pm$0.04	&	0.23$\pm$0.04	&	1.60$\pm$0.06	&	1.54$\pm$0.06	&	2.66$\pm$0.22	&	13.85$\pm$0.07	&	3.14$\pm$0.07 &    $<$0.24	\\
10 & 220.8	&	9.27$\pm$0.06	&	0.13$\pm$0.03	&	0.11$\pm$0.03	&	0.73$\pm$0.06	&	0.55$\pm$0.04	&	0.60$\pm$0.19	&	4.39$\pm$0.04	&	1.20$\pm$0.04 &    $<$0.1	\\
11 & 276	&	4.17$\pm$0.08	&	$<$0.18 	&	$<$0.13		&	0.30$\pm$0.09	&	$<$0.17		&	0.30$\pm$0.13	&	1.52$\pm$0.04	&	0.48$\pm$0.04 &    $<$0.09	\\

\hline
N$^{*}$ & R & [Fe\,{\sc ii}] & Pa$\beta$ & [Si\,{\sc x}] & [Fe\,{\sc ii}] & [Si\,{\sc vi}]  & H$_2$ & Br$\gamma$ & [Ca\,{\sc viii}] & [Si\,{\sc vii}] \\
 & (pc)   & 1.257~$\mu$m & 1.282~$\mu$m & 1.43~$\mu$m & 1.644~$\mu$m & 1.963~$\mu$m & 2.1218~$\mu$m & 2.1657~$\mu$m & 2.32~$\mu$m & 2.483~$\mu$m \\
\hline	

1 & -276	&	0.47$\pm$0.02	&	1.07$\pm$0.02	&	$<$0.08		&	0.38$\pm$0.02	&	0.27$\pm$0.06	&	0.11$\pm$0.01	&	0.16$\pm$0.01	&	$<$0.07	        &	0.41$\pm$0.17	\\
2 &-220.8	&	0.86$\pm$0.04	&	2.15$\pm$0.03	&	$<$0.1		&	0.70$\pm$0.02	&	0.81$\pm$0.06	&	0.28$\pm$0.01	&	0.43$\pm$0.01	&	0.12$\pm$0.02	&       1.05$\pm$0.12	\\
3 &-165.6	&	1.02$\pm$0.04	&	3.96$\pm$0.03	&	$<$0.2		&	0.93$\pm$0.03	&	1.51$\pm$0.08	&	0.72$\pm$0.02	&	0.92$\pm$0.02	&	0.22$\pm$0.02	&	2.08$\pm$0.09	\\
4 &-110.4	&	1.78$\pm$0.07	&	6.33$\pm$0.06	&	$<$0.3		&	1.70$\pm$0.07	&	1.84$\pm$0.13	&	2.07$\pm$0.03	&	1.80$\pm$0.03	&	0.32$\pm$0.03	&	2.32$\pm$0.13	\\
5 &-55.2	&	3.87$\pm$0.13	&	9.20$\pm$0.10	&	1.58$\pm$0.16	&	3.70$\pm$0.11	&	2.77$\pm$0.25	&	3.51$\pm$0.05	&	3.01$\pm$0.06	&	1.29$\pm$0.03	&	4.07$\pm$0.19	\\
6 & 0   	&	6.07$\pm$0.24	&	13.80$\pm$0.18	&	4.26$\pm$0.24	&	7.05$\pm$0.20	&	6.71$\pm$0.58	&	4.78$\pm$0.11	&	6.20$\pm$0.13	&	4.17$\pm$0.06	&	13.40$\pm$0.41	\\
7 & 55.2	&	3.66$\pm$0.12$^{**}$   &12.64$\pm$0.14	&	1.41$\pm$0.13	&	4.69$\pm$0.15	&	4.44$\pm$0.36	&	2.50$\pm$0.06	&	3.18$\pm$0.08	&	1.45$\pm$0.04	&	5.83$\pm$0.26	\\
  &             &       1.49$\pm$0.08   &                       &                       &                       &                       &                       &                       &                       & \\
8 & 110.4	&	3.86$\pm$0.09   &	10.81$\pm$0.10	&	$<$0.55		&	4.29$\pm$0.14	&	2.94$\pm$0.17	&	1.64$\pm$0.04	&	2.25$\pm$0.05	&	0.57$\pm$0.05	&	3.27$\pm$0.15	\\
  &             &       1.43$\pm$0.09   &                       &                       &                       &                       &                       &                       &                       & \\
9 & 165.6	&	3.71$\pm$0.09   &	6.97$\pm$0.07	&	$<$0.22		&	3.82$\pm$0.06	&	1.23$\pm$0.12	&	1.20$\pm$0.02	&	1.51$\pm$0.03	&	0.19$\pm$0.05	&	1.11$\pm$0.15	\\
  &             &       0.76$\pm$0.13   &                       &                       &                       &                       &                       &                       &                       & \\
10 & 220.8	&	1.5$\pm$0.06    &	2.35$\pm$0.05	&	$<$0.15		&	2.05$\pm$0.05	&	0.37$\pm$0.10	&	0.65$\pm$0.02	&	0.60$\pm$0.03	&	$<$0.06 	&	0.36$\pm$0.10	\\
  &             &       0.58$\pm$0.07   &                       &                       &                       &                       &                       &                       &                       & \\
11 & 276	&	0.68$\pm$0.05   &	0.76$\pm$0.04	&	$<$0.15		&	0.82$\pm$0.03	&	0.10$\pm$0.05	&	0.24$\pm$0.01	&	0.18$\pm$0.02	&	$<$0.06 	&	0.19$\pm$0.08	\\
  &             &       0.29$\pm$0.05   &                       &                       &                       &                       &                       &                       &                       & \\
\hline

\multicolumn{11}{l}{$^*$ Aperture number along the slit as identified in Figure~\ref{fig:slitpos}}\\
\multicolumn{11}{l}{$^{**}$ When two entries are listed, the top one is the flux of the blue component and the bottom one that of the red component.}\\
\end{tabular}
\end{table*}

\subsection{Internal extinction distribution}
\label{ext}

The host galaxy of NGC\,4388 is a nearly edge-on spiral with a dust lane
crossing the nuclear region. Therefore,  dust
obscuration plays an important role in the interpretation of
the galaxy structure. Optical imaging by \citet{Falcke+} 
shows two wedges of reduced emission, demonstrating
the presence of obscuring dust bands along the disk of this
galaxy right into the nuclear region.  

It is then necessary to quantify the amount
of extinction towards the nucleus and in the circumnuclear region of 
NGC\,4388 in order to determine the true luminosity of both the active nucleus
and the NLR. Large and small values have been reported in the 
literature for this source. \citet{Phillips+}, for example, reported an 
$E(B-V)$ = 0.5 mag based on the H$\alpha$/H$\beta$ ratio for the nucleus. 
Later, \citet{Colina92} found $E(B-V)$ = 0.2 and 0.6 mag in 
regions SW and NE of the nucleus, respectively, and a nuclear 
extinction of at least 0.3 mag. \citet{Petitjean+} found an
$E(B-V)$ = 0.32 also from the Balmer decrement. However, few works on NGC\,4388 
in the literature have addressed this issue using NIR diagnostic lines, 
probably because the lack of simultaneous observations covering at least 
two extinction sensitive lines.

The numerous H\,{\sc i} and forbidden lines observed simultaneously 
in the spectra of NGC\,4388, spanning a large interval in wavelength, allowed us
to evaluate, for the first time, the intrinsic extinction affecting the nuclear
and off-nuclear gas  within the central $\sim$550 parsecs by means of several indicators.
For this purpose, we used the Pa$\beta$/Br$\gamma$, Pa$\gamma$/Br$\gamma$
and Pa$\delta$/Br$\gamma$ flux line ratios and the expressions:

\begin{eqnarray}\label{eq:red1}
 E\left(B-V\right)_{Pa\beta/Br\gamma}  = 5.22 \times log\left(\frac{5.88}{F_{Pa\beta}/F_{Br\gamma}}\right)
\end{eqnarray}

\begin{eqnarray}\label{eq:red2}
 E\left(B-V\right)_{Pa\gamma/Br\gamma}  =  3.46 \times log\left(\frac{3.22}{F_{Pa\gamma}/F_{Br\gamma}}\right)
\end{eqnarray}

\begin{eqnarray}\label{eq:red3}
 E\left(B-V\right)_{Pa\delta/Br\gamma}  =  2.84 \times log\left(\frac{2.0}{F_{Pa\delta}/F_{Br\gamma}}\right)
\end{eqnarray}

where F$_{Pa\beta}$, F$_{Pa\gamma}$ F$_{Pa\delta}$ and F$_{Br\gamma}$ are the observed
 emission lines fluxes of Pa$\beta$, Pa$\gamma$, Pa$\delta$ and Br$\gamma$, respectively,
listed in Table~\ref{tab:fluxes}. In Equations~\ref{eq:red1} to~\ref{eq:red3},
the extinction law of \citet{Cardeli+} was adopted.
The intrinsic ratios  F$_{Pa\beta}/$F$_{Br\gamma}$ = 5.88; Pa$\gamma$/Br$\gamma$ = 3.22
and Pa$\delta$/Br$\gamma$ = 2, corresponding to 
case~B recombination \citep{Osterbrock06}, were employed.

\begin{figure}
\includegraphics[width=90mm]{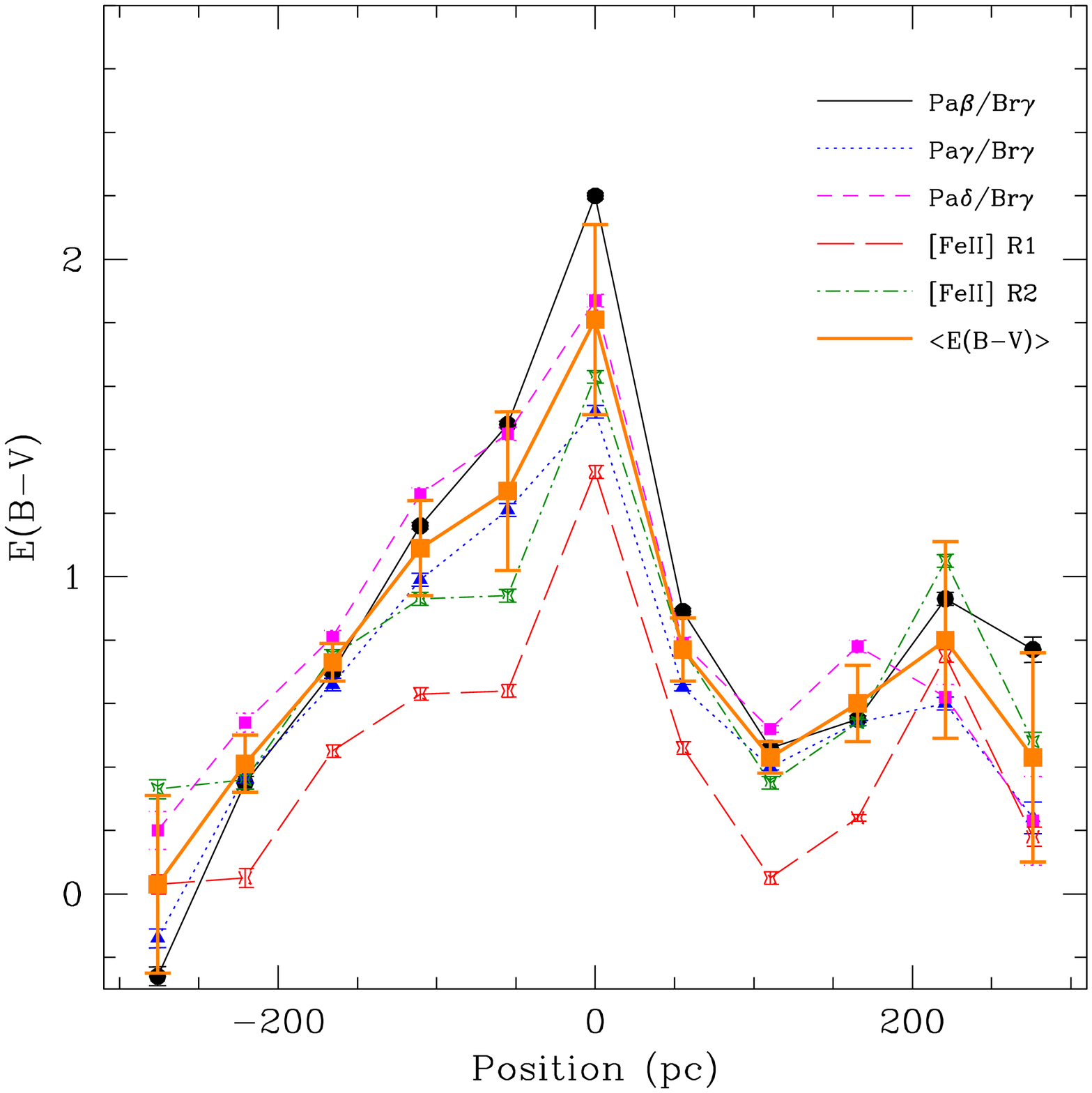}
\caption{Values of E(B-V) derived from different H\,{\sc ii} line flux ratios as
well as from the [Fe\,{\sc ii}]\,1.257\,$\mu$m/1.644\,$\mu$m flux ratio using an
intrinsic value of 1.25 (R1, long-dashed line) and 1.36 (R2, short-dashed-dotted line) 
for the nuclear and off-nuclear apertures of NGC\,4388. The thick full line is the average 
extinction along the spatial direction. See text for further details.}
\label{fig:reddening}
\end{figure}

Figure~\ref{fig:reddening} and columns~2 to~4 of Table~\ref{tab:reddening}  
show the values of E(B-V) found for the nuclear and off-nuclear apertures using 
the three different  H\,{\sc i} line ratios described above. 
Our results show that the 
region covered by the nuclear spectrum displays the highest amount of extinction, 
E(B-V) $\sim$1.9 mag. To the NE, from the nucleus to distances of 
$\sim$150~pc, the extinction remains large, $\gtrsim$1 mag. Farther out, 
it drops considerably, to E(B-V) = 0.6, becoming negligible
at $\sim$240~pc. The large reddening found in the centre and the inner 150~pc 
to the NE is consistent with the system of dust lanes seen in 
the optical images of \citet{Pogge88} that obscures the northern portion of the 
galaxy. To the SW, right after
crossing the nuclear region, the extinction drops steeply, reaching
E(B-V)$\sim$0.5 at 150~pc. Farther out, it rises again, to 0.87 mag at 240~pc. 

The very good match between the E(B-V) found from different indicators in most 
apertures indicates the consistency of our approach. Although some discrepancies are noted at some specific
locations, overall differences of less than $\sim$0.2 mag between the minimum and 
maximum values of E(B-V) were observed.  

In addition to the H\,{\sc i} lines, there are a good number of [Fe\,{\sc ii}] 
lines in the NIR region that originate from radiative transitions involving
levels within the 3$d^6$4$s ^4D$ multiplet. Having more than one
line from the same upper level is useful because  their intrinsic line ratio
becomes insensitive to density and tempeture effects. Thus, they can be used as a
dust extinction diagnostic. 
The strongest lines that fullfil this requirement and were detected in all
apertures are 1.257~$\mu$m and 1.644~$\mu$m. The intrinsic 
$j$1.257~$\mu$m/$j$1.644~$\mu$m ratio is estimated to be 1.25 with an accuracy
of $\sim$20\% \citep{Bautista+}. Extinction due to dust would decrease 
this ratio.

We employed the [Fe\,{\sc ii}] 1.257~$\mu$m/1.644~$\mu$m
flux ratio  measured along the different apertures in order to map the
extinction affecting the region where that ion is formed. As for the 
H\,{\sc i} lines, the law of \citet{Cardeli+} was adopted. The
derived equation for E(B-V) employed is,

\begin{eqnarray}\label{eq:red4}
 E\left(B-V\right)_{\rmn{[Fe\,II]}} =  8.22 \times log\left(\frac{1.25}{F_{1.257~\mu m}/F_{1.644~\mu m}}\right)
\end{eqnarray}

where F$_{1.257~\mu m}$ and F$_{1.644~\mu m}$ are the observed fluxes of 
[Fe\,{\sc ii}]\,1.257$\mu$m and [Fe\,{\sc ii}]\,1.644$\mu$m, respectively, listed 
in Table~\ref{tab:fluxes}. In the apertures where two components were detected
in the former line we opted to sum up their fluxes. Although the 
presence of splitted lines may suggest separate kinematic systems, we are
primarily assessing here extinction due to dust between the NLR and the observer. 
As both components are in the same line of sight, it is consistent to use the
total flux observed.  The values of E(B-V) derived using this indicator 
are listed in Column~5 of Table~\ref{tab:reddening} and plotted in Figure~\ref{fig:reddening}
with a dashed red line.

\begin{table*}  
 \centering
 \caption{E(B-V) derived in the nuclear and off-nuclear apertures in NGC\,4388 using the emission
line flux ratios quoted in columns 2 to 6. Average values found are in Column 7. \label{tab:reddening}}
  \begin{tabular}{@{}lcccccc@{}}
  \hline
R     &   Pa$\beta$/Br$\gamma$ & Pa$\gamma$/Br$\gamma$ & Pa$\delta$/Br$\gamma$ & [Fe\,{\sc ii}] (R1)$^1$ &
[Fe\,{\sc ii}] (R2)$^2$ & $<$E(B-V)$>$ \\
\hline
-276	& -0.26$\pm$0.03&	-0.14$\pm$0.03	&	0.20$\pm$0.06	&	0.03$\pm$0.03	& 0.33$\pm$0.03	& 0.03$\pm$0.28 \\
-220.8	& 0.35$\pm$0.02	&	0.37$\pm$0.02	&	0.54$\pm$0.03	&	0.05$\pm$0.03	& 0.36$\pm$0.03	& 0.41$\pm$0.09 \\
-165.6	& 0.70$\pm$0.01	&	0.66$\pm$0.02	&	0.81$\pm$0.02	&	0.45$\pm$0.02	& 0.75$\pm$0.02	& 0.73$\pm$0.06 \\
-110.4	& 1.16$\pm$0.01	&	0.99$\pm$0.02	&	1.26$\pm$0.02	&	0.63$\pm$0.02	& 0.93$\pm$0.02	& 1.09$\pm$0.15 \\
-55.2	& 1.48$\pm$0.01	&	1.21$\pm$0.02	&	1.45$\pm$0.02	&	0.64$\pm$0.02	& 0.94$\pm$0.02	& 1.27$\pm$0.25 \\
0	& 2.20$\pm$0.01	&	1.52$\pm$0.02	&	1.87$\pm$0.02	&	1.33$\pm$0.02	& 1.63$\pm$0.02	& 1.81$\pm$0.30 \\
55.2	& 0.89$\pm$0.01	&	0.65$\pm$0.01	&	0.79$\pm$0.02	&	0.46$\pm$0.02	& 0.77$\pm$0.02	& 0.77$\pm$0.10 \\
110.4	& 0.46$\pm$0.01	&	0.40$\pm$0.01	&	0.52$\pm$0.01	&	0.05$\pm$0.02	& 0.35$\pm$0.01	& 0.43$\pm$0.05 \\
165.6	& 0.55$\pm$0.01	&	0.54$\pm$0.01	&	0.78$\pm$0.02	&	0.24$\pm$0.01	& 0.54$\pm$0.02	& 0.60$\pm$0.12 \\
220.8	& 0.93$\pm$0.02	&	0.60$\pm$0.02	&	0.62$\pm$0.04	&	0.75$\pm$0.02	& 1.05$\pm$0.04	& 0.80$\pm$0.31 \\
276	& 0.77$\pm$0.04	&	0.24$\pm$0.05	&	0.23$\pm$0.14	&	0.18$\pm$0.03	& 0.48$\pm$0.14	& 0.43$\pm$0.33 \\
\hline
\multicolumn{7}{l}{$^1$ Values determined using an intrinsic line ratio [Fe\,{\sc ii}]\,1.257~$\mu$m/1.644~$\mu$m of 1.25 \citep{Bautista+}}\\
\multicolumn{7}{l}{$^2$ Values determined using an intrinsic line ratio [Fe\,{\sc ii}]\,1.257~$\mu$m/1.644~$\mu$m of 1.36 \citep{Bautista98}}\\
\end{tabular}
\end{table*}

Although the overall shape of the dust distribution profile found for
iron is very similar to that of hydrogen (see Figure~\ref{fig:reddening}), our results point out that the 
reddening determined from the former, when adopting an intrinsic line ratio of 1.25, 
is considerably smaller than that of the latter. In the nucleus, for example,
the difference in E(B-V) reaches 0.9 magnitudes between both indicators. 
Similar estimates in the literature have found that the E(B-V) determined from
the iron lines is significantly larger than that from hydrogen 
\citep{riffel+06,Mazzalay07,martins+13}. 
However, note that these works employed an intrinsic line ratio 
[Fe\,{\sc ii}]\,1.257$\mu$m/[Fe\,{\sc ii}]\,1.644$\mu$m 
of 1.36, as originally proposed by \citet{Bautista98}. Column~6 of 
Table~\ref{tab:reddening} lists the values of E(B-V) derived from Equation~\ref{eq:red4} 
but adopting an intrinsic line flux ratio of 1.36.  It can be seen that the difference
between the iron and hydrogen extinction decreases considerably. Indeed, it becomes
indistinguishable at some positions. Note that the latter intrinsic 
flux ratio is still within the 20\% uncertainty determined by \citet{Bautista+}.
Because the results found for [Fe\,{\sc ii}] using an intrinsic flux ratio
of 1.36 better agree with that of H\,{\sc ii}, we opted for keeping these
latter values.  

Based on the E(B-V) values listed in columns~2 to~4 and colum~6 of Table~\ref{tab:reddening}, we 
determined the average extinction for the inner 560~pc
of NGC\,4388. In this calculation we employ the results found for the three H\,{\sc i} 
ratios and that of iron assuming an intrinsic flux ratio of 1.36. The average extinction, 
at each aperture is listed in the last 
column of Table~\ref{tab:reddening} and also plotted in Figure~\ref{fig:reddening}. 
The results confirm that the central regions of NGC\,4388 are dusty, with the
dust distributed inhomogeneously, peaking at the nucleus with an E(B-V) of 
1.81$\pm$0.3 mag. It then decreases both to the NE and SW although it remains high
mainly to the NE. To the SW at 220~pc from the nucleus we see a secondary peak that reaches
0.85$\pm$0.3 mag. Thanks to the reduced sensitivity
of the NIR to dust, we are able to see, for the first time, emission from
the NE side of the galaxy not observed before in the optical

\subsection{Emission line ratios and excitation structure}

The rich nuclear and extended emission line spectra of NGC\,4388 allow us to study 
how the gas ionization varies with distance to the central source as well as to gather
information about the mechanisms that power these lines. Results obtained by means of 
imaging and long-slit spectroscopy reveal that line ratios like [Fe\,{\sc ii}]/Pa$\beta$ and 
H$_2$/Br$\gamma$ may vary dramatically on scales of 100~pc near the nuclei of 
Seyfert galaxies \citep{Mazzalay07,Knop+,Ramos-Almeida06,Mezcua+}. 
The analysis of these line ratios points to partially ionized 
regions created by X-ray photoionization from the nucleus or shocks from the interaction 
of outflowing gas, or both. This complex scenario has been confirmed in the last decade 
thanks to the gain in angular resolution provided by adaptive optics (AO) observations. 
The distinct flux distributions and kinematics of the H$_2-$ and [Fe\,{\sc ii}]$-$ emitting gas
in nearby AGNs show that the former is more restricted to the plane of the galaxy,
with the nuclear disc being fed by gas coming from the outer regions. 
The  [Fe\,{\sc ii}] traces the outflows related to radio jets, evidenced by the highest
velocity dispersion values 
(up to 150~km\,s$^{-1}$) and the highest blueshifts and redshifts of up to 500\,km s$^{-1}$ of 
these lines when compared to the stellar rotation velocity or velocity dispersion of
other emission lines \citep{Riffel11,Riffel13,Mazzalay15}. 

Most of the studies (including the ones with NGC\,4388) aimed at studying the
excitation mechanisms leading to the observed emission line spectrum are restricted 
to the [Fe\,{\sc ii}] and H$_2$ lines, and very few works have traced the gas distribution 
using emission lines that are usually the most prominent ones in the NIR region. 
This is the case of [S\,{\sc iii}]\,$\lambda$9531 and He\,{\sc i}\,1.083~$\mu$m. 
The former can be considered as the equivalent of [O\,{\sc iii}]\,$\lambda$5007
because of the similarity in the ionization potential of both lines. The latter
is emitted in a 2\,$^3 P$ to 2\,$^3S$ transition and because of its small 
energy change ($\Delta E$ = 1.14 eV), it is readily collisionally excited from the 
metastable 2\,$^3S$ triplet state. For that reason, it can potentially be a useful 
indicator of the density of the NLR.  
 
\begin{figure}
\includegraphics[width=90mm]{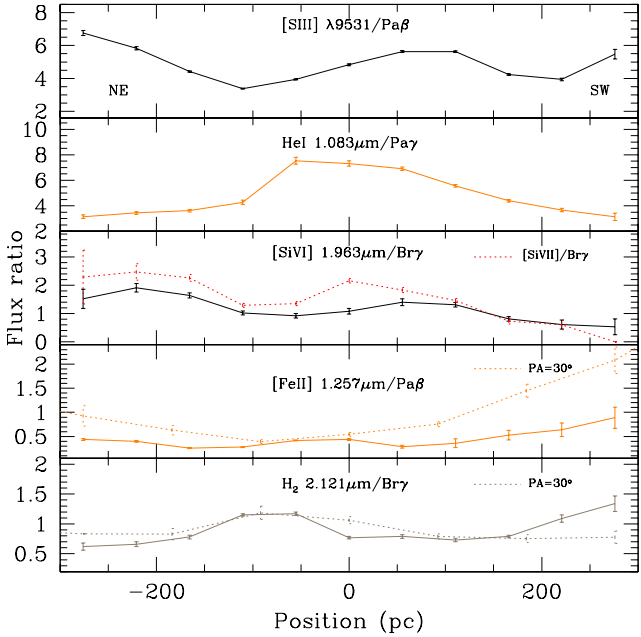}
\caption{Flux line ratios of the most conspicuous emission lines observed
in NGC\,4388. For [Fe\,{\sc ii}]\,1.257~$\mu$m/Pa$\beta$ and 
H$_2$~2.1218~$\mu$m/Br$\gamma$ values reported in the literature by 
Knop et al. (2002) using a PA=30$\degr$ are plotted for comparison purposes.}
\label{fig:ratios}
\end{figure}

Using the fluxes listed in Table~\ref{tab:fluxes} we have plotted in Figure~\ref{fig:ratios}
the line flux ratios [S\,{\sc iii}]/Pa$\beta$, He\,{\sc i}/Pa$\beta$, [Si\,{\sc vi}]/Br$\gamma$,
[Si\,{\sc vii}]/Br$\gamma$, [Fe\,{\sc ii}]/Pa$\beta$, and H$_2$/Br$\gamma$. For 
the [Fe\,{\sc ii}]~1.257\,$\mu$m line, only the flux of the blue component was employed.
As will be shown in Sect~\ref{sec:kinematics}, the gas emitting that component follows very closely
the kinematics exhibited by Pa$\beta$, specially in the SW side, suggesting that these two lines are
emitted by the same parcel of gas.
In spite of the strong extinction affecting the circumnuclear region of NGC\,4388, the ratios are formed by 
lines that are close enough in wavelength so that they are nearly or totally independent of 
the presence of dust. Except for the final two, these ratios have never
been presented before for this source.

Overall, Figure~\ref{fig:ratios} confirms that the gas distribution in the inner
500~pc of NGC\,4388 is highly inhomogeneous, with at least two regions of 
enhanced high-ionization emission lines: one in the centre, and another 
at $\sim$250~pc NE of the nucleus. Both, the [S\,{\sc iii}]/Pa$\beta$ and the 
silicon lines ([Si\,{\sc vi}] and [Si\,{\sc vii}]) display high ratios at these locations,
with values in the off-nuclear position as high or higher than those in the nucleus.
To the SW, the silicon lines drop considerably in intensity relative to Br$\gamma$
from 100~pc of the nucleus to 280~pc.
 
These results are consistent with the optical
maps of the central region of this source presented by \citet{Falcke+}.  
Their uppermost right panel of Figure~12 (and partially reproduced here in the right panel
of Figure~\ref{fig:slitpos}) displays WFPC2/HST H$\alpha$ emission for the inner
12$\arcsec$. In this image, it can be seen
a bright emission coincident with the nucleus and a faint cloud at 
RA = 12$^h$~14$^m$~76$^s$; DEC = 14$\degr$~56$\arcmin$~21$\arcsec$
that should be enhaced in the NIR because of the reduced sensitivity to dust. 
We associate the later region to the high-ionized peak detected at $\sim$250~pc NE 
of the nucleus.

The behaviour of the He\,{\sc i} line relative to hydrogen contrasts with 
what is seen with the higher ionization lines.
It displays a flat peak in the central 100~pc and then drops steeply
to the NE, reaching less than half the peak value at 270~pc. To the SW, the
line ratio also decreases but more smoothly. If the line emissivity of
He\,{\sc i}~1.083~$\mu$m is strongly dependent on the gas density, the 
results show a much larger gas density in the central region of NGC\,4388. 

The line ratios [Fe\,{\sc ii}]/Pa$\beta$ and H$_2$/Br$\gamma$, plotted in
the last two panels (from top to bottom) of Figure~\ref{fig:ratios}, confirm
previous results presented by \citet{Knop+} using $J-$ and $K-$band
spectroscopy. They employed a slit position angle of 30$\degr$, very close 
to the orientation of the elongated, jetlike radio structure at PA=24$\degr$. Their results
are also shown in Figure~\ref{fig:ratios} with dotted lines for comparison. 
For [Fe\,{\sc ii}]/Pa$\beta$, both this work and Knop's et al. data
shown an increase in the values of the ratio towards the SW, 
starting from the position where double-peak iron lines show up,
suggesting an extra-source of line excitation.  \citet{Knop+} do not
reported splitted iron lines. However, in their
figure~13 it can be noticed broad assymetric iron lines SW of the nucleus.
 
We interpret the enhancement of the [Fe\,{\sc ii}]/Pa$\beta$ ratio to the SW as direct 
evidence of shocks produced by the interacion between the radio-jet and the 
ISM gas. The larger value of that ratio, relative to the one found in
the nucleus, is more pronounced in the \citet{Knop+} data than in ours.  
This is consistent with the fact that the slit in Knop's et al. observations was 
nearly aligned to the elongated radio structured to the SW. 
To the NE, our data show a subtle decrease in the values of the ratio and 
then an increase 200~pc NE and outwards.  
The relationship betwen [Fe\,{\sc ii}] and radio-emission is well-know since \citet{forbes+}.
They found that the strength of the former in the central regions of AGNs is tightly correlated 
with the 6-cm radio emission. This correlation has been widely confirmed by means of
AO NIR integral field unit (IFU) observations in several AGNs \citep{Riffel11, Riffel15, Riffel15b} 
at spatial scales of a few parsecs. All these works clearly evidence that the interaction of 
the radio jet with the gas produces wings in the [Fe\,{\sc ii}] line profiles at locations 
around the radio hotspots. In single-Gaussian fits, this effect appears as an enhancement of
$\sigma$, similar to what is observed in the [Fe\,{\sc ii}] lines to the NE 
(See Sect~\ref{sec:kinematics}).

The line ratio H$_2$/Br$\gamma$ shows values $<$1 in most of the apertures, including 
the nucleus. The exception is the region between 50~pc and 100~pc NE of the AGN, when it
increases to $\sim$1.2 and at 200~pc SW and outwards, where it reaches 1.3. Our results
agree to those reported by \citet{Knop+} and \citet{Greene14}. They can be understood if we consider that most
of the H$_2$ is located in a nuclear disk with a PA of 90$\degr$. As our slit is positioned
along the edge of the ionization cone, both above and below the disk, the values of the
ratio are low within the cone (H$_2$/Br$\gamma <$1) and larger in the regions dominated
by the disk (H$_2$/Br$\gamma >$1).

The results discussed above highlights the 
very complex nature of the nuclear and circumnuclear region of NGC\,4388. 
Our data reveal what is probably one of the best pieces of evidence of the 
intricated mixture of an AGN, a radio-jet, dust and circumnuclear gas.
The detection of splitted [Fe\,{\sc ii}] lines SW of the nucleus 
points out to this scenario. Shocks produced by the interaction between 
the jet and the NLR gas enhanced this emission.
Observations at superior angular resolution as provided by 
adaptive optics and sub-arsecond radio observations of the jet with VLA
would be helpful to unveil at greater detail this turbulent cauldron.
 
\section{Stellar and gas kinematics of NGC\,4388}
\label{sec:kinematics}

The velocity and excitation structure of the NLR of NGG\,4388 
is widely known for being very complex \citep{Corbin+88,Veilleux91} 
both at optical and radio wavelengths. It has been described by 
rotation plus outflow by several authors \citep[e.g.,][]{Veilleux99,Veilleux01,Greene14}. 
The radio-emission morphology suggests a collimated
AGN-driven outflow reaching $\sim$200\,pc south of the nucleus \citep[e.g.,][]{Stone+,Falcke+}.
The extended optical emission has two components:
One associated with the galactic disc and
another reaching 50\arcsec\ ($\sim$4 kpc) above the galactic plane in the form
of two opposite cones \citep{Pogge88,Corbin+88,Falcke+,Veilleux99}.
The current scenario for this object is that the north edge of the galaxy 
is the near side. We see primarily the south ionization cone coming roughly 
towards us and the northeast ionization cone directed away from us. In the 
optical region, the northeat cone shows up only when it is no longer obscured by the disk. 
The reduced sensitivity of the NIR to dust allows that both the
northern and southern components of the ionization cone can be observed.  

The GNIRS data set allows us to extend our understanding of the stellar and gas kinematics 
of NGC~4388 by constructing position-velocity (PV) diagrams for extended line emission 
covering a wide range of species and  ionization potentials.
As shown by Figure~\ref{fig:perfil_luz}, the [S\,{\sc iii}]\,$\lambda$9531,
[S\,{\sc viii}]\,$\lambda$9913, He\,{\sc i}\,1.083~$\mu$m, He\,{\sc ii}\,1.012$\mu$m, 
[Fe\,{\sc ii}]\,1.257$\mu$m, Pa$\beta$, H$_2$\,2.122~$\mu$m, 
[Si\,{\sc vi}]\,1.963~$\mu$m, Br$\gamma$, [Si\,{\sc vii}]\,2.48~$\mu$m lines
extend all across the inner 560~pc in the NE-SW direction mapped by our slit, 
providing unique evidence of the gas kinematics around the 
nucleus. In particular, this list includes three very high-ionization lines: 
[Si\,{\sc vi}]~1.963\,$\mu$m, [Si\,{\sc vii}]~2.48\,$\mu$m and [S\,{\sc viii}]\,0.991$\mu$m. 
To the best of our knowledge the latter two have not yet been used for this  
kind of analysis of this source. 
In absorption, the CO band heads provide information about the resolved stellar kinematics.

For each of the above emission lines
we measured their centroid position in each aperture by fitting a Gaussian function to the 
observed profile using the LINER routine as described in \S\ref{lines}. To construct the PV diagram 
for the stellar component of NGC\,4388 we used the penalized Pixel-Fitting (pPXF) 
method of \citet{Cappellari+}
to fit the $^{12}$CO and $^{13}$CO  stellar absorption band heads around
2.3\,$\mu$m and obtain the line-of-sight velocity distributions (LOSVD) of
the stars. The pPXF outputs the radial velocity (V$_*$), stellar velocity
dispersion ($\sigma_*$), and higher order Gauss-Hermite moments ($h_{3*}$
and  $h_{4*}$), as well as the uncertainties for each parameter.
As stellar template spectra we used those of the Gemini library of late
spectral type stars observed with the GNIRS IFU and  NIFS \citep{Winge09},
which contains the spectra of 60 late type stars. The spectral resolution
of the stellar templates (3.2\AA\ at 2.3\,$\mu$m) is better than that of
our data. Therefore, we degraded the stellar templates to the same resolution as
that of NGC\,4388 before running the pPXF to measure the LOSVD.

We also derived the velocity dispersion $\sigma$ for [S\,{\sc iii}], He\,{\sc i}, 
[Fe\,{\sc ii}], Pa$\beta$, [Si\,{\sc vi}], H$_2$ as well as for the stellar continuum.  
For the emission lines, the velocity
dispersion was derived from the full-width at half maximum (FWHM) found from the
Gaussian fit when measuring the line centroid along the different
apertures. This parameter was left to vary freely during the spectral fits, corrected in 
quadrature for the instrumental broadening, and then transformed into velocity 
dispersion using the relationship $\sigma$=FWHM/2.35. 

\begin{figure}
\includegraphics[width=90mm]{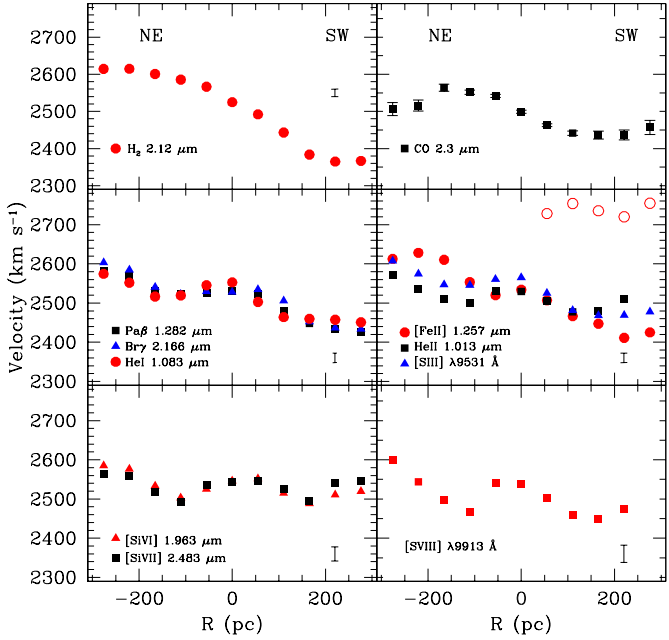}
\caption{Position-velocity (PV) diagrams for the different emission lines detected in NGC\,4388. 
The top right panel also shows the PV diagram for the stellar component
derived from the CO stellar absorption lines at 2.3~$\mu$m. The bar in each panel represents the largest
uncertainty in the velocity, usually observed in the apertures close to the slit edges
(apertures 1 and 11, centred at $\pm$276~pc NE and SW from the nucleus, respectively.) The two panels at the
bottom of the Figure map the velocity of the high-ionization lines. For the [Fe\,{\sc ii}]
line (middle-right panel), full circles represent the blue component while open circles are for
the red component.}
\label{fig:rotation}
\end{figure}

The resulting PV diagrams, shown in Figure~\ref{fig:rotation}, 
exhibit the complex kinematics of NGC\,4388 
already noted in optical and previous NIR spectroscopy of this source.
In all emission lines except the coronal lines, a velocity amplitude
of $\sim$135~km~s$^{-1}$ is clearly detected along the slit. However, only 
H$_2$~2.121~$\mu$m displays an organized velocity pattern, following a curve
consistent with disk rotation. The amplitude of the velocity curve is 
largest in this line, reaching $\sim$250~km~s$^{-1}$. This value is slightly larger than
the one ($\sim$175~s$^{-1}$) found by \citet{vlaan+}, who also obtained a rotation-like
curve for this same line but at a different PA (90$\degr$).

The stellar component also displays rotation, but it is much flatter than that of the molecular
gas, with a maximum amplitude of $V_{\rmn max}~\sim$60~km\,s$^{-1}$ at $\sim$150~pc
from the nucleus. An organized pattern is also seen in the velocity dispersion of the stellar component, 
consistent with disk rotation and in agreement with that presented by \citet{Greene14}.
While this is apparent in H$_2$, the stars are considerably dynamically hotter than the molecular gas,
with the latter concentrated in a thinner disk than the former. 
The stellar $\sigma$ also shows a feature not present in $H_2$: a drop in the dispersion field 
between 150-200~pc NE of the centre. At that location, a sudden decrease of 
nearly 50~km\,s$^{-1}$ is observed. 
This feature is also reported by \citet{Greene14}. Because of their 2D view,
they also found, on the same scale, a jump in velocity,
PA, and ellipticity, which they attributed to evidence of a transition from an
inner disk to more bulge-dominated kinematics. For the particular case of
NGC\,4388, the most likely explanation of these observations is that there is a disk 
within the inner 180~pc embedded in the larger-scale bulge/bar.

From our data, the galaxy recession velocity derived for H$_2$~2.121~$\mu$m 
(V=2525$\pm$5~km~s$^{-1}$) 
in the nuclear aperture matches well, within uncertainties, the value reported 
in the literature using H\,{\sc i} 21~cm observations: 2524$\pm$1~km\,s$^{-1}$ 
\citep{Lu+}. The best-fit systemic velocity for the stellar component is 2500$\pm$4\,km~s$^{-1}$.
This is 25~km\,s$^{-1}$ smaller than the velocity found from the molecular gas. The reason for this 
difference is not clear but it is probably due to the fact that our data were taken
with the slit at just one PA, far from the minor axis of the galaxy.

The observed recession velocities for Pa$\beta$ and Br$\gamma$ (in the nuclear aperture) are also in 
very good agreement (2530$\pm$5~km~s$^{-1}$ and 2528$\pm$7~km~s$^{-1}$, respectively) with
the systemic velocity of NGC\,4388 derived from H$_2$~2.121~$\mu$m. 
For the remaining lines, redshifts relative to the systemic velocity at the galaxy nucleus
were measured. The largest redshift is that of [S\,{\sc iii}]~0.953~$\mu$m, 
40~km~s$^{-1}$ followed by He\,{\sc i}~1.083~$\mu$m (28~km~s$^{-1}$). The high ionization lines 
[Si\,{\sc vi}]~1.963\,$\mu$m, [Si\,{\sc vii}]~2.48\,$\mu$m and [S\,{\sc viii}]\,0.991$\mu$m show, on average, 
redshifts of 22~km~s$^{-1}$. The velocity amplitude measured in the
inner 560~pc for these three lines is
small: 66~km~s$^{-1}$, 18~km~s$^{-1}$ and 69~km~s$^{-1}$, respectively.

Figure~\ref{fig:rotation} shows that 
all emission lines but [Fe\,{\sc ii}] are blueshifted NE of 
the nucleus and redshifted SW of it relative to H$_2$~2.121~$\mu$m.
Moreover, the high-ionized gas is non-rotation dominated.
The case of [Fe\,{\sc ii}] is
more complex. It is the only line that clearly display splitted profiles.
The blue component follows closely the disk rotation
with local perturbations at some positions  (full circles in the middle-right panel of
Figure~\ref{fig:rotation}). The red peak that shows up to the SW (open circles) 
exhibit no sign of rotation 
and is strongly redshifted relative to both the molecular and ionized
gas across the
different apertures where it is detected.

\begin{figure}
\includegraphics[width=85mm]{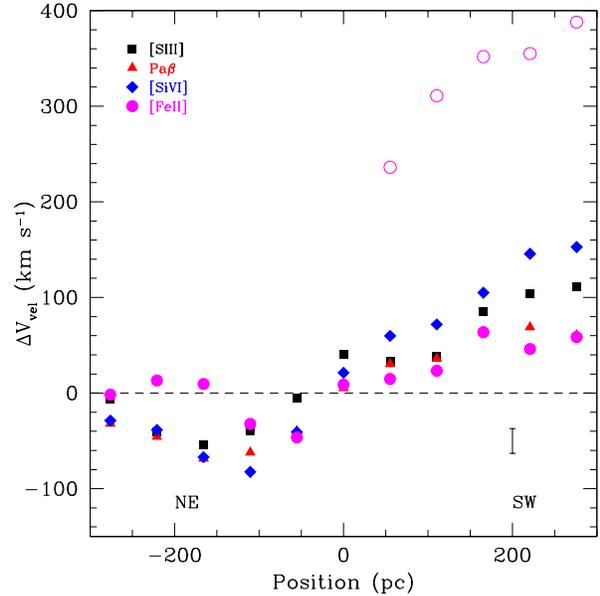}
\caption{Difference between pure gas rotation (assumed to be represented by the H$_2$ position-velocity curve 
of Figure~\ref{fig:rotation}) and the gas velocity observed for [S\,{\sc iii}] (black squares), Pa$\beta$ (red triangles),
and [Si\,{\sc vi}] (full diamonds). For [Fe\,{\sc ii}], full and open circles are for the blue and red components of the 
line, respectively. The bar at the bottom right represents the
typical error bar of the measurements. The dashed line is for reference and indicates gas moving along with H$_2$. }
\label{fig:residuals}
\end{figure}

Assuming that the galaxy rotation in the FOV covered by our
slit is well represented by the H$_2$~2.121~$\mu$m rotation
curve, differences 
in velocity between the ionized and molecular gas at a given position
can be interpreted as a departure of the ionized gas from the rotation of the galaxy's disk. 
This is illustrated in Figure~\ref{fig:residuals}, where it is
seen that [S\,{\sc iii}], [Si\,{\sc vi}] H\,{\sc i} and the red component 
of [Fe\,{\sc ii}] largely depart from pure disk rotation. The case of
the later line is extreme. The residual velocity 
increases steeply from 240~km\,s$^{-1}$ at 55~pc SW of the nucleus, 
to 400~km\,s$^{-1}$ at 276~pc. The clouds 
emitting the red component of [Fe\,{\sc ii}] are clearly part of a distinct 
system, moving away from the observer. We propose here that this system is produced 
by interactions between the radio-jet and the NLR gas. This is consistent
with the radio-emission morphology, which suggests a collimated
AGN-driven outflow reaching $\sim$200~pc south of the nucleus
\citep{Stone+, Falcke+}.


The residual velocity exhibited by the
high-ionized gas in Figure~\ref{fig:residuals} confirms that the bulk of that
emission is out of the nuclear disk, very likely distributed along the 
ionization cone. This hypothesis is consistent with the [Si\,{\sc vi}]~1.963~$\mu$m IFU
maps presented by \citet{Greene14} in their Figures~5 and~7, where it can
be seen that this emission runs preferentially at an angle of $\approx 30\degr$,
tracing the general orientation of the radio jet.  
The gas velocity for this line measured from our data at the different positions 
along the slit (PA=64$\degr$) matches that presented in figure~8 of \citet{Greene14}.
This result, in combination with the excellent agreement between the PV curves 
of the coronal lines in Figure~\ref{fig:rotation} supports that the bulk of this 
emission is oriented in the direction of the radio-jet, with very little
gas in the nuclear disk. 

Mid- and low-ionization lines ([S\,{\sc iii}], Pa$\beta$ and [Fe\,{\sc ii}])
show a more complex kinematics, reflecting the fact that they should be 
produced by gas distributed both along the disk and along the ionization cone. The blue
peak of [Fe\,{\sc ii}], for example, follows closely the disk rotation of H$\_2$
to the NE and up to 100~pc SW of the nucleus. Farther out to the SW
we see predominantly iron gas that lies along the ionization cone.
In contrast, [S\,{\sc iii}] is primarily distributed along the ionization cone, although
some sulfur gas should lie in the inner 100~pc of the nuclear disk. 
This is also the case for Pa$\beta$.

The blueshifted emission shown by the ionized gas relative to the molecular gas
to the NE (see Figure~\ref{fig:residuals}) along with the redshifted ionized
emission SW of the nucleus, i.e., diametrically opposite to the blueshifted northeast 
complex, points out to a bipolar outflow of extraplanar gas. To the NE, bulk of this component is 
approaching us while to the SW it is is moving away from us. Line splitting to
the southwest clouds may reflect regions of high dissipation of
kinetic energy caused by the interaction/deflection of the
outflowing gas with ambient NLR gas.
We are thus seeing in detail the inner 600~pc of the well-known outflow system of ionized
material that extends to kiloparsec scales in this object.

\begin{figure}
\includegraphics[width=85mm]{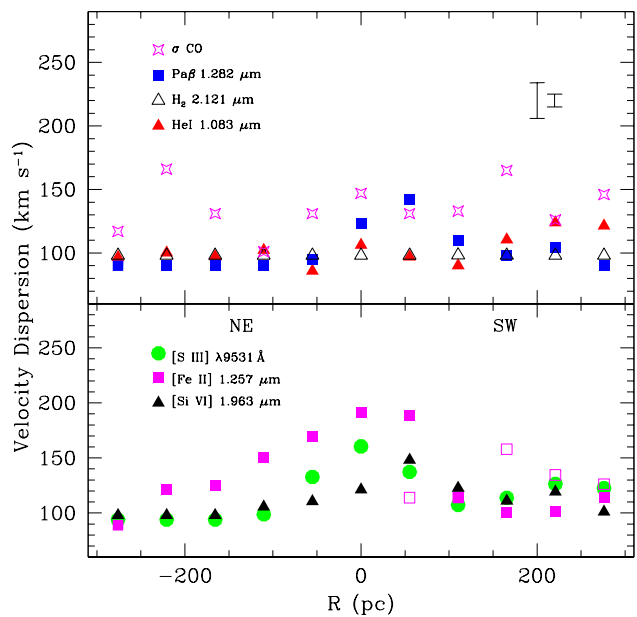}
\caption{Velocity dispersion derived for the different emission lines observed as well as for
the stellar CO band heads. The instrumental
velocity resolution is 90~km\,s$^{-1}$, meaning that H$_2$ is not resolved spectroscopically in any
apertures. The bars at the right top of the plot indicate the maximum and average errors (1-$\sigma$)
derived from the measurements. For [Fe\,{\sc ii}] (bottom panel), full squares are for the
blue component of the line while open squares are for the red component. At 110~pc SW, both 
components have the same width.}
\label{fig:sigma}
\end{figure}

The velocity dispersion of the gas, $\sigma_{\rmn gas}$, 
covers a broad range of values (see Figure~\ref{fig:sigma}), with the largest ones
ocurring in or around the nucleus.  [Fe\,{\sc ii}]\,1.257$\mu$m  
is the line that displays the largest $\sigma_{\rmn gas}$, reaching
190\,km~s$^{-1}$ at the galaxy centre. Moreover, it sistematically
displays the largest values of velocity dispersion to the NE. 
[S\,{\sc iii}] also displays a maximum of $\sigma_{\rmn gas}$ at the galaxy centre
while $\sigma_{\rmn gas}$ for [Si\,{\sc vi}] and Pa$\beta$ peaks at 55~pc
SW of the nucleus. A similar result was also found for [Si\,{\sc vii}],
Pa$\gamma$ and Pa$\delta$, although for clarity they were omitted from Figure~\ref{fig:sigma}). 
Note that 55~pc SW of the nucleus coincides with the location where [Fe\,{\sc ii}]\,1.257$\mu$m 
splits into two components, suggesting that the kinematics of the ionized gas
from this position and outwards is dominated by the outflow component.
As was already pointed out in the literature \citep[e.g.][and references therein]{Riffel11, Riffel15, Riffel15b}, 
the interaction of the radio jet with the gas typically produces splitted [Fe\,{\sc ii}] 
profiles or an enhancement of $\sigma$ in single-Gaussian fits at locations 
around the radio hotspots. Both effects are observed for [Fe\,{\sc ii}],
giving further support to the role of shocks as an additional source of gas
excitation in the central few hundred parsecs of NGC\,4388.

In contrast to the ionized gas, molecular hydrogen displays velocity dispersion
values of $\sim$100~km\,s$^{-1}$ across all the FOV of the slit. This reflects the
fact that the bulk of this emission is concentrated in the nuclear disk. 
He\,{\sc i}~1.083~$\mu$m is the only ionized line that shows low values of $\sigma_{\rmn gas}$,
at the limit of the spectral resolution. However, we noted an increase in the
velocity dispersion at 166~pc SW of the nucleus and outwards.

No trend is observed between the velocity dispersion 
and the ionization potential of the lines. A close inspection to
Figure~\ref{fig:sigma} shows that He\,{\sc i} is spectroscopically 
unresolved at all apertures while Pa$\beta$ is only resolved in the
nucleus and at 55~pc SW of it.  
[Si\,{\sc x}], the line with the largest ionization potential of all detected 
in NGC\,3288 (see Table~\ref{tab:iplines}), is narrower than [Si\,{\sc vi}] (100~km\,s$^{-1}$ and 
120~km\,s$^{-1}$, respectively). In contrast, [Fe\,{\sc ii}] is the broadest ionized 
line not only at the nucleus but also to the NE, being resolved in nearly all apertures. 
Assuming that the gas is virialized, the lack of any correlation between
IP and width of the lines implies that the NLR gas we see in NGC\,4388 is no longer afected by
the potential well of the central supermassive black hole and instead 
primarily governed by the gravitational potential of the stars. Additional sources of line broadening 
such as shocks/outflows should become important, particularly
in the circumnuclear region. \citet{Rodriguez11}, in their study of the NIR
properties of 51 AGNs, already noticed the lack of that correlation in some
sources while in others it is clearly present.  

In summary, the kinematics of the stellar and emission gas within the
central 560~pc of NGC\,4388 reveals the presence of a nuclear disk where
the stars and molecular gas are distributed. In adition, a bipolar outflow of ionized gas is found
to be distributed mostly outside the nuclear disk, along the direction of the radio-jet. 
We also found evidence of an an additional structure, probably a shell of
gas, not reveal before and traced by [Fe\,{\sc ii}], very likely due to direct
interaction between the NLR gas and the radio-jet.

\section{The high ionization line emitting gas}
\label{sec:coronal}

Figure~\ref{fig:XD_allspectra} shows that NGC\,4388 displays 
a remarkable high-ionization emission line spectrum, with 
prominent lines seen not only in the the nuclear but also 
in the off-nuclear apertures. In the wavelength interval covered by GNIRS, 
lines of [S\,{\sc viii}]~0.991\,$\mu$m, [S\,{\sc ix}]~1.252\,$\mu$m, 
[Si\,{\sc x}]~1.43\,$\mu$m, [Si\,{\sc vi}]~1.963\,$\mu$m, 
[Al\,{\sc ix}]~2.045\,$\mu$m, [Ca\,{\sc viii}]~2.32\,$\mu$m
and [Si\,{\sc vii}]~2.48\,$\mu$m were detected, some of them
reported for the first time here. \citet{Lutz+}  had already reported 
the detection of [Si\,{\sc ix}]\,3.94~$\mu$m in this object.
All these lines span a large interval of
ionization potentials (IPs), from 127.7~eV ([Ca\,{\sc viii}]) up to
351.1~eV ([Si\,{\sc x}]), see Table~\ref{tab:iplines}). 
The spatial distribution of [S\,{\sc ix}],
[Al\,{\sc ix}] and [Si\,{\sc x}], the lines with the largest IPs, is
essentially point-like, while [S\,{\sc viii}], [Si\,{\sc vi}] and [Si\,{\sc vii}] extend 
all over the inner $\sim$560~pc region covered by the GNIRS slit.
Few objects in the literature reveal such an spatially extended coronal 
line spectrum. To be best of our knowledge, the only comparable examples
are NGC\,1068 \citep{Mazzalay13} and Mrk\,78 \citep{Ramos-Almeida06}.

\begin{table}
\centering
\caption{Wavelengths, ionization potential (IP) and critical densities of the NIR coronal lines detected in NGC\,4388}
\label{tab:iplines}
\begin{tabular}{@{}lcc}
\hline
Ion & IP & log $n_{\rmn{e}}$ \\
 & (eV) & (cm$^{-3}$) \\
\hline
\lbrack\mbox{S\,{\sc viii}}]\,9913~\AA\ & 280.9 & 10.6 \\
\lbrack\mbox{S\,{\sc ix}}]\,1.252~$\mu$m & 328.8 & 9.4 \\
\lbrack\mbox{Si\,{\sc x}}]\,1.423~$\mu$m & 351.1 & 8.8 \\ 
\lbrack\mbox{Si\,{\sc vi}}]\,1.963~$\mu$m & 166.8 & 8.8 \\
\lbrack\mbox{Al\,{\sc ix}}]\,2.045$\mu$m & 284.6 & 8.3 \\
\lbrack\mbox{Ca\,{\sc viii}}]\,2.321~$\mu$m & 127.7 & 7.9 \\
\lbrack\mbox{Si\,{\sc vii}}]\,2.48~$\mu$m & 205 & 7.30 \\
\lbrack\mbox{Si\,{\sc ix}}]\,3.94~$\mu$m & 303 & 6.32\\
\hline
\end{tabular}
\end{table}
  
The conspicuous coronal line (CL) spectrum in the NIR contrasts strongly with the modest
high-ionization spectrum in the optical region.
\citet{Pogge88} and \citet{Veilleux91}, for example, did not detect 
[Fe\,{\sc vii}]\,$\lambda$6087 (IP = 97~eV), a CL that is 
typically strong in AGN coronal line emitters, although \citet{Colina92} does report it
in his integrated nuclear spectrum. Moreover, the SDSS spectrum
of this object reveals the presence of [Fe\,{\sc vii}]
at $\lambda$5721 and $\lambda$6084 and very likely, 
[Fe\,{\sc x}]\,$\lambda$6374. 

To the best of our knowledge, this is the first simultaneous 
detection of all the above NIR lines in NGC\,4388. The fact that 
three Silicon lines ([Si\,{\sc vi}], [Si\,{\sc vii}] and [Si\,{\sc x}])
and two Sulfur lines ([S\,{\sc viii}] and [S\,{\sc ix}])
with different degrees of excitation are present in this AGN,  
allow us to construct line ratios between ions of the same atom 
that are independent of the metallicity 
of the gas. Very few sources with a positive detection of at 
least four out of the five lines above are found in the literature 
\citep{Oliva94,riffel+06,Ramos-Almeida09,martins+10,Mason15}. 

All the above makes NGC\,4388 an optimal target to study the relationship
between coronal emission and the physical conditions suitable for their formation.
It is known that extended, soft X-ray emission coincident in extension and 
overall morphology with the [O\,{\sc iii}]\,$\lambda$5007 emission is observed in this object
\citep{Bianchi+}. Does this high-energy emission provide sufficient photons to ionize
the gas and produce the extended coronal line spectrum observed? 
\citet{Rodriguez11}, for instance, found that coronal line 
emission becomes stronger with increasing nuclear X-ray emission 
(soft and hard). This result would indicate photoionization as the dominant excitation 
mechanism for the high-ionization lines. However, this trend holds only when considering 
Type\,1 sources alone; it gets weaker or vanishes when including Type\,2 sources, very 
likely because the X-ray emission measured in the latter is not the intrinsic 
ionizing continuum.

\begin{figure}
\includegraphics[width=90mm]{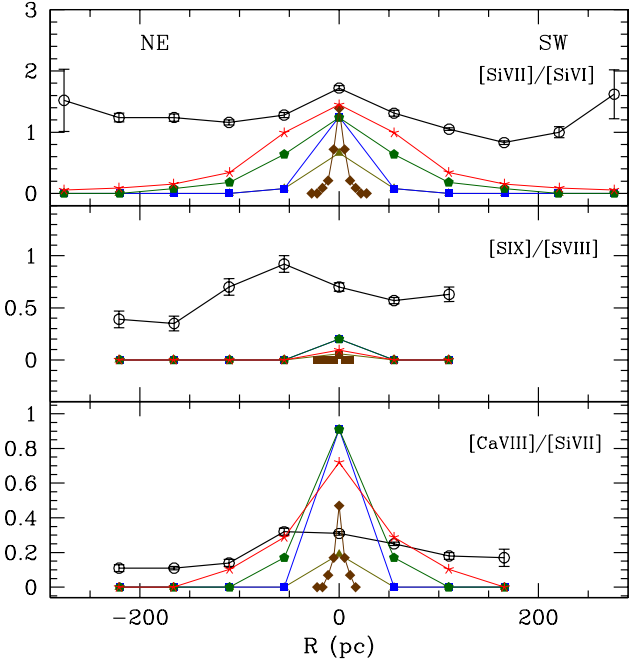}
\caption{Line flux ratios [Si\,{\sc vii}]/[Si\,{\sc vi}] (upper panel), [S\,{\sc ix}]/[S\,{\sc viii}]
(middle panel) and [Ca\,{\sc viii}]/[Si\,{\sc vii}] (bottom panel) vs distance from the
active nucleus. Observations are marked with open circles. They were corrected for extinction according to the values of E(B-V) listed in 
Column~7 of Table~\ref{tab:reddening} and assuming the law of \citet{Cardeli+}. 
Full symbols are model predictions of CLOUDY for gas clouds with no
dust and density $n_{\rm H}$ = 500~cm$^{-3}$ (stars), $n_{\rm H}$ = 10$^{3}$\,cm$^{-3}$ (pentagons), 
$n_{\rm H}$ = 10$^{4}$\,cm$^{-3}$ (squares)
and $n_{\rm H}$ = 10$^{5}$\,cm$^{-3}$ (diamonds). See text for other parameters employed in the models.}
\label{fig:lineratios}
\end{figure}

Given the difference in ionization potential between the coronal lines
listed in Table~\ref{tab:iplines}, their flux ratios are
useful to map the ionization structure and mechanisms powering them.
Fig.~\ref{fig:lineratios} shows, in open circles and from top to bottom, the observed [Si\,{\sc vii}]/[Si\,{\sc vi}], 
[S\,{\sc ix}]/[S\,{\sc viii}] and [Ca\,{\sc viii}]/[Si\,{\sc vii}] line flux ratios
in the different apertures where these lines were detected at the 3-$\sigma$ level. 
Note that the first two ratios are insensitive to abundace effects while the last one
was proposed by \citet{Ferguson+} as a reliable abundance indicator. The three line flux
ratios shown for each aperture were corrected for extinction assuming the E(B-V) listed in 
Column 7 of Table~\ref{tab:reddening} and the law of \citet{Cardeli+}.

For [Si\,{\sc vii}]/[Si\,{\sc vi}] (upper panel of Fig.~\ref{fig:lineratios}),
the largest value in the ratio is found in the nuclear region,
where it reaches 1.7$\pm$0.04. Outside the nucleus, to the NE, the
distribution is nearly flat, with values close to 1.25. Towards the SW, this ratio
decreases smoothly from the centre, reaching $\sim$1 at 220 pc.
Overall, [Si\,{\sc vii}]/[Si\,{\sc vi}] varies from $\sim$0.8 to $\sim$1.7
in the inner 550~pc from the nucleus.
The [S\,{\sc ix}]/[S\,{\sc viii}] ratio (middle panel of Figure.~\ref{fig:lineratios}) 
peaks at $\sim$55~pc NE of the NIR nucleus, where a line flux ratio of 0.9$\pm$0.2 
is measured. Further to the NE, it decreases sharply, down to 0.4$\pm$0.07 at $\sim$166~pc. 
In the opposite direction, it decreases smoothly, down to $\sim$0.6 at 150~pc from the peak.
The distribution of values for this ratio is clearly assymetric. Moreover, [S\,{\sc ix}] is 
significatively more extended towards the NE than to the SW. It is observed up to distances
of $\sim$220~pc NE of the centre. In contrast, it is detected up to $\sim$110~pc to the SW.

The bottom panel of Fig.~\ref{fig:lineratios} displays [Ca\,{\sc viii}]/[Si\,{\sc vii}].
As with the previous line ratio, it peaks at 50~pc NE of the nucleus, with a value of 0.32$\pm$0.02. 
Farther out, it decreases sharply, falling to about one third of the peak value and 
remaining nearly constant up to $\sim$220~pc NE, where [Ca\,{\sc viii}] is no longer detected. 
Towards the SW, the values of that line ratio at the different apertures decrease smoothly 
relative to the peak, reaching 0.17$\pm$0.05 at 166~pc. This ratio is also distributed 
assymetrically.

\citet{Ferguson+} presented the results of a large number
of photoionization simulations of coronal emission lines in AGNs, with the ionization
parameter $U$(H) being the fundamental parameter of their
models. They found that CLs form at distances from just outside
the broad-line region to $\sim400L^{1/2}_{43.5}$ pc, where $L_{43.5}$ 
is the ionizing luminosity in units of 10$^{43.5}$ ergs s$^{-1}$, 
in gas with ionization parameter -2.0$<$ log $U$(H) $<$ 0.75. 
This suggests that CLs form close to the nucleus in high-density gas 
and further out from the nucleus in lower density gas. The models provide 
the peak equivalent width of each line. Since that quantity is referenced
to the same point in the incident continuum, ratios between the equivalent width
of different lines should indicate grossly their expected relative strengths. 

Using Table~1 of \citet{Ferguson+}, a [Si\,{\sc vii}]/[Si\,{\sc vi}] ratio 
of 1.2 is predicted. This value is about 40\% smaller  
than the one observed in nucleus of NGC\,4388 and is similar to the ratios
observed at distances larger than 100~pc. However, Ferguson's et al. models
refer to the peak in the equivalent width distribution, which occurs
at distances of a few parsecs from the central source and gas density
$>10^6$ cm$^{-3}$. At distances of about 100\,pc, the predicted equivalenth 
width of [Si\,{\sc vii}] drops to just a few angstroms, implying in
weak emission lines, close to the detection limit.
This strongly contrasts with our observations at these distances, with equivalent 
widths of tens of Angstroms. Sulfur and calcium coronal lines 
are not that bright as those of silicon but the gas that emits them in NGC\,4388 also extends
to scales of a few hundred of parsecs.  

To further investigate whether photoionisation by the central source
can be responsible for the observed coronal line ratios, we followed the 
multi-cloud approach presented by \citet{kracre00}. To this aim, we
generated a grid of models using CLOUDY (version C013.03, \citet{fer13+}).   

The input to the models include the gas density, $n_{\rm H}$; the distance 
of the clouds to the nucleus, $R$; the AGN luminosity, the spectral energy
distribution of the ionizing radiation, the elemental abundances,
the dust/gas ratio, and the column density of the
emission-line clouds. Solar abundaces from \citet{grevesse10}
were employed in all cases. The numerical abundances relative to hydrogen are as
follows: He = 8.51 $\times ~10^{-2}$,
C = 2.69 $\times ~10^{-4}$, O = 4.9 $\times~10^{-4}$, N = 6.76 $\times ~10^{-5}$, 
Ne = 8.51 $\times ~10^{-5}$, S = 1.32 $\times ~10^{-5}$, Si = 3.24$\times ~10^{-5}$, 
Mg = 3.98 $\times ~10^{-5}$, and Fe = 3.16$\times ~10^{-5}$. No other values of 
abundances were used as no reliable indicators of that quantity
exist using NIR lines. Moreover, evidence of gas with solar abundances in
this source has been reported using optical and X-ray observations \citep{Yoshida04,shirai08+}.

The ionizing continuum employed was similar to that deduced by \citet{mafer87}. It is meant 
to represent a typical radio-quiet AGN continuum and consists of several broken power-laws of the
form $f_{\nu} \propto \nu^{\alpha}$ with $\alpha$ taking different values according to the
wavelength range. The intrinsic luminosity of NGC\,4388 above the
Lyman limit, 1.26 $\times~10^{44}$ erg\,s$^{-1}$, was estimated by \citet{vasude13}.
This value is typical of Seyfert\,1 nuclei \citep{pier94}, and was employed in all models.
Clouds with densities  $n_{\rm H} $=$ 500, 10^{3}, 10^{4}, 10^{5}$ and $10^{6}$ cm$^{-3}$
were considered at distances $R$ varying from a fraction of a parsec to $\sim$270~pc
from the nucleus, covering the range of distances mapped by our observations.

\begin{figure}
\includegraphics[width=85mm]{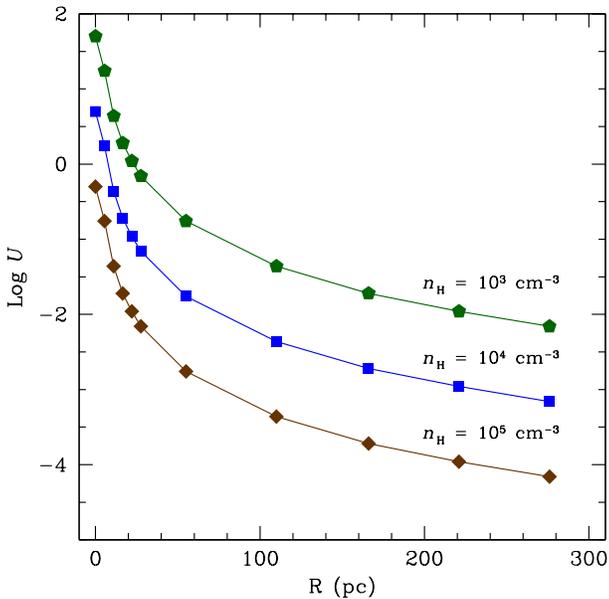}
\caption{Variation of the log of the ionization parameter, Log~$U$, with
the distance from the centre of the AGN for clouds of density
$n_{\rm H} = 10^{3}, 10^{4}$ and $10^{5}$ cm$^{-3}$. The symbols are the same
employed in Fig~\ref{fig:lineratios}. }
\label{fig:ioniza_var}
\end{figure}

Figure~\ref{fig:lineratios} displays the model predictions for clouds with gas density
$n_{\rm H} $= 500 cm$^{-3}$ (stars), $10^{3}$ cm$^{-3}$ (full pentagons), 
$10^{4}$ cm$^{-3}$ (full squares)
and 10$^{5}$ cm$^{-3}$ (full diamonds) for the three coronal line ratios
already described at the different apertures 
extracted in NGC~4388.  
Results for clouds of $n_{\rm H}$ = $10^{6}$ cm$^{-3}$ are not shown 
as they produce coronal lines only for the nuclear aperture. Full triangles show 
the predictions for clouds of $n_{\rm H}$ = 10$^{4}$ cm$^{-3}$,
assuming depletions of elements from gas phase with a size distribution and abundance
similar to the ISM of our Galaxy.  Figure~\ref{fig:ioniza_var} shows the variation of the 
ionization parameter $U$ with the distance from the centre for clouds of density 
$n_{\rm H}$ = $10^{3}$ cm$^{-3}$, $n_{\rm H}$ = 10$^{4}$ cm$^{-3}$ and
$n_{\rm H}$ = 10$^{5}$ cm$^{-3}$. The symbols are the same as those displayed in 
Figure~\ref{fig:lineratios}.

\begin{table}
\caption{Predicted emission line ratios from model components, composite and observations for the nuclear aperture}
\label{tab:mod_nucleus}
\begin{tabular}{@{}lcccc}
\hline
Line ratio & Cloud A$^1$ & Cloud B$^2$ & Composite$^3$ & Observed$^4$  \\
\hline 
\lbrack\mbox{S\,{\sc iii}}]/Pa$\beta$        & 0.83  &	0.73 &	0.78	& 10.95$\pm$0.18 \\
\lbrack\mbox{S\,{\sc viii}}]/Pa$\beta$       & 0.12  &	0.14 &	0.13	& 0.28$\pm$0.02 \\
\lbrack\mbox{S\,{\sc ii}}]/Pa$\beta$         & 0.07  &	0.06 &	0.07	& 0.83$\pm$0.05 \\
He\,{\sc i}/Pa$\beta$                        & 6.44  &  7.1  &  6.8    & 4.84$\pm$0.11 \\
\lbrack\mbox{S\,{\sc ix}}]/Pa$\beta$         & 0.02  &	0.01 &	0.02	& 0.19$\pm$0.01 \\
\lbrack\mbox{Fe\,{\sc ii}}]/Pa$\beta$         & 0.12  &   0.0 &  0.06   & 0.46$\pm$0.02 \\
\lbrack\mbox{Si\,{\sc x}}]/Pa$\beta$           & 0.06  & 0.0 &	0.03	& 0.26$\pm$0.01 \\
\lbrack\mbox{Si\,{\sc vi}}]/Br$\gamma$       &  2.25 &	2.89 &	2.57	& 2.51$\pm$0.09 \\ 
\lbrack\mbox{Si\,{\sc vii}}]/Br$\gamma$      & 2.8   &	4.02 &	3.41	& 4.31$\pm$0.11 \\
\lbrack\mbox{Ca\,{\sc viii}}]/Br$\gamma$     & 2.54  &	1.9 &	2.22	& 1.34$\pm$0.05 \\
\lbrack\mbox{Si\,{\sc vii}}]/[Si\,{\sc vi}] &  1.25  & 1.39 &	1.32	& 1.7$\pm$0.04 \\
\lbrack\mbox{S\,{\sc ix}}]/[S\,{\sc viii}]  &  0.2   & 0.07 &	0.14	& 0.7$\pm$0.04 \\
\lbrack\mbox{Si\,{\sc x}}]/[Si\,{\sc vi}]   &  0.16  & 0.0 &	0.08	& 0.55$\pm$0.05 \\
\lbrack\mbox{Ca\,{\sc viii}}]/[Si\,{\sc vii}]  & 0.91 & 0.47 &	0.69	& 0.31$\pm$0.01 \\
\hline
\multicolumn{5}{l}{$^1$ $U$ = 0.7; $R$ =  3.2~pc; $n_{\rm H}$ = $10^{4}$ cm$^{-3}$.}\\
\multicolumn{5}{l}{$^2$ $U$ = -0.3; $R$ =  3.2~pc; $n_{\rm H}$ = $10^{5}$ cm$^{-3}$.} \\
\multicolumn{5}{l}{$^3$ 50\% cloud A, 50\% cloud B.}\\
\multicolumn{5}{l}{$^4$ Dereddened; $E$($B-V$) = 1.81$\pm$0.3; Pa$\beta$ = 5.5$\pm 0.06 \times 10^{-14}$~erg\,cm$^{-2}$\,s$^{-1}$.}\\
\end{tabular}
\end{table}

\begin{table}
\caption{Predicted emission line ratios from model components, composite and observations at 55~pc NE and SW of the AGN}
\label{tab:mod_55pc}
\begin{tabular}{@{}lcccc}
\hline
Line ratio & Cloud A$^1$ & Cloud B$^2$ & Composite$^3$ & Observed$^{4,5}$  \\
\hline 
\lbrack\mbox{S\,{\sc iii}}]/Pa$\beta$	       &  2.12	& 2.35	& 2.19	 & 7.35$\pm$0.09 \\
                                                &        &       &        & 7.19$\pm$0.12 \\
\lbrack\mbox{S\,{\sc viii}}]/Pa$\beta$	       & 0.15	&  0.05	&  0.14	 & 0.21$\pm$0.02\\
                                                &        &       &        & 0.2$\pm$0.02 \\
\lbrack\mbox{S\,{\sc ii}}]/Pa$\beta$	       & 0.06	& 0.07	& 0.06	& 0.79$\pm$0.02\\
                                                &        &       &        & 0.56$\pm$0.01\\
He\,{\sc I}/Pa$\beta$	                        & 4.14	& 4.62	& 4.28	& 4.84$\pm$0.11\\
                                                &        &       &        & 4.11$\pm$0.14 \\
\lbrack\mbox{S\,{\sc ix}}]/Pa$\beta$       	& 0.0	& 0.0	& 0.0	& 0.21$\pm$0.02\\
                                                &        &       &        & 0.10$\pm$0.01\\
\lbrack\mbox{Fe\,{\sc ii}}]/Pa$\beta$	        & 0.15	& 0.16	& 0.15	& 0.46$\pm$0.02\\
                                                &        &       &        & 0.41$\pm$0.02\\ 
\lbrack\mbox{Si\,{\sc vi}}]/Br$\gamma$	        & 3.35	& 4.0	& 3.55	& 2.11$\pm$0.06\\
                                                &        &       &        & 2.85$\pm$0.11 \\
\lbrack\mbox{Si\,{\sc vii}}]/Br$\gamma$	        & 3.32	& 2.57	& 3.10	& 2.70$\pm$0.04\\
                                                &        &       &        & 3.74$\pm$0.13 \\
\lbrack\mbox{Ca\,{\sc viii}}]/Br$\gamma$	& 0.96	& 0.43	&  0.8	& 0.86$\pm$0.02\\
                                                &        &       &        & 0.94$\pm$0.04 \\
\lbrack\mbox{Si\,{\sc vii}}]/[Si\,{\sc vi}]    &  1.0   & 0.64  & 0.89   & 1.28$\pm$0.04 \\
                                               &        &       &        & 1.31$\pm$0.05 \\
\lbrack\mbox{S\,{\sc ix}}]/[S\,{\sc viii}]     &  0.0   & 0.0   & 0.0	 & 0.92$\pm$0.08 \\
                                               &        &       & 	 & 0.57$\pm$0.03 \\
\lbrack\mbox{Ca\,{\sc viii}}]/[Si\,{\sc vii}]  &  0.29  & 0.17  & 0.25	 & 0.32$\pm$0.02 \\
                                               &        &       &        & 0.25$\pm$0.01 \\
\hline
\multicolumn{5}{l}{$^1$ $U$ = -0.46; $R$ =  55~pc; $n_{\rm H}$ = 500 cm$^{-3}$.}\\
\multicolumn{5}{l}{$^2$ $U$ = -0.76; $R$ =  55~pc; $n_{\rm H}$ = $10^{3}$ cm$^{-3}$.} \\
\multicolumn{5}{l}{$^3$ 70\% cloud A, 30\% cloud B.}\\
\multicolumn{5}{l}{$^4$ The first entry corresponds to the NE aperture while the second one}\\
\multicolumn{5}{l}{~~~~to the SW aperture.}\\
\multicolumn{5}{l}{$^5$ Dereddened; $E$($B-V$)$_{NE}$ = 1.27$\pm$0.25; $E$($B-V$)$_{SW}$ = 0.77$\pm$0.1.}\\
\multicolumn{5}{l}{~~~~ Pa$\beta_{NE}$ = 2.34$\pm 0.03 \times 10^{-14}$~erg\,cm$^{-2}$\,s$^{-1}$}\\
\multicolumn{5}{l}{~~~~ Pa$\beta_{SW}$ = 2.31$\pm 0.04 \times 10^{-14}$~erg\,cm$^{-2}$\,s$^{-1}$.} \\
\end{tabular}
\end{table}

An inspection to the [Si\,{\sc vii}]/[Si\,{\sc vi}] line ratio in Figure~\ref{fig:lineratios} 
shows that all models, regardless of the density, peak at a similar value of that
ratio ($\sim$1.4). The main difference among them is on the size of the emission region
as well as on the value of the ionization parameter. High-density
clouds ($n_{\rm H} > 10^{4}$ cm$^{-3}$) produce very compact [Si\,{\sc vii}] and [Si\,{\sc vi}] 
emission regions, peaking at the nucleus and extending only to the central few tens of parsecs.
The ionization parameter for these clouds is high but still within the
expected conditions of the NLR. In contrast, lower density clouds
($n_{\rm H}$ = 500 and $10^{3}$ cm$^{-3}$) emit silicon up to distances of a few hundred of parsecs. 
However, in the innermost few parsecs, the ionization parameter for these later
clouds is rather high, up to two orders of magnitude higher than that expected for NLR clouds.

We found that a suitable combination of clouds of density 
$n_{\rm H}$ = 500, 10$^{3}, 10^{4}$ and $10^{5}$ cm$^{-3}$
are able to reproduce the observed values (open circles) within the central 55~pc from the nucleus. 
At larger distances, these clouds do not produce enough [Si\,{\sc vii}] to sustain the large values
of that ratio
compatible to the observations. At 110~pc, for instance, low-density clouds ($n_{\rm H} \leq 10^{3}$
cm$^{-3}$) can account for only 30\% of the observed ratio while at 160~pc and farther away,  
they are responsible for less than 10\% of the observed values. 
Note that we have assumed that the clouds in all apertures are 
not coplanar. Therefore, no screening of one cloud by the other takes place. This warrants
that all clouds are illuminated by the same AGN continuum. We also assume that the coronal 
lines are emitted preferencially by these clouds. This may not be the case for mid- and low-ionization 
lines, for which clouds not facing directly the central source may also contribute to the
observed flux.

\begin{table}
\centering
\caption{Predicted emission line ratios from model components, composite and observations at 110~pc NE and SW of the AGN}
\label{tab:mod_110pc}
\begin{tabular}{@{}lcccc}
\hline
Line ratio & Cloud A$^1$ & Cloud B$^2$ & Composite$^3$ & Observed$^{4,5}$  \\
\hline 
\lbrack\mbox{S\,{\sc iii}}]/Pa$\beta$	       & 3	& 3.9	& 3.09	 & 5.74$\pm$0.06	\\
                                               &        &       &        & 6.51$\pm$0.13\\
\lbrack\mbox{S\,{\sc viii}}]/Pa$\beta$	       & 0.09	& 0.0	& 0.09	 &  0.17$\pm$0.05	\\
                                               &        &       &        & 0.10$\pm$0.01\\
\lbrack\mbox{S\,{\sc ii}}]/Pa$\beta$	       &  0.06	& 0.09	& 0.06	 & 0.44$\pm$0.01	\\
                                               &        &       &        & 0.22$\pm$0.03\\
He\,{\sc I}/Pa$\beta$	                       & 3.22	& 3.9	& 3.29	 & 2.32$\pm$0.06	\\
                                               &        &       &        & 3.19$\pm$0.14\\
\lbrack\mbox{S\,{\sc ix}}]/Pa$\beta$	       & 0.0	& 0.0	& 0.0	& 0.15$\pm$0.03	\\
                                               &        &       &        & 0.08$\pm$0.01\\
\lbrack\mbox{Fe\,{\sc ii}}]/Pa$\beta$	       & 0.24	& 0.31	& 0.25	 & 0.32$\pm$0.03	\\
                                               &        &       &        & 0.49$\pm$0.02\\
\lbrack\mbox{Si\,{\sc vi}}]/Br$\gamma$	       & 3.6	& 2.7	& 3.51	 & 2.20$\pm$0.14	\\
                                               &        &       &        & 2.82$\pm$0.32\\
\lbrack\mbox{Si\,{\sc vii}}]/Br$\gamma$	       & 1.24	& 0.49	& 1.17	 & 2.56$\pm$0.18	\\
                                               &        &       &        & 2.96$\pm$0.36\\
\lbrack\mbox{Ca\,{\sc viii}}]/Br$\gamma$	& 0.13	& 0.0	& 0.12	 & 0.36$\pm$0.04	\\
                                               &        &       &        & 0.53$\pm$0.06\\
\lbrack\mbox{Si\,{\sc vii}}]/[Si\,{\sc vi}]    &  0.34  & 0.18  & 0.32   & 1.16$\pm$0.04 \\
                                               &        &       &        & 1.05$\pm$0.2 \\
\lbrack\mbox{S\,{\sc ix}}]/[S\,{\sc viii}]     &  0.0   & 0.0   & 0.0	 & 0.70$\pm$0.08 \\
                                               &        &       & 	 & 0.63$\pm$0.07 \\
\lbrack\mbox{Ca\,{\sc viii}}]/[Si\,{\sc vii}]  &  0.1   & 0.0   & 0.09	 & 0.14$\pm$0.02 \\
                                               &        &       &        & 0.18$\pm$0.02 \\
\hline
\multicolumn{5}{l}{$^1$ $U$ = -1.06; $R$ =  110~pc; $n_{\rm H}$ = 500 cm$^{-3}$.}\\
\multicolumn{5}{l}{$^2$ $U$ = -1.36; $R$ =  110~pc; $n_{\rm H}$ = $10^{3}$ cm$^{-3}$.} \\
\multicolumn{5}{l}{$^3$ 90\% cloud A, 10\% cloud B.}\\
\multicolumn{5}{l}{$^4$ The first entry corresponds to the NE aperture while the second one}\\
\multicolumn{5}{l}{~~~~to the SW aperture.}\\
\multicolumn{5}{l}{$^5$ Dereddened; $E$($B-V$)$_{NE}$ = 1.09$\pm$0.15; $E$($B-V$)$_{SW}$ = 0.43$\pm$0.05}\\
\multicolumn{5}{l}{~~~~ Pa$\beta_{NE}$ = 1.44$\pm 0.02 \times 10^{-14}$~erg\,cm$^{-2}$\,s$^{-1}$}\\
\multicolumn{5}{l}{~~~~ Pa$\beta_{SW}$ = 1.5$\pm 0.03 \times 10^{-14}$~erg\,cm$^{-2}$\,s$^{-1}$.} \\
\end{tabular}
\end{table}

CLOUDY predictions for [S\,{\sc ix}]/[S\,{\sc viii}] are shown in the middle 
panel of Figure~\ref{fig:lineratios}. It can be seen that all models, regardless 
of the density, underpredict the observations. 
The observed extinction corrected ratio at the nucleus is 0.7$\pm$0.04 
while clouds with $n_{\rm H}$ = 10$^{4}$ cm$^{-3}$ predict a ratio of 0.2.
Moreover, none of the clouds considered are able to form [S\,{\sc ix}] 
and [S\,{\sc viii}] at distances larger than 55~pc and 110~pc, respectively, from the centre. 
This strongly contrasts with our observations, as we detect [S\,{\sc viii}] 
as far as 221~pc NE and SW of the nucleus. The gas emitting 
[S\,{\sc ix}] is less extended than that of [S\,{\sc viii}] to the SW but 
yet it is detected up to 110~pc from the AGN. Thus, clouds with 
suitable physical conditions for the NLR fail at reproducing our
data. It is important to mention that the values for that
ratio in NGC\,4388 do not appear to be unusual among AGN. For example, NGC\,1068 shows sulfur coronal 
line ratios rather similar to those of NGC\,4388 in its nuclear and extended NLR \citep{martins+10}. 
Moreover, the observed sulfur ratios in NGC\,4388 are also well within the distribution of values
for that ratio presented by \citet{Rodriguez11} in their study of coronal lines in 54~AGNs. 
Thus, sulfur coronal line emission posses a challenge to photoionization by the central
source.

Model results for [Ca\,{\sc viii}]/[Si\,{\sc vii}]
show that clouds with $n_{\rm H} \sim 10^{4-5}$ cm$^{-3}$ reproduce the data points at 
the nucleus while lower density clouds dominate the emission of highly-ionized calcium 
farther out. Note, however, that CLOUDY is unable
to produce [Ca\,{\sc viii}] at distances larger than 110~pc while that ion is
detected up to 220~pc NE and 160~pc SW in NGC\,4388.
 \citet{Ferguson+} proposed the [Ca\,{\sc viii}]/[Si\,{\sc vii}] line flux ratio as a suitable 
abundance indicator. Assuming solar abundances, the models reproduce consistently 
the observations up to the point where [Ca\,{\sc viii}] is theoretically produced.   
Optical and X-ray observations \citep{Yoshida04,shirai08+} had already pointed out 
solar metallicity in the nucleus of NGC\,4388. Our results support these previous findings.

\begin{figure*}
\includegraphics[width=140mm]{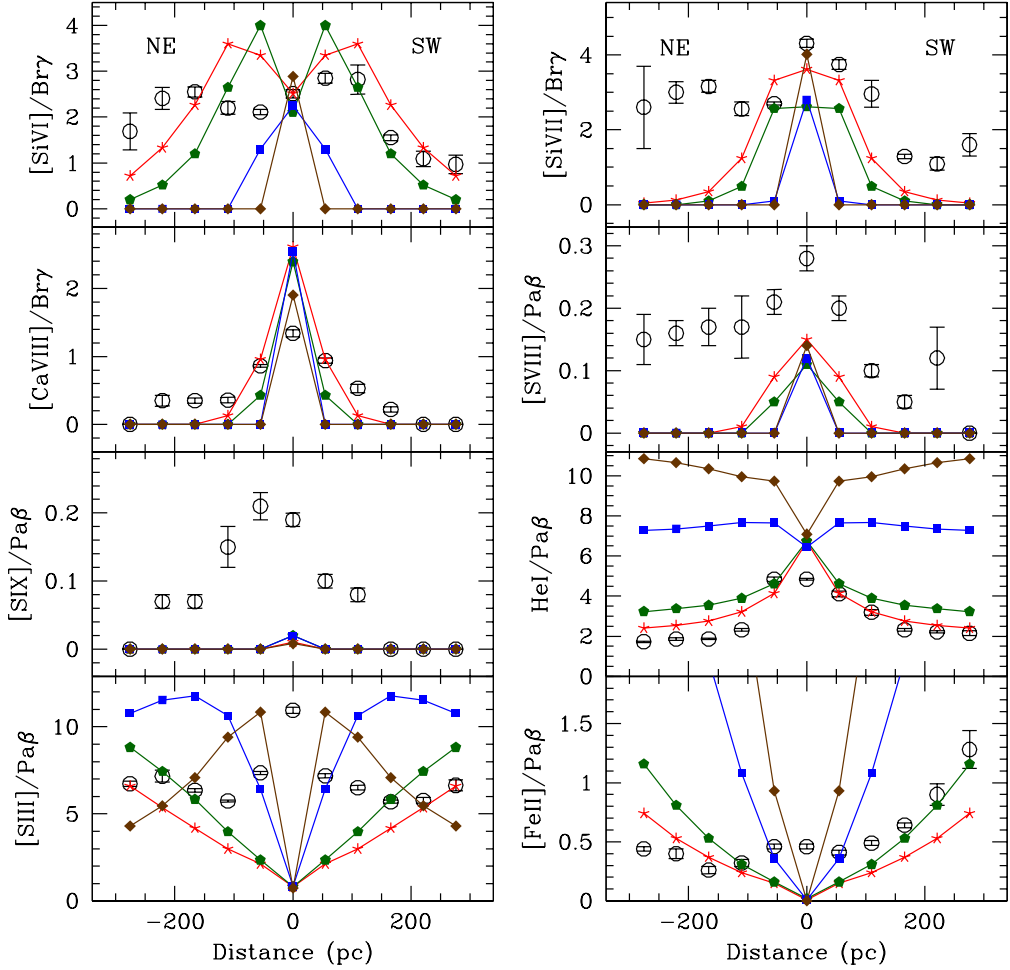}
\caption{Emission line flux ratios vs distance from the active nucleus. Observations are marked with
open circles. They were corrected for extinction according to the values of E(B-V) listed in 
Column~7 of Table~\ref{tab:reddening} and assuming the \citet{Cardeli+} law. 
Full symbols are model predictions of CLOUDY for gas clouds with no
dust and density $n_{\rm H}$ = 500~cm$^{-3}$ (stars), $n_{\rm H}$ = 10$^{3}$\,cm$^{-3}$ (pentagons), 
$n_{\rm H}$ = 10$^{4}$\,cm$^{-3}$ (squares)
and $n_{\rm H}$ = 10$^{5}$\,cm$^{-3}$ (diamonds).
See text for other parameters employed in the models.}
\label{fig:models_obs}
\end{figure*}

In order to check the consistency of our results, we have compared
high- mid- and low-ionization lines to hydrogen recombination lines.
The results are shown in Figure~\ref{fig:models_obs}, where eight different ratios are
plotted. Observed data are represented by open circles while the models follow the same convention
employed in Figure~\ref{fig:lineratios}. A rapid inspection to Figure~\ref{fig:models_obs}  
reveals that for most coronal lines the models strongly underpredict the observations for apertures 
$\sim$150~pc and farther away from the centre. At these same positions low- and mid-ionization 
lines are well covered by CLOUDY.

It is possible, by suitably combining clouds of different physical conditions
weighted by their contribution to the total flux, to fit the observed emission line ratios 
displayed in Figures~\ref{fig:lineratios} and~\ref{fig:models_obs} at each aperture. 
This multi-component approach is based on the evidence that components of 
different densities at the same radial distances exist in the NLR. It has been succesfully 
employed in the literature \citep{schulz+,kracre00} to model the NLR and extended NLR of AGNs. 
Tables~\ref{tab:mod_nucleus} to~\ref{tab:mod_276pc} 
list the predicted line ratios at the nucleus and at the off-nuclear apertures.
The composite flux was obtained after combining the output of two given pairs of models plotted in 
Figures~\ref{fig:lineratios} and~\ref{fig:models_obs}. For simplicity, we have limited ourselves to a
two-component model (named Cloud~A and Cloud~B) although we are aware that
this is a degenerate problem, so that multiple solutions for each aperture are possible. 
The main goal here is to illustrate the range of physical conditions necessary to 
to fit the coronal and low- to mid-ionization lines. 

An inspection to Table~\ref{tab:mod_nucleus} shows that clouds of $n_{\rm H}$ = 10$^{4}$ cm$^{-3}$
and 10$^{5}$ cm$^{-3}$, located at $\sim$3~pc from the centre, are able to reproduce the 
line ratios measured in the nucleus of NGC\,4388. The good agreement found between model 
predictions and observations for most line ratios points out that photoionization by the AGN is the 
main source of excitiation for the nuclear gas.  

Farther out, at 55~pc from the centre (see Table~\ref{tab:mod_55pc}), photoionization by the AGN is still able 
to reproduce most of the observations. However, the density of the clouds responsible for the production
of the emission lines must be lower than that in 
the nucleus, $n_{\rm H}~ \leq$ 10$^{3}$ cm$^{-3}$. Indeed, a larger contribution of clouds with 
$n_{\rm H}$ = 500 cm$^{-3}$ (70\% by weight) was found. As can be observed in
Figure~\ref{fig:models_obs}, clouds with $n_{\rm H}~\geq ~10^4$ cm$^{-3}$ produce negligible or
no coronal lines at all at that position. At 110~pc from the AGN, the bulk of clouds responsible for 
the production of high-ionization lines is dominated by low-density clouds, of $n_{\rm H}$ $\approx$ 500 cm$^{-3}$. 
This result is shown in Table~\ref{tab:mod_110pc}. It can also be seen that CLOUDY underpredicts
by a factor of 2-3 the observations of CLs, pointing out that either a more complex approach is needed
or that photoionization by the AGN starts failing at sustaining the production of high-ionization 
lines. Low-ionization lines, in contrast, are well reproduced by the models. Their production
are largely favored by clouds of $n_{\rm H}~ \leq$ 10$^{3}$ cm$^{-3}$, giving further support to
our results. Recall that [S\,{\sc iii}]\,9531~\AA\ and He\,{\sc i}~1.083~$\mu$m are the strongest
lines detected in all apertures.

Tables~\ref{tab:mod_166pc} to~\ref{tab:mod_276pc} show that at 166~pc and farther away from the 
AGN, photoionization by the central source alone is no longer able to reproduce the observations
of high-ionization lines: CLOUDY 
predicts line ratios that are one order of magnitude smaller than the observations or it is unable
to produce them. Indeed, at 221~pc, 
clouds with $n_{\rm H}~ \leq$ 10$^{3}$ cm$^{-3}$ are unable to
produce coronal lines while they still sustain the production of low- to mid-ionization lines.  
We also tested models with clouds of gas density as low as $n_{\rm H}$ = 200 cm$^{-3}$. 
This later value was derived by \citet{Colina92} from the optical [S\,{\sc ii}]~$\lambda\lambda$6717,6731
doublet in NGC\,4388 for a projected distance of 380~pc from the nucleus and agrees with the density found by
\citet{Petitjean+} at a similar distance from the central source. 
Although such clouds do produce coronal lines, the predicted line ratio [Si\,{\sc vii}]/[Si\,{\sc vi}]
is 0.21. That it, a factor $\sim$5 lower than observed.

\begin{table}
\centering
\caption{Predicted emission line ratios from model components, composite and observations at 166~pc NE and SW of the AGN}
\label{tab:mod_166pc}
\begin{tabular}{@{}lcccc}
\hline
Line ratio & Cloud A$^1$ & Cloud B$^2$ & Composite$^3$ & Observed$^{4,5}$  \\
\hline 
\lbrack\mbox{S\,{\sc iii}}]/Pa$\beta$	       & 4.2	& 6	& 4.68	 & 6.35$\pm$0.10	\\
                                               &        &       &        & 5.70$\pm$0.09\\
\lbrack\mbox{S\,{\sc viii}}]/Pa$\beta$	       & 0.01	& 0.0	& 0.01	 & 0.17$\pm$0.03	\\
                                                &        &       &        & 0.05$\pm$0.01\\
\lbrack\mbox{S\,{\sc ii}}]/Pa$\beta$	       & 0.08	& 0.0	& 0.10	&  0.37$\pm$0.05	\\
                                               &        &       &        &  0.64$\pm$0.03\\
He\,{\sc I}/Pa$\beta$	                       & 2.8	& 3.60	& 3.04	 & 1.87$\pm$0.04	\\
                                               &        &       &        &  2.32$\pm$0.06\\
\lbrack\mbox{S\,{\sc ix}}]/Pa$\beta$	       & 0.0	&  0.0	& 0.0	&  0.07$\pm$0.01	\\
                                               &        &       &        & \\
\lbrack\mbox{Fe\,{\sc ii}}]/Pa$\beta$	       & 0.4	& 0.54	& 0.44	& 0.26$\pm$0.03	\\
                                               &        &       &        &  0.64$\pm$0.02\\
\lbrack\mbox{Si\,{\sc vi}}]/Br$\gamma$	       & 2.3	& 1.20	& 1.97	 & 2.55$\pm$0.11	\\
                                               &        &       &        &  1.55$\pm$0.06\\
\lbrack\mbox{Si\,{\sc vii}}]/Br$\gamma$	       & 0.3	& 0.10	& 0.24	& 3.16$\pm$0.16	\\
                                               &        &       &        &  1.29$\pm$0.06\\
\lbrack\mbox{Ca\,{\sc viii}}]/Br$\gamma$       & 0.0	& 0.0	& 0.0	& 0.35$\pm$0.04	\\
                                               &        &       &        &  0.22$\pm$0.04\\
\lbrack\mbox{Si\,{\sc vii}}]/[Si\,{\sc vi}]    &  0.16  & 0.08  & 0.14   & 1.24$\pm$0.06 \\
                                               &        &       &        & 0.83$\pm$0.03 \\
\lbrack\mbox{S\,{\sc ix}}]/[S\,{\sc viii}]     &  0.0   & 0.0   & 0.0	 & 0.35$\pm$0.07 \\
                                               &        &       & 	 & ... \\
\lbrack\mbox{Ca\,{\sc viii}}]/[Si\,{\sc vii}]  &  0.0   & 0.0   & 0.0	 & 0.11$\pm$0.01 \\
                                               &        &       &        & 0.17$\pm$0.05 \\
 \hline
\multicolumn{5}{l}{$^1$ $U$ = -1.42; $R$ =  166~pc; $n_{\rm H}$ = 500 cm$^{-3}$.}\\
\multicolumn{5}{l}{$^2$ $U$ = -1.72; $R$ =  166~pc; $n_{\rm H}$ = $10^{3}$ cm$^{-3}$.} \\
\multicolumn{5}{l}{$^3$ 70\% cloud A, 30\% cloud B.}\\
\multicolumn{5}{l}{$^4$ The first entry corresponds to the NE aperture while the second one}\\
\multicolumn{5}{l}{~~~~to the SW aperture.}\\
\multicolumn{5}{l}{$^5$ Dereddened; $E$($B-V$)$_{NE}$ = 0.73$\pm$0.06; $E$($B-V$)$_{SW}$ = 0.60$\pm$0.12}\\
\multicolumn{5}{l}{~~~~ Pa$\beta_{NE}$ = 6.84$\pm 0.08 \times 10^{-15}$~erg\,cm$^{-2}$\,s$^{-1}$}\\
\multicolumn{5}{l}{~~~~ Pa$\beta_{SW}$ = 10.9$\pm 0.13 \times 10^{-15}$~erg\,cm$^{-2}$\,s$^{-1}$.} \\
\end{tabular}
\end{table}

\begin{table}
\caption{Coronal line ratios from model and observations at 221~pc NE and SW of the AGN}
\label{tab:mod_221pc}
\begin{tabular}{@{}lcccc}
\hline
Line ratio & Cloud A$^1$ &  Cloud B$^2$ & Composite$^3$ & Observed$^{4,5,6}$  \\
\hline 
\lbrack\mbox{S\,{\sc iii}}]/Pa$\beta$	       & 5.40	& 7.40	& 6.20	& 7.17$\pm$0.34	\\
                                               &        &      &        & 5.78$\pm$0.16\\
\lbrack\mbox{S\,{\sc viii}}]/Pa$\beta$	       & 0.0	& 0.0	& 0.0	& 0.16$\pm$0.02	\\
                                               &        &      &        & 0.12$\pm$0.05\\
\lbrack\mbox{S\,{\sc ii}}]/Pa$\beta$	       & 0.10	& 0.20	& 0.14	& 0.45$\pm$0.12	\\
                                               &        &      &        & 0.34$\pm$0.08\\
He\,{\sc I}/Pa$\beta$	                       & 2.53	& 3.40	& 2.88	& 1.86$\pm$0.07	\\
                                               &        &      &        & 2.22$\pm$0.06\\
\lbrack\mbox{S\,{\sc ix}}]/Pa$\beta$	       & 0.0	& 0.0	& 0.0	& 0.10$\pm$0.02	\\
                                               &        &      &        & ... \\
\lbrack\mbox{Fe\,{\sc ii}}]/Pa$\beta$	       & 0.53	& 0.80	& 0.64	& 0.40$\pm$0.04	\\
                                               &        &      &        & 0.9$\pm$0.09\\
\lbrack\mbox{Si\,{\sc vi}}]/Br$\gamma$	       & 1.30	& 0.50	& 0.98	& 2.41$\pm$0.24	\\
                                               &        &      &        & 1.09$\pm$0.17\\
\lbrack\mbox{Si\,{\sc vii}}]/Br$\gamma$	       & 0.10	& 0.0	& 0.06	& 3.00$\pm$0.29	\\
                                               &        &      &        & 1.09$\pm$0.17\\
\lbrack\mbox{Ca\,{\sc viii}}]/Br$\gamma$	& 0.0	& 0.0	& 0.0	& 0.35$\pm$0.08	\\
                                               &        &      &        & ... \\
\lbrack\mbox{Si\,{\sc vii}}]/[Si\,{\sc vi}]    &  0.09  & 0.0  & 0.06   & 1.24$\pm$0.07 \\
                                               &        &      &       & 1.0$\pm$0.1 \\
\lbrack\mbox{S\,{\sc ix}}]/[S\,{\sc viii}]     &  0.0   & 0.0  & 0.0    & 0.39$\pm$0.08 \\
                                               &        &      &       & ... \\
\lbrack\mbox{Ca\,{\sc viii}}]/[Si\,{\sc vii}]  &  0.0   & 0.0  & 0.0    & 0.11$\pm$0.01 \\
                                               &        &      &       &  ... \\
\hline
\multicolumn{5}{l}{$^1$ $U$ = -1.66; $R$ =  221~pc; $n_{\rm H}$ = 500 cm$^{-3}$.}\\
\multicolumn{5}{l}{$^2$ $U$ = -1.96; $R$ =  221~pc; $n_{\rm H}$ = $10^{3}$ cm$^{-3}$.}\\
\multicolumn{5}{l}{$^3$ 60\% cloud A, 40\% cloud B.} \\ 
\multicolumn{5}{l}{$^4$Only [Si\,{\sc vii}], [Si\,{\sc vi}] and [S\,{\sc ix}] are detected to the SW at this} \\
\multicolumn{5}{l}{~~~~position. For that reason, only [Si\,{\sc vii}]/[Si\,{\sc vi}] is shown.}\\
\multicolumn{5}{l}{$^5$ The first entry corresponds to the NE aperture while the second one}\\
\multicolumn{5}{l}{~~~~to the SW aperture.}\\
\multicolumn{5}{l}{$^6$ Dereddened; $E$($B-V$)$_{NE}$ = 0.41$\pm$0.09; $E$($B-V$)$_{SW}$ = 0.8$\pm$0.3. }\\
\multicolumn{5}{l}{~~~~ Pa$\beta_{NE}$ = 2.90$\pm 0.07 \times 10^{-15}$~erg\,cm$^{-2}$\,s$^{-1}$}\\
\multicolumn{5}{l}{~~~~ Pa$\beta_{SW}$ = 4.31$\pm 0.05 \times 10^{-15}$~erg\,cm$^{-2}$\,s$^{-1}$.} \\
\end{tabular}
\end{table}

The inclusion of dusty clouds (although appealling for several
reasons) does not help to improve the fits. The model that 
takes them into account, shown in Figure~\ref{fig:lineratios}, produces too 
little coronal line emission that its contribution appear to be negligible in the
central apertures. We also tested models with other values of densities for that
type of clouds with similar
results as those shown here. \citet{Ferguson+} had already demonstrated  that
dust in regions of large $U$ absorbs much of the
incident continuum. This greatly reduces the strength of the coronal lines,
implying that they must be formed in nearly dust-free gas.
Thus, under this scenario, most of the extinction affecting the observed spectra 
should be produced by dust located outside of the CLR.

\begin{table}
\caption{Coronal line ratios from model and observations at 276~pc NE and SW of the AGN}
\label{tab:mod_276pc}
\begin{tabular}{@{}lcccc}
\hline
Line ratio & Cloud A$^1$ &  Cloud B$^2$ & Composite$^3$ & Observed$^{4,5,6}$  \\
\hline 
\lbrack\mbox{S\,{\sc iii}}]/Pa$\beta$	       & 6.61	& 8.81	& 7.05	& 6.73$\pm$0.15	\\
                                               &        &      &        & 6.64$\pm$0.32\\
\lbrack\mbox{S\,{\sc viii}}]/Pa$\beta$	      &  0.0	& 0.0	& 0.00	& 0.15$\pm$0.04	\\
                                               &        &       &    & ... \\
\lbrack\mbox{S\,{\sc ii}}]/Pa$\beta$	      &  0.13	& 0.23	& 0.15	& 0.28$\pm$0.06	\\
                                               &        &      &        & ... \\
He\,{\sc I}/Pa$\beta$	                      & 2.4	& 3.23	& 2.57	& 1.72$\pm$0.04	\\
                                               &        &      &        & 2.15$\pm$0.12\\
\lbrack\mbox{Fe\,{\sc ii}}]/Pa$\beta$	      &  0.74	& 1.16	& 0.82	& 0.44$\pm$0.02	\\
                                               &        &      &        & 1.28$\pm$0.16\\
\lbrack\mbox{Si\,{\sc vi}}]/Br$\gamma$	      & 0.72	& 0.2	& 0.62	& 1.69$\pm$0.4	\\
                                               &        &      &        & 0.97$\pm$0.2\\
\lbrack\mbox{Si\,{\sc vii}}]/Br$\gamma$	      & 0.04	& 0.0	& 0.03	& 2.6$\pm$1.1	\\
                                               &        &      &        & 1.6$\pm$0.3\\
\lbrack\mbox{Si\,{\sc vii}}]/[Si\,{\sc vi}]	& 0.06	&  0.0	& 0.05	& 1.5$\pm$0.51	 \\
                                               &        &      &        & 1.6$\pm$0.4 \\
\hline
\multicolumn{5}{l}{$^1$ $U$ = -2.16; $R$ =  276~pc; $n_{\rm H}$ = 500 cm$^{-3}$.}\\
\multicolumn{5}{l}{$^2$ $U$ = -1.96; $R$ =  276~pc; $n_{\rm H}$ = $10^{3}$ cm$^{-3}$.}\\
\multicolumn{5}{l}{$^3$ 80\% cloud A, 20\% cloud B.} \\ 
\multicolumn{5}{l}{$^4$[Si\,{\sc vii}], [Si\,{\sc vi}] and [S\,{\sc viii}] are the only CLs detected at this} \\
\multicolumn{5}{l}{~~~~position. Thus, only line ratios involving these lines are shown.}\\
\multicolumn{5}{l}{$^5$ The first entry corresponds to the NE aperture while the second one}\\
\multicolumn{5}{l}{~~~~to the SW aperture.}\\
\multicolumn{5}{l}{$^6$ Dereddened; $E$($B-V$)$_{NE}$ = 0.41$\pm$0.09; $E$($B-V$)$_{SW}$ = 0.8$\pm$0.3. }\\
\multicolumn{5}{l}{~~~~ Pa$\beta_{NE}$ = 1.07$\pm 0.02 \times 10^{-15}$~erg\,cm$^{-2}$\,s$^{-1}$}\\
\multicolumn{5}{l}{~~~~ Pa$\beta_{SW}$ = 1.09$\pm 0.05 \times 10^{-15}$~erg\,cm$^{-2}$\,s$^{-1}$.} \\
\end{tabular}
\end{table} 

We should also note that the line flux ratio [Si\,{\sc x}]/[Si\,{\sc vi}] 
(or [Si\,{\sc x}]/Pa$\beta$)
is strongly underestimated by the models (see Table~\ref{tab:mod_nucleus}).
Recall that [Si\,{\sc x}] is detected only in the nuclear aperture.
For the assumed luminosity and form
of the ionizing continuum, sufficient [Si\,{\sc x}] relative to [Si\,{\sc vi}]
is produced if the emitting clouds are located 0.3~pc or closer to the AGN.
That is, a factor of 10 closer to the central source than the value
assumed.
This implies that the bulk of [Si\,{\sc x}] is produced in the boundaries of the 
broad line region and the torus.  
The [Si\,{\sc x}]/[Si\,{\sc vi}] ratio is also a
robust indicator of the form of the input ionizing continuum because of the large difference in
ionization potential of the lines involved ($\Delta$IP = 184~eV). A better
fit to the observations can be obtained if the clouds emitting [Si\,{\sc x}]  
are illuminated by a harder continuum than that of the other coronal lines.

Although it is clear that photoinization by the central source 
accounts for most of the observed coronal line strengths not only
in the nucleus but in regions as far as 110~pc away of it, at larger distances the 
model outputs largely underpredicts our observations. This suggests
that additional mechanisms should be present to enhace the high-ionization spectrum observed. 
In the previous sections, 
we found that the high-ionization gas is primarily distributed along
the elongated nuclear radio structure, tracing an outflow that 
extends up to 200~pc SW of the nucleus. Moreover, 
the presence of splitted components, such as the ones we
detected in [Fe\,{\sc ii}] evidences interaction between the radio-jet
and the ambient gas. \citet{Greene14}
presented examples of [Si\,{\sc vi}]~1.963~$\mu$m emission line profiles along 
the ionised cone, giving additional support to this scenario.
It then makes sense to consider the role of 
shocks produced by interactions between the radio jet or a radially 
accelerated outflow and the ISM to enhance the high-ionization lines
strengths in the off-nuclear apertures.  Note, for example, that 
\citet{kracre00} showed the necessity of fast 
shocks ($\sim$ 1000\,km\,s$^{-1}$) at distances larger than 100~pc
from the nucleus in NGC\,1068 to explain the high [Ne\,{\sc v}] and [Fe\,{\sc vii}]
strengths relative to H$\beta$ measured in that object. No suitable combination 
of clouds powered by photoinisation by the central source could explain 
the observed ratios.

The models of \citet[][CV01 hereafter]{Contini+}
are useful to test the effects of shocks coupled
to photoionization by the central source; even in the presence of shocks, the effect of the
central continuum cannot be ignored. In their modelling, CV01
considered that the clouds are moving outwards from the galaxy centre, 
with the shock front forming on the outer edge of the cloud, whereas 
the ionizing radiation reaches the opposite edge that faces the active centre. 
The ionization due to both the primary radiation (from the central source)
and the diffuse radiation generated by the free-free and free-bound
transitions of the shocked and photoionized gas, as well as the collisional
ionization, are all accounted for. The shock velocity $V_{\rm s}$ and
the ionizing flux from the central source at the Lyman limit reaching
the cloud, $F_{\rm h}$ (in units of cm$^{-2}$ s$^{-1}$ eV$^{-1}$), are the main 
input parameters. Other parameters include the pre-shock density, $n_0$, and
the pre-shock magnetic field, $B0$.

Tables~1 to~12 of CV01 show that shock-dominated clouds
($F_{\rm h}$ = 0) with shock velocities in the range 300 $-$ 1500 km\,s$^{-1}$
strongly favour the production of coronal lines. They predict [Si\,{\sc vii}]/[Si\,{\sc vi}] ratios 
between 1.5 and 3 and 0.4 $<$ [S\,{\sc ix}]/[S\,{\sc viii}] $<$ 1
(models 56 and 70 for example). 
For all these clouds, solar metallicity, $n_{0}$ = 300\,cm$^{-3}$ and B$_0$ = 10$^{-4}$ Gauss
were adopted. When coupled to the presence of the radiation
field from the central source (e.g. log $F_{\rm h}$ = 12, model 62),
the value of these two ratios ($\sim$0.8 and $\sim$0.1, respectively) are still
consistent with our observations.  

Other models that consider the effects of shocks provide
additional support to the above results. \citet{Allen+}
published a library of fully radiative shock models that include 
both the radiative shock and its photoionized precursor. 
Similar to the models of CV01, the shock velocity, the pre-shock density 
and the intensity of the magnetic field are the main input parameters. 
Model predictions for the composite shock+precursor structure for the 
[Si\,{\sc vii}]/[Si\,{\sc vi}] ratio are found to vary from 0.5 to 1.2 
for velocities between 500 and 1000 km\,s$^{-1}$, respectively, and 
$n0$ of 100 cm$^{-3}$. The shock velocities are significantly higher
than those found by \citet{Ciroi+} in their modelling of the
optical spectra of the SW cone of NGC\,4388. Composite models that 
account for the combined effects of photoionization and shocks
were able to reproduce the observed, high [O\,{\sc iii}]\,$\lambda$5007/H$\beta$ ratio, 
marginally consistent with simple photoionization.

The alignment between radio-jets and the morphology of the 
ionized gas has also been observed in other Seyfert galaxies by
means of 3D IFU observations \citep{MullerS+11,Mazzalay13,Riffel15,Riffel15b}. 
One example is NGC\,1068, well-known for displaying a prominent radio-jet in the nuclear 
and circumnuclear region, with most of the NLR ionized gas following the 
morphology and structure of the radio-emission \citep{axon98,exposito11,Mazzalay13}. 
\citet{geballe09} had already found enhanced coronal line emission where 
the radio jet changes direction, while \citep{Mazzalay13} showed the
necessity of shocks to reproduce the enhanced coronal line ratios 
as far as 150~pc from the AGN.
The case for NGC\,4388 is not much different.
The observed coronal emission follows that of the radio-jet, as observed
here from our data and from \citet{Greene14}.
Moreover, outflowing ionized gas has been systematically reported
in this object. For the data presented here,  evidence
of outflowing gas is easily seen in Figure~\ref{fig:residuals}, when
the disk rotation is subtracted from the observed velocity radial 
velocity measured in the coronal lines. Blueshifted gas is clearly
seen in the NE while redshifted gas is detected towards the SW.
The outflowing component is the largest for the high-ionization
lines and enhanced towards the SW, where the extinction is 
significantly reduced.   

We should keep in mind, though,
that the shock model coupled to radiation from the central source predictions 
provide us with only a first-order
approximation. In order to have a full description of the CL region based
on model-fitting, we should suitably combine clouds under different
physical conditions (i.e., clouds with different $V_s$ and $n_0$) with
the proper weight so that the emission lines and the observed continuum
can be reproduced \citep{Contini98,Contini03,ROA05}. 
This is far beyond the scope of this paper
and is left for a future publication. What is most important here
is that we have collected solid evidence of the necessity of shocks 
coupled to photoinization by the AGN in
order to reproduce CL flux ratios at distances
as far as a few hundred parsecs, where photoionization by
the central source predicts no or very faint CL emission. Unrealistic 
physical conditions need to be assumed for the gas if we want to
explain the observed high-ionization lines in terms of
only one mechanism. 
 
\section{Concluding Remarks}
\label{sec:final}

We have carried out the first spectroscopic analysis of the nuclear and
circumnuclear region of the Seyfert~2 galaxy NGC\,4388 covering
simultaneously the wavelength interval 0.84$-$2.45~$\mu$m, part of it never
being reported in the literature. The lower sensitivy of NIR to dust
allowed us to unveil the north-east region of this source, traditionally
hidden at optical wavelengths, in detail and using a wide variety 
of emission lines both of low- and high-ionization. 

Our data reveal that NGC\,4388 displays an outstanding
emission line spectrum with prominent lines of [S\,{\sc iii}]\,$\lambda\lambda$9068,9531~\AA,
He\,{\sc i}\,1.083~$\mu$m, [Fe\,{\sc ii}]\,1.257$\mu$m, Pa$\beta$, 
H$_2$\,2.122~$\mu$m, [Si\,{\sc vi}]\,1.963~$\mu$m and [Si\,{\sc vii}]\,2.48~$\mu$m. 
They all extend from the nucleus to the NE and SW ends of the slit.
Along the spatial direction, the NLR gas has a complex, irregular
structure, with at least two knots of emission, with the brightest one 
coinciding with the peak of the continuum light distribution. The asymmetry 
of the light profile distributions, with excess emission 
towards the SW compared with the NE, 
is most pronounced at shorter wavelengths. Some of the
structures detected are seen at the first time in our
data.

We determined the average extinction for the inner 560~pc
of NGC\,4388 from both the H\,{\sc i} and [Fe\,{\sc ii}] lines. 
The results confirm that the central regions of NGC\,4388 are dusty, with the
dust distributed inhomogeneously, peaking at the nucleus with an E(B-V) of 
1.81$\pm$0.3 mag. It then decreases both to the NE and SW although it remains high
mainly to the NE, consistent with the fact that this inner region
is seen through the galaxy disk. To the SW at 220~pc from the nucleus we see a highly dusty
region with E(B-V) reaching 0.8$\pm$0.3 mag.

The emission line ratios [S\,{\sc iii}]/Pa$\beta$,
[Si\,{\sc vi}]/Br$\gamma$ and [Si\,{\sc vii}]/Br$\gamma$
show that in the inner 500~pc of NGC\,4388 the gas distribution is highly 
inhomogeneous, with at least two regions of enhanced high-ionization emission lines: 
one in the centre, and another at $\sim$250~pc NE of the nucleus. A third peak of enhanced ionization
is observed to the SW, and is evident in the [S\,{\sc iii}]/Pa$\beta$
and [Fe\,{\sc ii}]/Pa$\beta$ line ratios. The enhancement in ionization
suggests strong interaction between the radio-jet 
and the circumnuclear NLR gas.

We are able to analyze the gas kinematics based on several lines of moderate to very 
high excitation. The results show that there is a significant parcel of gas that is 
not rotation-dominated and located out of the disk of the galaxy. High-ionization 
lines display flat PV curves, suggesting that the bulk of this
emission consists of outflowing gas. Double-peak [Fe\,{\sc ii}]
lines SW of the nucleus give additional support to this picture. 
Only the molecular Hydrogen and the stellar component
follow a rotation pattern consistent with disk rotation.  

The forbidden high-ionization line spectrum observed in NGC\,4388 is outstanding, 
not only because of the strength of the emission lines but also because of the size of
the emission region where it is detected. [Si\,{\sc vi}] and [Si\,{\sc vii}],
for example, fill up the field-of-view covered by the slit, that is, nearly 
600~pc along the spatial direction. Few AGNs in the literature are reported to display such a large
coronal line region.  These observations cannot be explained solely as due to 
photoinization by radiation from the central engine. Models that consider only this 
excitation mechanism fail at reproducing off-nuclear line ratios. 
Therefore, an additional source of gas excitation must be present. Models 
that include the effects of shocks coupled to photoionization by the central source are
able to account for the observations. This result is supported by 
observational evidence that suggests interactions between the radio jet and the 
ISM in the central few hundred parsecs of this AGN.

The picture that emerges from our observations highlights the 
very complex nature of the nuclear and circumnuclear region of NGC\,4388. 
This AGN is one of the best pieces of evidence of the 
intricate mixture of an AGN, a radio-jet, dust and a rich ISM.
Determining the precise contribution of each of these components to the
observed spectrum would require observations at superior angular resolution 
as provided by adaptive optics and sub-arsecond observations of the radio
jet with VLA.

\section*{Acknowledgments}

We thank to an anonymous Referee for his/her useful comments and suggestions
that helped to improve this manuscript.
Based on observations obtained at the Gemini Observatory which is operated by the Association of 
Universities for Research in Astronomy, Inc., under a cooperative agreement with 
the NSF on behalf of the Gemini partnership: the National Science Foundation 
(United States), the National Research Council (Canada), CONICYT (Chile), the 
Australian Research Council (Australia), Minist\'{e}rio da Ci\^{e}ncia, Tecnologia e 
Inovac\~{a}o (Brazil) and Ministerio de Ciencia, Tecnolog\'{i}a e Innovaci\'{o}n Productiva 
(Argentina). {\it ARA} acknowledges the Conselho Nacional de Desenvolvimento Cient\'ifico
e Tecnol\'ogico (CNPq) for partial support to this work (grants 307403/2012-2 and 311935/2015-0). {\it RAR} 
acknowledges support from FAPERGS (project N0. 2366-2551/14-0) and CNPq 
(project N0. 470090/2013-8 and 302683/2013-5). LM thanks CNPq (grant 305291/2012-2) and FAPESP (2015/02984-3) for
partial financial support. CRA is supported by a Marie Curie Intra European Fellowship within the 7th
European Community Framework Programme (PIEF-GA-2012-327934). LH is supported by grant 
2016YFA0400702 from the Ministry of Science and Technology of China.

 

\begin{thebibliography}{99}

\bibitem[\protect\citeauthoryear{Allen et~al.}{2008}]{Allen+}
Allen, M. G., Groves, B. A., Dopita, M. A., Sutherland, R. S., \& Kewley, L. J. ApJS, 178, 20

\bibitem[\protect\citeauthoryear{Asari et~al.}{2007}]{asari+07}
Asari N.~V., Cid Fernandes, R., Stasi\'nska G., Torres-Papaqui J.~P., Mateus
  A., et~al. 2007, MNRAS, 381, 263

\bibitem[\protect\citeauthoryear{Axon et al.}{1998}]{axon98}
Axon, D. J., Marconi, A., Capetti, A., Macchetto, F. D., Schreier, E., et al. 1998, ApJL, 496, 75

\bibitem[\protect\citeauthoryear{Bautista et~al.}{2015}]{Bautista+}
Bautista, M. A., Fivet, V., Ballance, C., Quinet, P., Ferland, G. J., et~al. 2015, ApJS, in press.

\bibitem[\protect\citeauthoryear{Bautista \& Pradhan}{1998}]{Bautista98}
Bautista, M. A., \& Pradhan, A. K. 1998, ApJ, 492, 650

\bibitem[\protect\citeauthoryear{Bennert et~al.}{2006}]{bennert+}
Bennert, N., Jungwiert, B., Komossa, S., Hass, M., \& Chini, R. 2006, A\&A, 456, 953

\bibitem[\protect\citeauthoryear{Bianchi et~al.}{2006}]{Bianchi+} 
Bianchi, S., Guainazzi, M., \& Chiaberge, M. 2006, A\&A, 448, 499Vasudevan


\bibitem[\protect\citeauthoryear{Cappellari \& Emsellem}{2004}]{Cappellari+}
Cappellari, M., \& Emsellem, E. 2004, PASP, 116, 138

\bibitem[\protect\citeauthoryear{Cardelli et~al}{1989}]{Cardeli+} 
Cardelli J.~A., Clayton G.~C., \& Mathis J.~S. 1989, ApJ, 345, 245

\bibitem[\protect\citeauthoryear{Cid Fernandes et~al.}{2004}]{cid+04}
Cid Fernandes R., Gonz\'alez Delgado, R. M., Schmitt, H., Storchi-Bergmann, T., \& Martins, L.
 2004, ApJ, 605, 105

\bibitem[\protect\citeauthoryear{Cid Fernandes et~al.}{2005a}]{cid+05a}
Cid Fernandes, R., Mateus A., Sodr\'{e} L., Stasi$\backslash$'nska G., Gomes
  J.~M., 2005a MNRAS, 358, 363

\bibitem[\protect\citeauthoryear{Cid Fernandes et~al.}{2005b}]{cid+05b}
Cid Fernandes, R., Mateus A., Sodr\'{e} L., Stasi$\backslash$'nska G., Gomes
  J.~M., 2005b, MNRAS, 358, 363

\bibitem[\protect\citeauthoryear{Ciroi et~al.}{2003}]{Ciroi+}
Ciroi, S., Contini, M., Rafanelli, P., \& Richter, G. M. 2003, A\&A, 400, 85

\bibitem[\protect\citeauthoryear{Coelho}{2014}]{coelho14}
Coelho, P. R. T. 2014, MNRAS, 440, 1027 

\bibitem[\protect\citeauthoryear{Colina}{1992}]{Colina92} 
Colina, L. 1992, ApJ, 386, 59

\bibitem[\protect\citeauthoryear{Contini et al.}{1998}]{Contini98}
Contini, M., Prieto, M. A., Viegas, S. M. 1998, ApJ, 505, 621

\bibitem[\protect\citeauthoryear{Contini \& Viegas}{2001}]{Contini+}
Contini, M., \& Viegas, S. M. 2001, ApJs, 132, 211

\bibitem[\protect\citeauthoryear{Contini et al.}{2003}]{Contini03}	
Contini, M., Rodr\'{\i}guez-Ardila, A., \& Viegas, S. M. 2003, A\&A, 408, 101	

\bibitem[\protect\citeauthoryear{Corbin et~al.}{1988}]{Corbin+88}
Corbin, M. R., Baldwin, J. A., \& Wilson, A. S. 1988, ApJ, 334, 584

\bibitem[\protect\citeauthoryear{Emonts et al.}{2005}]{Emonts05}
Emonts, B. H. C., Morganti, R., Tadhunter, C. N., Oosterloo, T. A., Holt, J., et al. 2005, MNRAS, 362, 931

\bibitem[\protect\citeauthoryear{Exposito et al.}{2011}]{exposito11}
Exposito, J., Gratadour, D., Cl\'enet, Y., Rouan, D. 2011, A\&A, 533, 63

\bibitem[\protect\citeauthoryear{Fabian}{2012}]{Fabian12} 
Fabian, A. C. 2012, ARA\&A, 50, 455

\bibitem[\protect\citeauthoryear{Falcke et al.}{1998}]{Falcke+} 
Falcke, H., Wilson, A. S., \& Simpson, C. 1998, ApJ, 502, 199

\bibitem[\protect\citeauthoryear{Ferguson et~al.}{1997}]{Ferguson+} Ferguson, J. W., Korista, K. T., \& 
Ferland, G. J. 1997, ApJs, 110, 287

\bibitem[\protect\citeauthoryear{Ferland et al.}{2013}]{fer13+}
Ferland, G. J., Porter, R. L., van Hoof, P. A. M., Williams, R. J. R., Abel, N. P., et~al. 2013, RMxAA, 49, 137 

\bibitem[\protect\citeauthoryear{Forbes \& Ward}{1993}]{forbes+}
Forbes, D. A., \& Ward, M. J. 1993, ApJ, 416, 150

\bibitem[\protect\citeauthoryear{Geballe et~al.}{2009}]{geballe09}
Geballe, T. R.; Mason, R. E.; Rodr\'iguez-Ardila, A.; Axon, D. J.

\bibitem[\protect\citeauthoryear{Greene et~al.}{2013}]{Greene13}
Greene, J. E., Seth, A., den Brok, M., Braatz, J. A., Henkel, C., et~al. 2013, ApJ, 771, 121

\bibitem[\protect\citeauthoryear{Greene et~al.}{2014}]{Greene14}
Greene, J. E., Seth, A., Lyubenova, M., Walsh, J., van de Ven, G., et~al. 2014, ApJ, 788, 145

\bibitem[\protect\citeauthoryear{Grevesse et~al.}{2010}]{grevesse10}
Grevesse, N., Asplund, M., Sauval, A. J., \& Scott, P. 2010, Ap\&SS, 328, 179

\bibitem[\protect\citeauthoryear{Helou et~al.}{1981}]{Helou+}
Helou, G., Salpeter, E. E., \& Krumm, N. 1981, ApJS, 46, 267

\bibitem[\protect\citeauthoryear{Ho et~al}{1997}]{Ho97}
Ho, Luis C., Filippenko, A. V., Sargent, W. L. W. 1997, ApJS, 112, 315

\bibitem[\protect\citeauthoryear{Hummel \& Saikia}{1991}]{Hummel+} 
Hummel, E., \& Saikia, D. J. 1991, A\&A, 249, 43

\bibitem[\protect\citeauthoryear{Imanishi \& Alonso-Herrero}{2004}]{IAH04}
Imanishi, M., \& Alonso-Herrero, A. 2004, ApJ, 614, 1221

\bibitem[\protect\citeauthoryear{Ishibashi \& Fabian}{2012}]{IF12}
Ishibashi, W., \& Fabian, A. C. 2012, MNRAS, 427, 2998

\bibitem[\protect\citeauthoryear{Ishibashi et al.}{2013}]{ifc13}
Ishibashi, W., Fabian, A. C., \& Canning, R. E. A. 2013, MNRAS, 431, 2350

\bibitem[\protect\citeauthoryear{Iwasawa et al.}{2003}]{Iwasawa+} Iwasawa, K., Wilson, A. S., 
Fabian, A. C., \& Young, A. J. 2003, MNRAS, 345, 369

\bibitem[\protect\citeauthoryear{Kraemer \& Crenshaw}{2000}]{kracre00}
Kraemer, S. B., \& Crenshaw, D. M. 2000, ApJ, 544, 763

\bibitem[\protect\citeauthoryear{Knop et~al.}{2001}]{Knop+} 
Knop, R. A., Armus, L., Matthews, K., Murphy, T. W., \& Soifer, B. T. 200, AJ, 122, 764

\bibitem[\protect\citeauthoryear{Kuo et~al.}{2011}]{Kuo+}
Kuo, C. Y., Braatz, J. A., Condon, J. J., Impellizzeri, C. M. V., Lo, K. Y., et~al. 2011, ApJ, 727, 20 

\bibitem[\protect\citeauthoryear{Lu et~al.}{1993}]{Lu+}
Lu, N. Y., Hoffman, G. L., Groff, T., Roos, T., \& Lamphier, C. 1993, ApJS, 88, 383

\bibitem[\protect\citeauthoryear{Lutz et~al.}{2002}]{Lutz+}
Lutz, D., Maiolino, R., Moorwood, A. F. M., Netzer, H., Wagner, S. J., et~al. 2002, A\&A, 396, 439

\bibitem[\protect\citeauthoryear{Maraston}{2005}]{maraston05}
Maraston C. 2005, MNRAS, 362, 799

\bibitem[\protect\citeauthoryear{Martins et~al.}{2010}]{martins+10}
Martins L.~P., Riffel R., Rodr\'{\i}guez-Ardila A., Gruenwald R., \& de~Souza R. 2010, MNRAS, 406, 2168

\bibitem[\protect\citeauthoryear{Martins et~al.}{2013}]{martins+13}
Martins, L. P., Rodr\'{\i}guez-Ardila, A., Diniz, S., Gruenwald, R., de~Souza, R. 2013, MNRAS, 431, 1823

\bibitem[\protect\citeauthoryear{Mason et~al.}{2015}]{Mason15}
Mason, R., Rodr\'{\i}guez-Ardila, A., Martins, L., Riffel, R., Gonz\'alez Mart\'{\i}n, O., et~al. 2015, ApJS, 217, 13

\bibitem[\protect\citeauthoryear{Matt et~al.}{1994}]{Matt+}
Matt, G., Piro, L., Antonelli, L. A., Fink, H. H., Meurs, E. J. A., Perola, G. C. 1994, A\&A, 292, 13

\bibitem[\protect\citeauthoryear{Mathews \& Ferland}{1987}]{mafer87}
Mathews, W. G., \& Ferland, G. J. 1987, ApJ, 323, 456

\bibitem[\protect\citeauthoryear{Mazzalay \& Rodr\'{\i}guez-Ardila}{2007}]{Mazzalay07}
Mazzalay, X., \& Rodr\'{\i}guez-Ardila, A. 2007, A\&A, 463, 445

\bibitem[\protect\citeauthoryear{Mazzalay et~al.}{2013}]{Mazzalay13}
Mazzalay, X., Rodr\'{\i}guez-Ardila, A., Komossa, S., \& McGregor, P. J. 2013, MNRAS, 430, 2411

\bibitem[\protect\citeauthoryear{Mazzalay et~al.}{2015}]{Mazzalay15}
Mazzalay, X., Maciejewski, W., Erwin, P., Saglia, R. P., Bender, R., et~al. 2015, MNRAS, 438, 2036

\bibitem[\protect\citeauthoryear{Mezcua et~al.}{2015}]{Mezcua+}
Mezcua, M., Prieto, M. A., Fern\'andez-Ontiveros, J. A., Tristram, K., Neumayer, N., et al.  2015, MNRAS,
Accepted for publication, arXiv:1506.07289.

\bibitem[\protect\citeauthoryear{Morganti et al.}{2013}]{Morganti13}
Morganti, R., Frieswijk, W., Oonk, R. J. B., Oosterloo, T., Tadhunter, C. 2013, A\&A, 552L, 4

\bibitem[\protect\citeauthoryear{M\"uller-S\'anchez et al.}{2011}]{MullerS+11}   
M\"uller-S\'anchez, F., Prieto, M. A., Hicks, E. K. S., Vives-Arias, H., 
Davies, R. I., et~al. 2011, ApJ, 739, 69

\bibitem[\protect\citeauthoryear{Oliva et~al.}{1994}]{Oliva94}
Oliva, E., Salvati, M., Moorwood, A. F. M., Marconi, A. 1994, A\&A, 291, 18

\bibitem[\protect\citeauthoryear{Osterbrock \& Ferland}{2006}]{Osterbrock06}
Osterbrock, D. E., \& Ferland, G. J. $in$ Astrophysics of gaseous nebulae and active 
galactic nuclei, 2nd. ed. by D.E. Osterbrock and G.J. Ferland. Sausalito, CA. 
University Science Books, 2006

\bibitem[\protect\citeauthoryear{Petitjean \& Durret}{1993}]{Petitjean+} 
Petitjean, P., \& Durret, F. 1993, A\&A, 277, 365

\bibitem[\protect\citeauthoryear{Phillips \& Malin}{1982}]{Phillips+}
Phillips, M. M., \& Malin, D. F. 1982, MNRAS, 199, 905

\bibitem[\protect\citeauthoryear{Pier et~al.}{1994}]{pier94}
Pier, E. A., Antonucci, R., Hurt, T., Kriss, G., \& Krolik, J. 1994, ApJ, 428, 124

\bibitem[\protect\citeauthoryear{Pogge}{1988}]{Pogge88} 
Pogge, R. 1988, ApJ, 322, 702

\bibitem[\protect\citeauthoryear{Pogge \& Owen}{1993}]{Pogge93}
Pogge, R. W., \& Owen, J. M. 1993, OSU Internal Rep. 93-01

\bibitem[\protect\citeauthoryear{Ramos-Almeida et~al.}{2006}]{Ramos-Almeida06}
Ramos Almeida, C., P\'erez Garc\'ia, A. M., Acosta-Pulido, J. A. Rodr\'iguez Espinosa, J. M., Barrena, R.,
et~al. 2006, ApJ, 645, 148 

\bibitem[\protect\citeauthoryear{Ramos-Almeida et~al.}{2009}]{Ramos-Almeida09}
Ramos-Almeida, C., P\'erez Garcia, A. M., Acosta-Pulido, J. A. 2009, ApJ, 694, 1379

\bibitem[\protect\citeauthoryear{Rayner et~al.} {2009}]{rayner+09}
Rayner J.~T., Cushing M.~C., Vacca W.~D. 2009, ApJS, 185, 289

\bibitem[\protect\citeauthoryear{Reynolds \& Fabian}{1995}]{reyfan95}
Reynolds, C. S., \&  Fabian, A. C. 1995, MNRAS, 273, 1167

\bibitem[\protect\citeauthoryear{Riffel et~al.}{2006}]{riffel+06}
Riffel, R., Rodr\'{\i}guez-Ardila, A., Pastoriza, M. G. 2006, A\&A, 457, 61

\bibitem[\protect\citeauthoryear{Riffel et~al.}{2007}]{riffel+07}
Riffel, R., Pastoriza, M. G., Rodr\'{\i}guez-Ardila, A., Maraston, C. 2007, ApJL, 659, 103

\bibitem[\protect\citeauthoryear{Riffel et~al.}{2009}]{riffel+09}
Riffel R., Pastoriza M.~G., Rodr\'iguez-Ardila A., Bonatto C. 2009, MNRAS, 400, 273

\bibitem[\protect\citeauthoryear{Riffel \& Storchi-Bergmann}{2011}]{Riffel11} 
Riffel, Rogemar A., Storchi-Bergmann, T., 2011, ApJ, 411, 469

\bibitem[\protect\citeauthoryear{Riffel et~al.}{2013}]{Riffel13} 
Riffel, Rogemar A., Storchi-Bergmann, T., \& Winge, C. 2013, MNRAS, 430, 2249

\bibitem[\protect\citeauthoryear{Riffel et~al.}{2015a}]{Riffel15} 
Riffel, R., Mason, R. E., Martins, L. P., Rodr\'iguez-Ardila, A., Ho, L. C., et~al. 2015, MNRAS, 450, 3069 

\bibitem[\protect\citeauthoryear{Riffel et~al.}{2015b}]{Riffel15b}
Riffel, Rogemar A., Storchi-Bergmann, Thaisa; \& Riffel, Rog\'erio, MNRAS, 451, 3587

\bibitem[\protect\citeauthoryear{Rodr\'{\i}guez-Ardila et~al.}{2005}]{ROA05}
Rodr\'{\i}guez-Ardila, A., Contini, M., Viegas, S. M. 2005, MNRAS, 357, 220

\bibitem[\protect\citeauthoryear{Rodr\'{\i}guez-Ardila et~al.}{2006}]{ROA06}
Rodr\'iguez-Ardila, A., Prieto, M. A., Viegas, S., Gruenwald, R. 2006, ApJ, 653, 1098

\bibitem[\protect\citeauthoryear{Rodr\'{\i}guez-Ardila et~al.}{2011}]{Rodriguez11}
Rodr\'{\i}guez-Ardila, A., Prieto, M. A., Portilla, J. G., \& Tejeiro, J. M. 2011, ApJ, 743, 100

\bibitem[\protect\citeauthoryear{Santini et~al.}{2012}]{Santini12}
Santini, P., Rosario, D. J., Shao, L., Lutz, D., Maiolino, R., et al. 2012, A\&A, 540, 109

\bibitem[\protect\citeauthoryear{Schlegel et~al.}{1998}]{Schlegel+}
Schlegel, D. J., Finkbeiner, D. P., Davis, M. 1998, ApJ, 500, 525

\bibitem[\protect\citeauthoryear{Schmitt et~al.}{2003}]{Schmitt+}
Schmitt, H. R., Donley, J. L., Antonucci, R. J., Hutchings, J. B., \& Kinney, A. L. 2003, ApJS, 148, 327

\bibitem[\protect\citeauthoryear{Schulz \& Komossa.}{1993}]{schulz+}
Schulz, H., \& Komossa, S. 1993, A\&A, 278, 29

\bibitem[\protect\citeauthoryear{Shirai et~al.}{2008}]{shirai08+}
Shirai, H., Fukazawa, Y., Sasada, M., Ohno, M., Yonetoku, D. et~al. 2008, PASJ, 60, 263.

\bibitem[\protect\citeauthoryear{Stone et~al.}{1988}]{Stone+}
Stone, J. L. Jr., Wilson, A. S., Ward, M. J. 1988, ApJ, 330, 105

\bibitem[\protect\citeauthoryear{Storchi-Bergmann et al.}{2010}]{Storchi-Bergmann10} 
Storchi-Bergmann, T., Lopes, R. D. Sim\~oes, McGregor, P. J., Riffel, Rogemar A., 
et~al. 2010, MNRAS, 402, 819

\bibitem[\protect\citeauthoryear{van der Laan et~al.}{2013}]{vlaan+}
van der Laan, T. P. R., Schinnerer, E., B\"oker, T., Armus, L. 2013, A\&A, 560, 99 

\bibitem[\protect\citeauthoryear{Vasudevan et~al.}{2013}]{vasude13}
Vasudevan, R. V., Brandt, W. N., Mushotzky, R. F., Winter, L. M., Baumgartner, W. H, et~al. 2013, ApJ, 763, 111.

\bibitem[\protect\citeauthoryear{Veilleux et~al.}{1991}]{Veilleux91}
Veilleux, S. 1991, ApJ, 369, 331 

\bibitem[\protect\citeauthoryear{Veilleux et~al.}{1999}]{Veilleux99} 
Veilleux, S., Bland-Hawthorn, J., \& Cecil, G. 1999, AJ, 118, 2108

\bibitem[\protect\citeauthoryear{Veilleux et~al.}{2001}]{Veilleux01}
Veilleux, S., Shopbell, P. L., \&  Miller, S. T. 2001, AJ, 121, 198

\bibitem[\protect\citeauthoryear{Contini \& Viegas}{2001}]{CV01}
Contini, M., \& Viegas, S. M. 2001, ApJS, 137, 75

\bibitem[\protect\citeauthoryear{Winge et al.}{2000}]{Winge00} 
Winge, C., Storchi-Bergmann, T., Ward, M. J., \& Wilson, A. S. 2000, MNRAS, 316, 1 

\bibitem[\protect\citeauthoryear{Winge et~al.}{2009}]{Winge09}
Winge, C., Riffel, R. A., Storchi-Bergmann, T. 2009, ApJS, 185, 186

\bibitem[\protect\citeauthoryear{Yoshida et~al.}{2002}]{Yoshida02}
Yoshida, M., Yagi, M., Okamura, S., Aoki, K., Ohyama, Y., et~al. 2002, ApJ, 567, 118

\bibitem[\protect\citeauthoryear{Yoshida et~al.}{2004}]{Yoshida04}
Yoshida, M., K., Ohyama, Y., Iye, M., Aoki, K., Kashikawa, N. et~al. 2004, ApJ, 127, 90


 
\end{thebibliography}
\end{document}